%% file: Gotbergetal18.tex
\documentclass{aa-package/aa}  

\usepackage{graphicx}
\usepackage[labelfont=bf,font=small]{caption}
\usepackage{subcaption}
\usepackage{amsmath}
\usepackage{amssymb}
\usepackage{txfonts}

\usepackage{fix2col}
\usepackage{ulem}

\usepackage{color}
\usepackage{colortbl} 
\definecolor{pinegreen}{RGB}{1, 121, 111}
\definecolor{salmon}{RGB}{255,160,122}
\usepackage[colorlinks,citecolor=blue]{hyperref}

\usepackage[hyphenbreaks]{breakurl}
\usepackage{natbib}

\definecolor{c1}{RGB}{91, 44, 111}
\definecolor{c2}{RGB}{13, 71, 161}
\definecolor{c3}{RGB}{14, 102, 85}

\definecolor{ylvacolour}{rgb}{.55,.4,.75}

\newcommand{\Msun}{\ensuremath{\,M_\odot}\xspace}
\newcommand{\Rsun}{\ensuremath{\,R_\odot}\xspace}
\newcommand{\Lsun}{\ensuremath{\,L_\odot}\xspace}
\newcommand{\kms}{\ensuremath{\,\rm{km}\,\rm{s}^{-1}}\xspace}
\newcommand{\Msunyr}{\ensuremath{\,M_\odot\,\rm{yr}^{-1}}\xspace}

\renewcommand\eqref[1]{\ifnum\ifhmode\spacefactor\else2000\fi>1000 Equation~\ref{#1}\else Eq.~\ref{#1}\fi}
\newcommand\eqreftwo[2]{\ifnum\ifhmode\spacefactor\else2000\fi>1000 Equations~\ref{#1} and \ref{#2}\else Eqs.~\ref{#1} and \ref{#2}\fi}

\newcommand\figref[1]{\ifnum\ifhmode\spacefactor\else2000\fi>1000 Figure~\ref{#1}\else Fig.~\ref{#1}\fi}
\newcommand\figreftwo[2]{\ifnum\ifhmode\spacefactor\else2000\fi>1000 Figures~\ref{#1} and \ref{#2}\else Figs.~\ref{#1} and \ref{#2}\fi}
\newcommand\figrefthree[3]{\ifnum\ifhmode\spacefactor\else2000\fi>1000 Figures~\ref{#1}, \ref{#2} and \ref{#3}\else Figs.~\ref{#1}, \ref{#2} and \ref{#3}\fi}

\newcommand\secref[1]{\ifnum\ifhmode\spacefactor\else2000\fi>1000 Section~\ref{#1}\else Sect.~\ref{#1}\fi}
\newcommand\secreftwo[2]{\ifnum\ifhmode\spacefactor\else2000\fi>1000 Sections~\ref{#1} and \ref{#2}\else Sects.~\ref{#1} and \ref{#2}\fi}
\newcommand\secrefthree[3]{\ifnum\ifhmode\spacefactor\else2000\fi>1000 Sections~\ref{#1}, \ref{#2} and \ref{#3}\else Sects.~\ref{#1}, \ref{#2} and \ref{#3}\fi}

\newcommand\appref[1]{\ifnum\ifhmode\spacefactor\else2000\fi>1000 Appendix~\ref{#1}\else Appendix~\ref{#1}\fi}

\newcommand\tabref[1]{\ifnum\ifhmode\spacefactor\else2000\fi>1000 Table~\ref{#1}\else Table~\ref{#1}\fi}
\newcommand\tabreftwo[2]{\ifnum\ifhmode\spacefactor\else2000\fi>1000 Tables~\ref{#1} and \ref{#2}\else Tables~\ref{#1} and \ref{#2}\fi}
\newcommand\tabrefthree[3]{\ifnum\ifhmode\spacefactor\else2000\fi>1000 Tables~\ref{#1}, \ref{#2} and \ref{#3}\else Tables~\ref{#1}, \ref{#2} and \ref{#3}\fi}

\newcommand{\YG}[1]{{{\color{black} #1}}}  

\newcommand{\rev}[1]{{\color{black} #1}}

\definecolor{darkcyan}{rgb}{0.0, 0.55, 0.55}
\definecolor{amethyst}{rgb}{0.6, 0.4, 0.8}

\usepackage{booktabs}    

\defcitealias{2017A&A...608A..11G}{Paper~I}
\usepackage{morefloats}

\bibpunct{(}{)}{;}{a}{}{,}

\begin{document} 

     \title{Spectral models for binary products: Unifying Subdwarfs and Wolf-Rayet stars as a sequence of stripped-envelope stars}
 \titlerunning{Unifying Subdwarfs and Wolf-Rayet stars as a sequence of stripped-envelope stars }

   \author{Y.~G\"{o}tberg$^{1,5}$, S.~E.~de Mink$^{1,5}$, J.~H.~Groh$^2$, T.~Kupfer$^3$, P.~A.~Crowther$^4$, E.~Zapartas$^1$ \and M.~Renzo$^{1,5}$}
   
   \authorrunning{G\"{o}tberg, De Mink \& Groh et al.}

   \institute{
            Anton Pannekoek Institute for Astronomy, University of Amsterdam, 1090 GE Amsterdam, The Netherlands \\
              \email{Y.L.L.Gotberg@uva.nl, S.E.deMink@uva.nl}
                       \and
   School of Physics, Trinity College Dublin, The University of Dublin, Dublin 2, Ireland
    \\
    \email{jose.groh@tcd.ie}
    \and
    Division of Physics, Mathematics and Astronomy, California Institute of Technology, Pasadena CA 91125, USA
    \and
    Department of Physics and Astronomy, University of Sheffield, Hounsfield Road, Sheffield S3 7RH, UK
    \and
    Kavli Institute for Theoretical Physics, University of California, Santa Barbara, CA 93106, USA
             }

   \date{\textit{Accepted for publication in A\&A}}
 
  \abstract
   {
Stars stripped of their hydrogen-rich envelope through interaction with a binary companion are generally not considered when accounting for ionizing radiation from stellar populations, despite the expectation that stripped stars emit hard ionizing radiation, form frequently and live $10 - 100$ times longer than single massive stars. We compute the first grid of evolutionary and spectral models specially made for stars stripped in binaries for a range of progenitor masses (2-20\Msun) and metallicities ranging from solar to values representative for pop II stars.

For stripped stars with masses in the range 0.3-7\Msun, we find consistently high effective temperatures (20 000-100 000 K, increasing with mass),  small radii (0.2-1\Rsun) and high bolometric luminosities, comparable to that of their progenitor before stripping. The spectra show a continuous sequence that naturally bridge subdwarf-type stars at the low mass end and Wolf-Rayet like spectra at the high mass end. For intermediate masses we find hybrid spectral classes showing a mixture of absorption and emission lines. These appear for stars with mass loss rates of $ 10^{-8}-10^{-6} \Msunyr$, which have semi-transparent atmospheres. At low metallicity, substantial hydrogen-rich layers are left at the surface and we predict spectra that resemble O-type stars instead.  
We obtain spectra undistinguishable from subdwarfs for stripped stars with masses up to 1.7\Msun, which questions whether the widely adopted canonical value of 0.47\Msun is uniformly valid. 

Only a handful of stripped stars of intermediate mass have currently been identified observationally. Increasing this sample will provide necessary tests for the physics of interaction, internal mixing and stellar winds. We use our model spectra to investigate the feasibility to detect stripped stars next to an optically bright companion and recommend systematic searches for their UV excess and possible emission lines, most notably HeII~$\lambda 4686$ in the optical and HeII~$\lambda 1640$ in the UV. Our models are publicly available for further investigations or inclusion in spectral synthesis simulations.
   } 


   \maketitle
%

\section{Introduction}
Massive stars are important for many fields of astrophysics, for example by providing mechanical, chemical and radiative feedback on galactic scales through stellar winds, outflows and supernovae \citep{2001PhR...349..125B, 2003ApJ...591..288H, 2011ARA&A..49..373B, 2014MNRAS.445..581H, 2016MNRAS.458..988R}, and as progenitors to neutron stars and black holes \citep[e.g.][]{1999ApJ...522..413F, 2011ApJ...730...70O, 2016ApJ...821...38S}. 
Observational surveys show that the vast majority of young massive stars orbit so close to a companion that interaction will be inevitable as the stars evolve and expand \citep{2007ApJ...670..747K, 2014ApJS..213...34K, 2012MNRAS.424.1925C, 2012Sci...337..444S, 2015A&A...580A..93D, 2017ApJS..230...15M, 2017A&A...598A..84A}. Binary interaction can therefore not be ignored when considering the evolution of massive stars, neither individually nor in stellar populations.  

Binary interaction can give rise to variety of exotic phenomena such as  X-ray binaries \citep{2006csxs.book..623T, 2017A&A...604A..55M} and double compact objects, which may emit a burst of gravitational waves when they coalesce \citep[see e.g.][]{2007PhR...442...75K, 2015ApJ...814...58D, 2017NatCo...814906S, 2017ApJ...846..170T}.  In addition to these very spectacular but rare phenomena, binary interaction also produces a variety of stellar objects that are expected to be rather common. These include (a) stripped-envelope stars that have lost most of their hydrogen-rich envelope through Roche-lobe overflow and (b) rejuvenated stars have accreted mass from their companion and (c) long-lived stellar mergers \citep[e.g.][]{van-Bever+1998, de-Mink+2014}. The stripped-envelope stars are arguably the best understood and are the primary topic of this work.

The most common type of interaction, expected for about a third of all massive stars, is mass transfer when the most massive star in the binary system crosses the Hertzsprung gap \citep[so called case~B mass transfer,][see also \citealt{2012Sci...337..444S}]{1967ZA.....65..251K}. 
%
%
After interaction, the hot helium core is exposed and left with only a thin layer of hydrogen on top \citep[see e.g.][]{2010ApJ...725..940Y, 2017ApJ...840...10Y, 2011A&A...528A.131C}. The main source of energy is fusion of helium through the triple alpha reaction in the center. This phase is long-lived, it accounts for about  $10\%$ of the total stellar lifetime.  Eventually these stars are expected to end their lives as stripped-envelope core-collapse supernovae, if they are massive enough. The high rate of such stripped-envelope supernovae, accounting for about a third of all core-collapse supernovae in volume-limited surveys  \citep[e.g.][]{2009ARA&A..47...63S, 2017ApJ...837..121G}, provides an independent indication that envelope stripping is a  common phenomenon \citep{1992ApJ...391..246P, 2013MNRAS.436..774E}. 

Stripped stars are thought to be very hot objects, emitting the majority of their photons in the extreme ultraviolet \citep{2017A&A...608A..11G}. This, in combination with the prediction that they form  frequently and are long-lived, make them interesting as stellar sources of ionizing radiation in nearby stellar population, but possibly also at high redshift during the epoch of reionization. To account for the effect of stripped stars on the integrated spectra of stellar populations, reliable models are needed for these stripped stars and their atmospheres. Atmosphere models are publicly available for stripped stars of very low mass, which are known as subdwarfs \citep[e.g.][]{2007MNRAS.380.1098H}. They are also available for the high mass end where stripped stars are indistinguishable from Wolf-Rayet stars \citep{2002A&A...387..244G}. However, atmosphere models for stripped stars in the intermediate mass regime, between $\sim 1 - 8$\Msun, have not been explored systematically. Exploring this part of the parameter space and providing a grid of appropriate models in this mass range is the primary aim of this work.

Many efforts have been made to model the radiation from populations containing single stars using spectral synthesis codes such as Starburst99 \citep{1999ApJS..123....3L, 2014ApJS..212...14L} and GALAXEV \citep{2003MNRAS.344.1000B}. Several groups have undertaken efforts to include the effects of binary interaction with increasing levels of sophistication over time.  These include the Brussels simulations \citep{2003A&A...400...63V, 2003A&A...400..429B, 2007ApJ...662L.107V} and the Yunnan simulations \citep{2004A&A...415..117Z, 2010Ap&SS.329...41H, 2010Ap&SS.329..277C, 2012MNRAS.421..743Z, 2012MNRAS.424..874L, 2015MNRAS.447L..21Z}. The BPASS models by \citet{2009MNRAS.400.1019E, 2012MNRAS.419..479E} and \citet{2017PASA...34...58E} have recently gained popularity by enforcing the link between stellar physics and cosmology. These simulations show that binary interaction can boost the ionizing output from stellar populations by about two orders of magnitude at an age of 30~Myr after starburst \citep{2016MNRAS.456..485S, 2016MNRAS.459.3614M}. 

Stars that are stripped in binaries are typically accounted for in the spectral population synthesis codes mentioned above. However, given the lack of applicable atmosphere models for stripped stars, various approximations have been used. These include the use of blackbody approximations and rescaling of spectral models that were originally intended for single stars, for example the use of rescaled models that were originally intended for Wolf-Rayet stars \citep{Hamann+2003, Sander+2015}. 


\rev{While theory predicts that many stripped stars in the mass range $\sim 1-8 \Msun$ exist with OB-type companions, observations have revealed only a handful objects so far.} Four subdwarf type stars with Be-type companions have been observed \citep[$\varphi$~Persei, FY~CMa, 59~Cyg, and 60~Cyg, see][respectively]{1998ApJ...493..440G, 2008ApJ...686.1280P, 2013ApJ...765....2P, 2017ApJ...843...60W}\footnote{The detection of another twelve candidate Be+sdO type systems has been reported by \citet{2018ApJ...853..156W} after finalizing this manuscript.}, and one more massive stripped star with a late B-type companion has been confirmed \citep[the quasi-WR star in HD~45166,][]{2005A&A...444..895S, 2008A&A...485..245G}. This scarcity of detected systems containing a stripped star poses an (apparent) paradox.  This may be simply explained as the result of biases and selection effects in the samples that are currently available, as argued by \citet{de-Mink+2014} and \citet{abels_paper}. However, the resolution of this paradox is not yet clear at present and requires a more careful assessment, which will require efforts on both the observational and theoretical side. In this work we will assess two promising strategies that can be used to increase the number of detected stripped stars. 

Increasing the sample of observed post-interaction binaries will allow also for valuable tests of the physics of binary interaction.  
Spectral features, orbital solutions and surface parameters (e.g., effective temperature, surface gravity and composition) of a large sample of stripped stars will provide insight in long-standing questions. Examples are: (1) how large is convective overshooting? \citep[e.g.][]{1976A&A....47..389M, 1997MNRAS.285..696S, 2007A&A...475.1019C}, (2) how efficient is mass transfer in binaries? \citep[e.g.][]{1981A&A...102...17P,2007A&A...467.1181D}, and (3) how does wind mass loss from hot and hydrogen-deficient stars work? \citep{2008A&ARv..16..209P, 2014ARA&A..52..487S, 2017A&A...607L...8V}. A sample of observed stripped stars also allows for an assessment of the initial conditions for populations of stars (for example binary fraction and the initial period and mass ratio distribution). 



The aim of this work is to provide tailor-made atmosphere models for stars that have been stripped by  interaction with a companion. We then aim to use these models to learn about their structure and spectral properties and assess observing strategies that can be used to increase the sample of known stripped stars. Their ionizing properties will be described and discussed in a companion paper.  This paper expands on the study of \citetalias{2017A&A...608A..11G} by covering a large range of mass, which overlaps with both subdwarfs and WR stars.  


We structure the paper as follows. In \secref{sec:modeling} we describe the binary stellar evolution code MESA, the radiative transfer code CMFGEN and how we compute spectra using the surface properties from the structure models. \secref{sec:evol_MESA} describes the evolutionary models of stripped stars. \secref{sec:spec_models} describes the spectral properties of stripped stars. In \secref{sec:observability} we use our models to asses observing strategies. In \secref{sec:uncertainties} we discuss the primary uncertainties. In \secref{sec:summary} we summarize and present our conclusions.


\section{Modeling}\label{sec:modeling}

The evolution of binary systems is complex and can occur through a variety of channels. 
We focus on the most common channel that produces long-lived stripped stars, since these may greatly affect the appearance of stellar populations by boosting the ionizing output. Stripped stars are created when the most massive star fills its Roche lobe shortly after completing its main sequence evolution, but still before the onset of helium burning. The type of mass transfer that follows is Case~B mass transfer in the original classification scheme by \citet{1967ZA.....65..251K}, see also \citet{1971ARA&A...9..183P}. The stripped stars resulting from this type of interaction are long lived since they have not yet completed central helium burning, which takes about 10\% of their total lifetime. 

To model the spectra of stars that have been stripped by a binary companion we use the same general approach as we used for \citetalias{2017A&A...608A..11G}. First, we follow the evolution of the progenitor star and its interaction with a companion using the stellar evolutionary code MESA, as described in \secref{sec:MESA_modeling}. This step provides us with a model for the interior structure of the stripped star and the surface properties. We then use these as input for radiative transfer simulations of the atmosphere, which allows us to compute the emerging stellar spectrum. For the second step we use the radiative transfer code CMFGEN, as described in \secref{sec:CMFGEN_modeling}. In \secref{sec:tailor}, we provide a brief description of how we connect the atmosphere models to the interior structure models.

\subsection{Evolution of the progenitor with MESA}\label{sec:MESA_modeling}

We use the open source stellar evolution code MESA \citep[version 8118,][]{2011ApJS..192....3P, 2013ApJS..208....4P, 2015ApJS..220...15P} to model the evolution of stars stripped through Roche-lobe overflow. This code solves the 1D equations of stellar structure for two stars in a binary system, while accounting for their evolutionary changes and interactions between. 
Here we briefly summarize the code and physical assumptions that we adopt. The setup of our grid of models and their initial conditions are described in \secref{sec:evol_MESA}. The code as well as our inlists, i.e.\ files that specify all the parameter settings that we used, are available online\footnote{\rev{The code is available at \url{mesa.sourceforge.net} and our inlists (input files with the parameter settings) at \url{mesastar.org}}.}.


 

\paragraph{Nuclear burning, mixing and diffusion}


We adopt a 49 isotope nuclear network (\texttt{mesa\_49.net}), which is appropriate for computations from the onset of central hydrogen burning until central carbon depletion, which we define as the moment in time when the central carbon mass fraction drops below $X_{\text{C, c}} = 0.02$. The evolution up to this stage corresponds to 99.9\% of the stellar lifetime, which is more than sufficient for the purposes of this work. Here, we are interested in the spectra of central helium burning stars and we have therefore ensured that all models have at least been completed until half-way helium burning $X_{\text{He, c}} = 0.5$, even though most of our models reach all the way to carbon depletion.

We account for mixing by convection using the mixing-length approach \citep{1958ZA.....46..108B} adopting a mixing-length parameter $\alpha _{\text{MLT}} = 2$. The precise choice of the value for this parameter has little effect for the convective regions in the deep interiors, where convective mixing is very efficient. It does matter for convective regions at or near the surface that are typical for cooler stars than we consider here. The value we adopt provides a decent fit for a solar model against the radius of the sun \citep[e.g.][]{1998MNRAS.298..525P}.  

We account for mixing in convective and semi-convective regions assuming that semiconvective mixing is efficient by setting $\alpha _{\text{sem}} = 1$ \citep{1983A&A...126..207L, 1991A&A...252..669L}. We also account for overshooting, i.e.\ mixing of regions above every convective burning region following \citet[][]{2011A&A...530A.115B}. These authors use the classical step overshooting formulation and find that an overshooting length of 0.335 pressure scale heights reproduces the width of the main sequence for an observed sample early-B type stars. The stars in their sample have inferred initial masses in the range 10--15\Msun, which makes their calibration a suitable choice for part of our model grid. For the lowest mass progenitors that we consider this overshooting parameter may lead to over estimating the core size (or equivalently, under estimating the initial progenitor mass that belongs to a stripped star with a given mass). We will return to this briefly in \secref{sec:observability}. 

We also account for the effects of thermohaline mixing \citep{1980A&A....91..175K} and rotational mixing \citep{2013ApJS..208....4P}, but find no indication that these processes are significant for the evolution of stripped stars. This is expected since the progenitor stars of the stripped stars that we consider are not expected to be fast rotators, neither before nor after the stripping process \citep[e.g.][]{2013ApJ...764..166D, 2015A&A...580A..92R} and they are not expected to develop inversions of the mean molecular weight gradient that can trigger thermohaline mixing. 

We take into account the effect of microscopic diffusion and gravitational settling following \citet{1994ApJ...421..828T} using atomic diffusion coefficients from \citet{1986ApJS...61..177P}. These effects are only expected to play a role for the lower mass stars in our grid, which produce compact and long-lived subdwarfs \citep{2016PASP..128h2001H}. We therefore consider this for our stellar evolution models with initial masses less than $M_{\text{init}} = 9 \Msun$. We use the iterative solver \texttt{ros2\_solver} \citep[for details see][]{2011ApJS..192....3P}.

We do not account for radiative levitation \citep[e.g.][]{1998ApJ...492..833R}. This has been proposed to be of importance to explain the very sensitive asteroseismological measurements for pulsating subdwarfs \citep{2008A&A...486L..39F}. However, this is not expected to have a major importance for the structure and thus neither for the temperature and radius. After the completion of the computations for this work, new updates have become available for the treatment of diffusion and levitation \citep{2017arXiv171008424P}. Test computations with the newer version of MESA show indeed that the effects are minimal (E.\ Bauer, priv.\ communication). We also do not account for any further forms of weak turbulent mixing, such as proposed by \citet{2011MNRAS.418..195H}. We do however artificially adopt a minimum floor for the helium abundance when computing the atmosphere models, consistent with observed abundances of subdwarfs (\secref{sec:tailor}).

\paragraph{Binary interaction}
We account for binary interaction as described in \citep{2015ApJS..220...15P}. We account for the effects  of tides  \citep{1981A&A....99..126H}. Roche-lobe overflow is treated with the implicit mass transfer scheme by \cite{1988A&A...202...93R}. We consider non-conservative mass transfer by following the response of the spin of the accreting star as it accretes mass and angular momentum \citep[e.g.][]{2013ApJ...764..166D}. When it it is spun up to critical rotation, we prevent further accretion. In practice this means that the secondary only accretes a very small fraction of the mass that is transferred \citep{1981A&A...102...17P} and that most of the mass is lost from the system. We assume that this material leaves the system as a fast outflow from the accreting star, by making the standard assumption that it has a specific angular momentum equal to that of the orbit of the accreting star (see Appendix A.3.3 of \citealt{2013ApJ...764..166D} and \citealt{2015ApJS..220...15P}). In a few cases where we experienced difficulties to converge the models, we replaced the secondary star by a point mass assuming conservative mass transfer. 

Although the efficiency of mass and angular momentum transfer and the response of the accreting star constitutes one of the primary uncertainties in binary evolutionary models \citep[e.g.][and references therein]{2007A&A...467.1181D}, this has almost no effect on the results presented here. The predicted properties of the resulting stripped stars considered in this study, are remarkably insensitive to the detailed assumptions regarding the treatment of mass transfer, see \citetalias{2017A&A...608A..11G} and references therein. (This may no longer hold at very low metallicities, see \citetalias{2017A&A...608A..11G} and \citealt{2017ApJ...840...10Y}.)


\paragraph{Stellar winds}
We account for the effects of stellar wind mass loss using the the wind schemes of \cite{2001A&A...369..574V} and \cite{1988A&AS...72..259D} for the progenitor stars as they evolve on the main sequence stars and early Hertzsprung gap. The assumptions for the stellar winds of the progenitor do not affect the results in this study, because they only change the mass and separation of the progenitor stars in the binary by at most a few percent in the mass range we consider here \citep{2017A&A...603A.118R}. The uncertain mass transfer rate and the efficiency of accretion have a much larger impact on our calculations, since they determine the outer structure and composition of the post-RLOF stripped star.
Conversely, the wind mass loss rate of the stripped star has an important impact.
Empirical constraints are still scarce since very few stripped stars have been observed so far. The situation will certainly improve in the near future as more stripped stars will be identified, but for now we have to rely on theoretical predictions and extrapolations of existing mass loss recipes. 

We adopt the following approach to account for stellar winds. For stripped stars with surface hydrogen mass fraction $X_{\text{H,s}} < 0.4$ and initial masses $M_{\text{init}} > 6\Msun$ we apply an extrapolation of the empirically derived wind mass loss scheme for Wolf-Rayet stars by \cite{2000A&A...360..227N}. This provides a good match with the observed stripped star in HD~45166 \citep{2008A&A...485..245G}. For stripped stars with surface hydrogen mass fraction $X_{\text{H,s}} > 0.4$ and initial masses $M_{\text{init}} > 6\Msun$ we use the main sequence mass loss scheme of \cite{2001A&A...369..574V}. For stars with initial masses $M_{\text{init}} < 6 \Msun$ we use the subdwarf mass loss scheme of \cite{2016A&A...593A.101K}. We apply the \citet{2016A&A...593A.101K} scheme when the stars have surface temperatures higher than 15\,000~K and radius smaller than 1.5\,\Rsun.   

We note that the mass loss rate for stripped stars is a source of uncertainty. See for example  \citetalias{2017A&A...608A..11G} for the effect of wind mass loss variation on the spectra. We also refer to \citet{2017A&A...607L...8V} who recently presented new theoretical predictions for a grid of stripped stars at fixed temperature. 
\rev{We shall return to this topic} in \secref{sec:uncertainties}.

\paragraph{Spatial and temporal resolution}
We vary the spatial and temporal resolution slightly during the stellar evolution to reach convergence as small changes can help the code to find a solution. In MESA jargon, we allow spatial resolution variations using $0.5 < $ \texttt{mesh\_delta\_coeff} $ < 2.5$, and temporal resolution variations using $10^{-5} < $ \texttt{varcontrol\_target} $ < 10^{-4}$. The variations are well within the default settings for massive stellar evolution in MESA \citep{2011ApJS..192....3P}.

\subsection{Stellar atmospheres with CMFGEN}\label{sec:CMFGEN_modeling} 

We model the spectra by using the publicly available\footnote{The code can be obtained from \url{http://kookaburra.phyast.pitt.edu/hillier/web/CMFGEN.htm} at the time of writing.} 1D radiative transfer code CMFGEN \citep[version of 2014,][]{1990A&A...231..116H, 1998ApJ...496..407H}. This code allows us to solve the equations for radiative transfer and statistical equilibrium for a spherically symmetric outflow. It accounts for effects of line blanketing and deviations from local thermodynamic equilibrium (non-LTE). CMFGEN has been originally designed to model the hot and (partially) optically-thick outflows in the atmospheres of O stars, WR stars and Luminous Blue Variables. Stripped stars cover similar temperatures and the more luminous stripped stars in our grid have WR properties. The code is therefore suitable for our purposes without major adaptations. 


%
We compute the atmospheric temperature and density structure as a function of radius from the stellar surface using a minimum of 40 mesh points, which is sufficient for convergence according to \citet{Hillier+1999}. We specify the luminosity, surface abundances and mass loss rate using the values that we derived with our MESA simulations and use CMFGEN to iteratively find a solution for the atmosphere that matches the structure model at an optical depth of $\tau=20$, as we describe in \secref{sec:tailor}.  We consider the atomic elements H, He, C, N, O, Fe and Si. We compute the spectral energy distribution as well as the normalized spectrum in the wavelength range $50$ \AA\ $< \lambda < 50\,000$~\AA .

We specify the mass-loss rate and velocity law above the sonic point. We assume that the wind velocity scales with the radius, $r$, according to a modified $\beta$-law (see manual at \url{http://kookaburra.phyast.pitt.edu/hillier/web/CMFGEN.htm}), which requires specifying the terminal wind speed, $v_{\infty}$, the stellar radius, $R_{\star}$ at an optical depth of $\tau = 100$, and a parameter $\beta$, which is set to 1. 
 We assume $v_{\infty} = 1.5 \times v_{\text{esc}}$, where $v_{\text{esc}}$ is the surface escape speed \citep{1999isw..book.....L}. This expression results in wind speeds between $1000 \kms \lesssim v_{\infty} \lesssim 2500$\kms at solar metallicity. 

We take into account the effect of wind clumping \citep{1988ApJ...335..914O} by using a volume filling factor $f_{\text{vol}} = 0.5$ for models with initial masses $M_{\text{init}} < 14 \Msun$ \citep[motivated by the parameters derived for the observed stripped star in the HD~45166 binary system,][]{2008A&A...485..245G}. For higher initial masses we use $f_{\text{vol}} = 0.1$ \citep[which is considered appropriate for WR stars,][]{Hillier+1999}. For more details we refer to \citet{2008A&A...485..245G}. 

We do not account for (soft) X-rays generated by shocks embedded in the wind \citep[e.g.][]{Cassinelli+2001}, since we do not expect these to be important for stripped stars, which generally have low mass loss rates, see however discussions by for example \citet{Feldmeier+1997}, \citet{Owocki+2013} and \citet{Cohen+2014}. 

To avoid convergence issues when computing the atmospheres for stars with extremely low mass loss rates, we enforce a minimum wind mass loss rate of $\dot{M}_{\text{wind, min}} \equiv 10^{-12} \Msunyr$ when computing our models for the atmospheres. This does not affect our results since such low wind mass loss rates do not affect the shape of the spectrum or the spectral features.  

We also encountered issues resulting from near zero helium abundances at the surface of our  MESA models for our lowest mass subdwarfs. These are the consequence of the treatment of gravitational settling in MESA, which causes helium to sink and hydrogen to float to the surface. There is substantial evidence for this process to be important, but the MESA models over estimate the effects since we are lacking a proper prescription for processes that can partially counteract the effects, such as for example weak turbulent mixing as proposed by \citep{2011MNRAS.418..195H}. Furthermore, such low helium abundances are not consistent with observations \citet{Edelmann+2003}. We therefore impose a minimum surface helium mass fraction in our CMFGEN simulations, a minimum of $\log _{10} (n_{\text{He}}/n_{\text{H}}) = -3$ for stripped stars with  $T_{\text{eff}} < 26\,000$~K and a minimum of $\log _{10} (n_{\text{He}}/n_{\text{H}}) = -2$ for stripped stars with $26\,000 < T_{\text{eff}} < 35\,000$~K. This follows the trend of observed helium abundances in the work by \citet{Edelmann+2003}. We find no differences in test runs with less stringent limits, since the helium features resulting from imposing this lower limit are very weak already.

\subsection{Connection between stellar structure and atmosphere}\label{sec:tailor}
 
\begin{figure}
\centering
\includegraphics[width=\hsize]{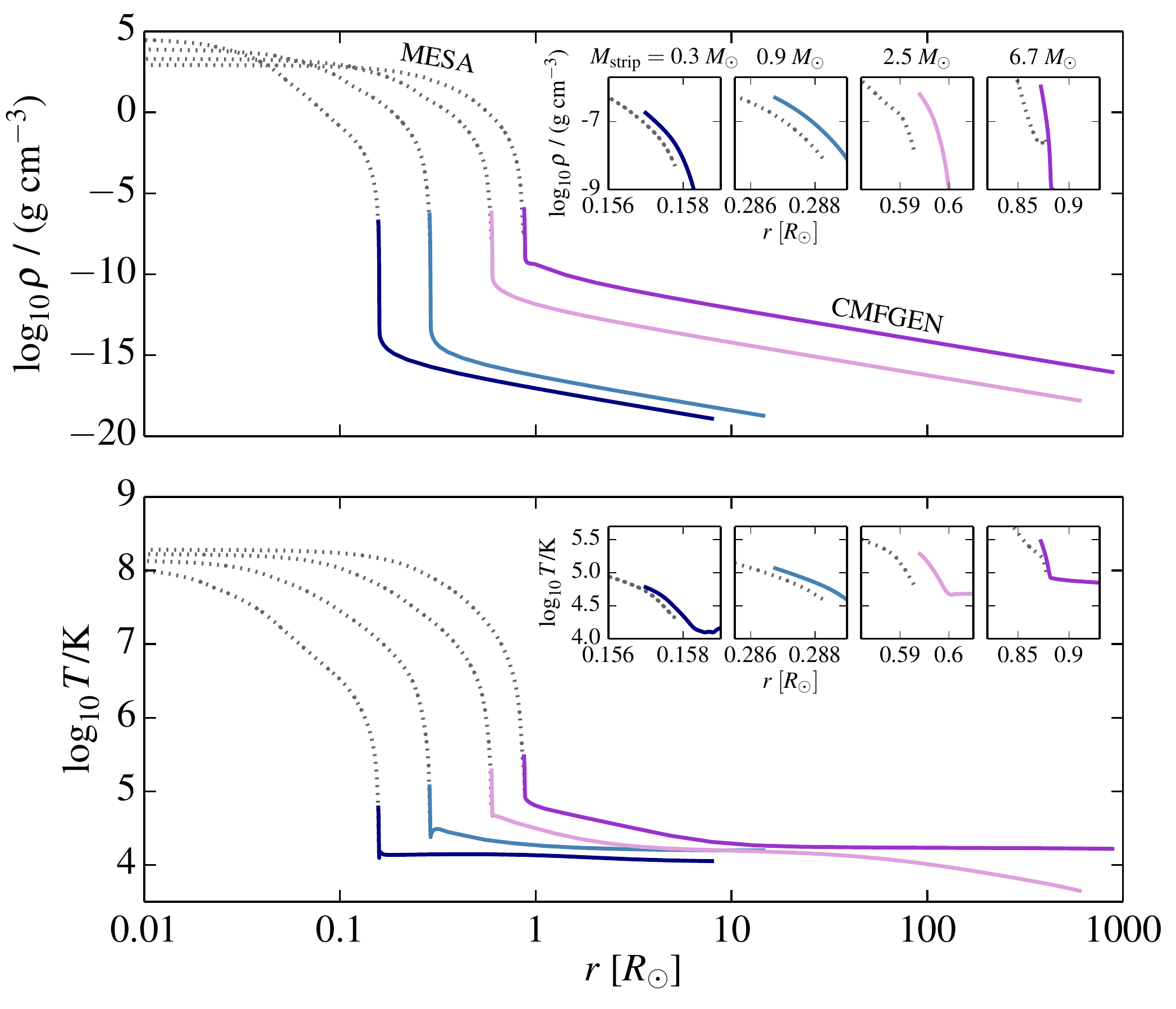}
\caption{Density (top panel) and temperature (bottom panel) structure shown with the radial coordinate for the four stripped stars with masses 0.3, 0.8, 2.5 and 6.7\Msun (corresponding to initial masses of 2, 4, 9 and 18\Msun respectively). Dotted lines show the interior structures computed with MESA. Solid lines show the atmospheres computed with CMFGEN. Inset panels show a zoom-in on the location where the structure and atmosphere models are connected. These models have solar metallicity.}
\label{fig:stitch}
\end{figure} 
 

To connect the structure models and atmosphere models we follow the approach by  \citet{2013A&A...558A.131G, 2014A&A...564A..30G}, see also \citetalias{2017A&A...608A..11G}. We first obtain stellar structure models for stripped stars that are undergoing central helium burning (\secref{sec:MESA_modeling}). Since stripped stars have nearly constant parameters during most of their helium burning phase (see \secref{sec:evol_MESA}), it is sufficient to extract only one structure model at a representative moment in time.  We take the structure models at  the moment when the central helium mass fraction drops below $X_{\text{He, c}} = 0.5$. 

We use the luminosity, surface abundances and mass loss rate given by the MESA structure model and use them as  fixed input conditions for the atmosphere calculations. We use the temperature ($T_{\text{eff, MESA}}$) as a starting condition for the temperature at an optical depth $\tau = 20$.  We then iteratively search for a solutions where the optical depth in the atmosphere as a function of radius $\tau(r)$ is such that  $\tau = 20$ occurs at a radius $r = R_{\star}$, where  $R_{\star}$ is the stellar surface radius given by the structure model.  


In \figref{fig:stitch} we show four examples of the transition between the stellar structure and atmosphere for stripped stars with mass $M_{\text{strip}} = $ 0.35, 0.8, 2.5 and 6.7\Msun . These models result from progenitor stars with zero-age main sequence masses of $M_{\text{init}} = $ 2, 4, 9 and 18\Msun. The models shown here are for solar metallicity. The dotted lines in \figref{fig:stitch} represent the interior structure models, while the solid lines show the atmospheric structure. 

The accuracy of the numerical solutions can be seen in the inset panels which show zoom-ins of the transition. The radial difference between the interior and the atmospheric structures are smaller than 0.001-0.003\,\Rsun. This corresponds to variations of only 0.2-0.3\,\% in the surface temperatures, which is more than accurate enough for our purposes.  


\section{Evolutionary models}\label{sec:evol_MESA}

In this section we describe the evolutionary models that are the basis for our spectral models. In this work we focus on the atmospheres and spectra, so we keep the discussion of the underlying evolutionary models concise. A more in depth discussion of the physical processes are given in \citetalias{2017A&A...608A..11G} and references therein. 


\subsection{Set-up of grids}

We create grids of binary evolutionary models following the evolution from the start of hydrogen burning up to at least half way central helium burning, but typically until central helium depletion. 
Our reference grid is computed assuming solar metallicity \citep[$Z = Z_{\odot} \equiv 0.014$,][]{2009ARA&A..47..481A}, but we also consider lower metallicities that are representative for local starforming dwarf galaxies as well as stellar populations at high redshift  ($Z = 0.006$, 0.002 and 0.0002).   
For each metallicity we compute models with  initial masses varying between 2.0 and 18.2\Msun, using more than twenty mass intervals that are equally spaced in $\log_{10}$. 
We use an initial mass ratio of $q = M_2/M_1 = 0.8$, where $M_1$ is the initial mass of the donor star and $M_2$ the initial mass of the accretor. We choose an initial orbital period such that mass transfer occurs early during the Hertzsprung gap passage (Case~B mass transfer).  
This results in stripped stars with masses between 0.35 and 7.9\Msun . 
\tabref{tab:params} provides an overview of the properties stripped stars in our solar metallicity grid.  Similar tables for the other metallicities can be found in \appref{app:metallicity} (\tabrefthree{tab:params_006}{tab:params_002}{tab:params_0002}).

\subsection{Evolutionary tracks in the HR diagram}

\begin{figure}
\centering
\includegraphics[width=\hsize]{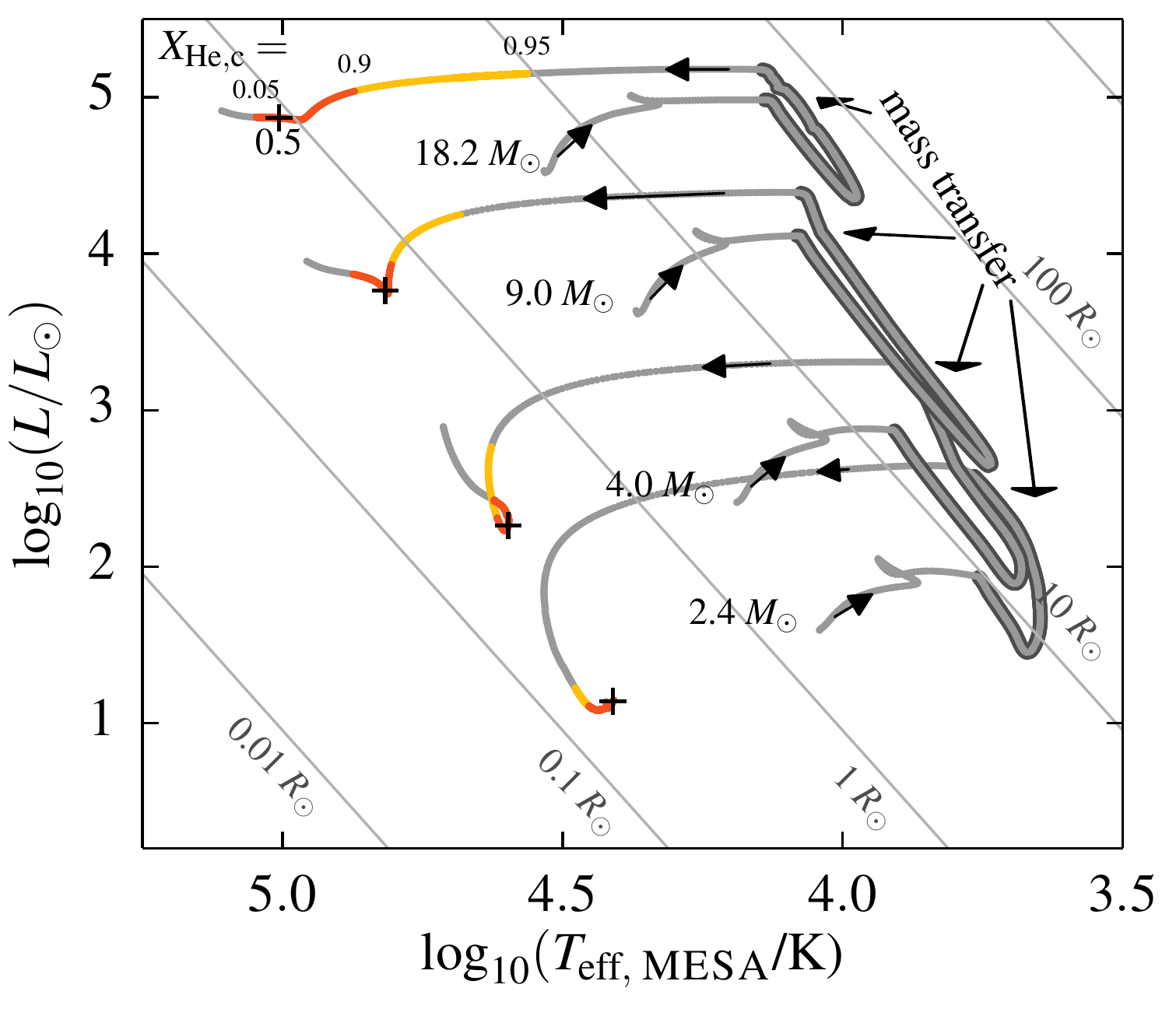}
\caption{Hertzsprung-Russell diagram showing the evolutionary tracks of models with initial masses of 2.4, 4, 9 and 18\Msun. Arrows along the tracks indicate the direction of evolution. The resulting stripped stars have masses of 0.44, 0.8, 2.5 and 6.7\Msun. Mass transfer is marked with \rev{an underlying thick, gray line}. Helium core burning is marked with yellow and dark orange color corresponding to the central helium mass fractions between $0.95 > X_{\text{He,c}} > 0.9$ and $0.9 > X_{\text{He,c}} > 0.05$ respectively. These models have solar metallicity. }
\label{fig:HRD_stitch}
\end{figure}

In \figref{fig:HRD_stitch} we show the Hertzsprung-Russell diagram with evolutionary tracks of the donor stars taken from four representative evolutionary models in our solar metallicity grid. The tracks start at the onset of hydrogen burning. Evolution proceeds with time in the direction of the black arrows indicated on the tracks. After completing the main sequence, the stars expand until they fill their Roche-lobe and mass transfer starts, which is marked in the diagram. During this phase, the star loses most of its hydrogen-rich envelope, which is about two thirds of the initial mass.  After that, mass transfer stops and the donor star contracts. It heats up and moves to the left part of the diagram. 

The yellow and dark orange part of the tracks in \figref{fig:HRD_stitch} mark the central helium burning phase. This phase starts before the star has fully contracted, as can be seen from the yellow part of the track, which corresponds to the early phases of central helium burning, where the helium mass fraction drops from $0.95 > X_{\text{He,c}} > 0.9$.  However, stripped stars spend most time at higher temperatures, in the part of the track that is marked in dark orange, which corresponds the phase where the central helium abundances range from $0.9 > X_{\text{He,c}} > 0.05$.  The luminosity and effective temperature remain nearly constant during the most of the helium core burning phase. The plus symbol indicates the moment where $X_{\text{He,c}} = 0.5$, which is when we extract the structure model that we use to construct the stellar atmospheres.  The evolutionary tracks are shown until central helium depletion, except for the lowest mass model which stops at $X_{\text{He,c}}= 0.48$ due to convergence issues.



\subsection{Properties of stripped stars during central helium burning}

\begin{figure}
\centering
\includegraphics[width=\hsize]{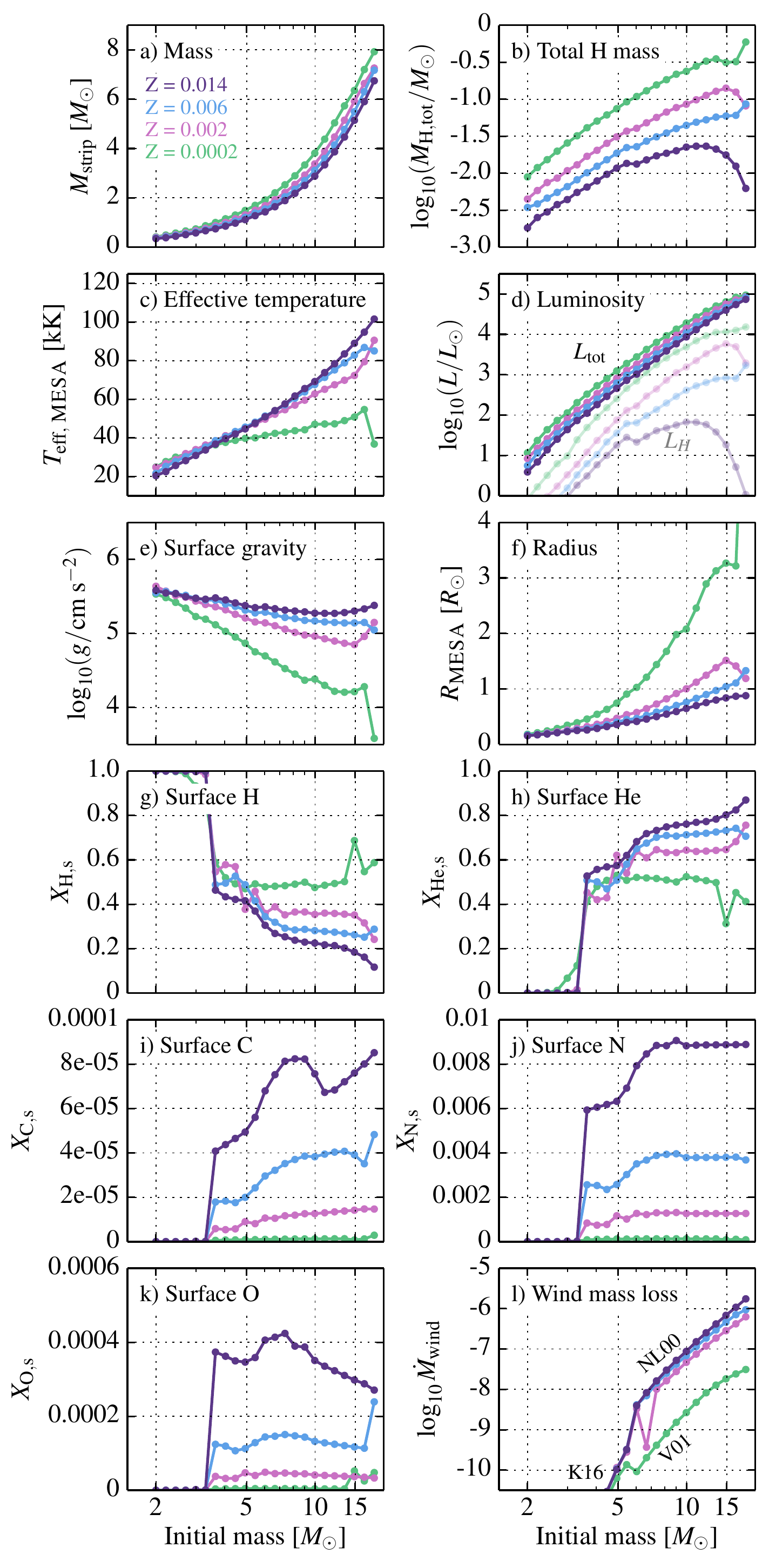}
\caption{Properties of stripped stars shown as a function of initial mass, for different metallicities. The parameters plotted here are derived from our evolutionary calculations with MESA, before inclusion of an atmosphere. Panel b) shows the total amount of pure hydrogen left after stripping, which remains in a layer consisting of helium and hydrogen at the surface. Panel d) shows the total luminosity $L_{\rm tot}$ as well as the luminosity resulting from the hydrogen burning shell alone  $L_{\rm H}$ in lighter colors. The labels in panel l) indicate the wind mass loss rates that are used, where NL00, V01 and K16 stand for \cite{2000A&A...360..227N}, \cite{2001A&A...369..574V} and \cite{2016A&A...593A.101K} respectively.}
\label{fig:prop}
\end{figure}

\begin{sidewaystable*}
\centering
\caption{Properties of stripped stars for solar metallicity ($Z = 0.014$). See online appendix for $Z = 0.0002, 0.002$ and 0.006.}
\label{tab:params}
{\small
\input{table_param.tex}
}
\tablefoot{This table shows parameters for stripped stars halfway through central helium burning (when central helium mass fraction reaches $X_{\text{He, c}} = 0.5$). We distinguish three groups: A: absorption line spectra, E: emission line spectra, and A/E: mixed absorption and emission line spectra. We display in order: initial mass ($M_{\text{init}}$), initial orbital period ($P_{\text{init}}$), mass of stripped star ($M_\star$), total mass of hydrogen ($M_{\text{H, tot}}$), luminosity ($L$), luminosity from hydrogen burning ($L_H$), surface temperature ($T_{\star}$), effective temperature ($T_{\text{eff}}$, defined at optical depth $\tau = 2/3$), effective surface gravity ($\log _{10} g_{\text{eff}}$), effective stellar radius ($R_{\text{eff}}$), surface hydrogen mass fraction ($X_{\text{H,s}}$), surface helium mass fraction ($X_{\text{He,s}}$), wind mass loss rate ($\dot{M}_{\text{wind}}$), (electron scattering related) Eddington factor ($\Gamma _e$), terminal wind speed ($v_{\infty}$) and emitted HI, HeI and HeII ionizing photon flux ($Q_0$, $Q_1$ and $Q_2$ respectively). Models which have a point-mass companion are marked with $^{x}$, models which do not reach carbon depletion are marked with $^{\dagger}$ and models which do not reach helium depletion are marked with $^{\ddagger}$. \rev{We have chosen not to display the final periods and final mass of the secondary as these are dependent on the mass transfer efficiency, which is an uncertain parameter. The effects on the structure and physical parameters of the stripped stars are negligible.} These models have solar metallicity, $Z = 0.014$.}
\end{sidewaystable*}

\figref{fig:prop} provides an overview of the stellar parameters and surface properties of stripped stars as a function of initial mass ($M_{\text{init}}$) when they are halfway core helium burning ($X_{\text{He,c}} = 0.5$). The lines with different colors show results for different metallicities, $Z = 0.014$ (purple), $Z = 0.006$ (blue), $Z = 0.002$ (pink) and $Z = 0.0002$ (green).  

\paragraph{Masses and remaining hydrogen layer}~\\

\noindent The masses of the stripped stars increase steeply with the mass of their progenitor. The steep rise reflects the fact that more massive progenitor stars have larger convective hydrogen burning cores, during their prior evolution as main sequence stars. They convert a larger fraction of their total initial mass into helium before filling their Roche-lobe and hence produce more massive stripped stars. 

We find a mild trend with metallicity, giving slightly higher mass stripped stars at lower metallicities (\figref{fig:prop}a).  For example, our 18\Msun progenitor model becomes a $\sim 8$\Msun stripped star in our simulations for low metallicity, while at solar metallicity the resulting stripped star is roughly 1\Msun less massive. This is in part because Roche-lobe overflow is less efficient at lower metallicity, as we discussed in \citetalias{2017A&A...608A..11G}, leaving a larger fraction of the hydrogen-rich envelope after mass transfer \citep[see also][]{2017ApJ...840...10Y}. 

A second reason for this trend has to do with the stellar wind. By the time the stripped stars are halfway their helium burning phase, stellar winds had time to remove the outermost layers for our most luminous and metal-rich models. The effects of stellar winds are, for example, responsible for the turn-over in the total remaining hydrogen mass that can be seen in panel b of \figref{fig:prop} for the highest metallicity models. 

The total amount of remaining hydrogen ranges between $\sim0.005$\Msun and $\sim 0.5$\Msun.  This is relevant in light of the question about the origin of the diversity observed in the early spectra of core collapse supernova. Depending on the amount of remaining hydrogen the supernova would be classified as type Type IIb or Ib \citep[][see also \citealt{2011MNRAS.414.2985D} and \citealt{2017ApJ...840...10Y}]{1997ARA&A..35..309F}. The values quoted here provide an upper limit to the amount of hydrogen that can still be present when the stripped star end its life and explodes. The final amount depends on whether or not the star experiences a second phase of mass transfer and wind mass loss in the final stages \citep{2010ApJ...725..940Y}.

\paragraph{Temperatures, Luminosity, Surface gravity and Radii}~\\

\noindent Stripped stars exhibit very high surface temperatures, $>20\, 000$\,K for the lowest mass models, reaching to 100\,000\,K for our highest mass models. Note that the temperatures quoted here result directly from our evolutionary calculations and characterize the surface of the star. For the most luminous and metal-rich stripped stars, we expect strong winds leading to an optically thick outflow. This slightly moves the photosphere outwards.
For stripped stars with initial masses less than about $5\Msun$, we find that the effective temperatures are almost independent of  metallicity. This is not true for stripped stars of higher mass, where we find substantial variations from 40\,000\,K for low metallicity up to 100\,000\,K for high metallicity (panel c, \figref{fig:prop}), as we already noted in \citetalias{2017A&A...608A..11G}. This is due to the remaining layer of hydrogen, which creates a modest size, low-mass envelope around our most metal poor models. 

Luminosity steeply increases with mass, ranging from sub-solar values for our lowest mass subdwarfs up to about $10^5 \Lsun$ for the most massive stars in our grid (panel d, \figref{fig:prop}). A rough approximation for mass-\rev{luminosity} relation, $L \propto M^{x}$ yields $x = 3.3$ for solar metallicity. This relation becomes less steep for lower metallicity, reaching down to $x = 3.0$.   

The main source of energy production in these stripped stars is nuclear fusion of helium into carbon and later oxygen. However, energy production by hydrogen burning can contribute in a shell around the core. This contributes up to about 30\% in our metal poor models, see the semi-transparent lines in panel d, \figref{fig:prop}. 

Stripped stars are further characterized by their high surface gravities and small sizes. The surface gravities range between $4.5 < \log _{10} (g/\text{cm\,s}^{-2}) < 6$ (panel e, \figref{fig:prop}). The radii are the largest for low metallicity models, reaching up to around 4\,\Rsun, while at high metallicity all models have radii below 1\,\Rsun (panel f, \figref{fig:prop}).

\paragraph{Surface abundances}~\\
  
\noindent The layers exposed at the surface of stripped stars originate from regions that were originally part of (or located just above) the convective core of the progenitor star. This makes the surface abundances of stripped stars potentially interesting as a way to probe these deep regions that are otherwise inaccessible, which may bear signatures of mixing processes that may occur in these layers. These layers have been partially processed by hydrogen burning through CN and CNO cycling. We therefore expect enhancements in helium and nitrogen, and depletions in hydrogen, carbon and oxygen with respect to their initial mixtures, for which we adopted scaled solar abundances \citep[cf.][]{1998SSRv...85..161G, 2009ARA&A..47..481A}. This is indeed what we find in most cases, as can be seen in panels g, h, i, j, and k of \figref{fig:prop}, although the abundances are further modified to additional physical processes that play a role.

The surface abundances of our lowest mass models are modified by the effects of gravitational settling, causing hydrogen to float to the surface and the heavier elements to sink. Our models with $M_{\text{init}} \lesssim 3.5 \Msun$ show surface hydrogen mass fraction reaching close to unity.  


\rev{The stripped stars with low metallicity show higher surface hydrogen mass fractions compared to the stripped stars with high metallicity. Conversely, the helium mass fractions on the surface are lower with the metallicity is low. This effect is particularly prominent in the higher mass models. The reason is that envelope stripping is less efficient at low metallicity, leaving a thicker layer of hydrogen-rich material at the surface \citepalias{2017A&A...608A..11G}. Additionally, in high mass and high metallicity models, wind mass loss removes the outer layers of the star, and thus also the hydrogen-rich material, leaving the stripped star completely hydrogen free.}


\paragraph{Wind mass loss rate}~\\
  
\noindent In panel l of \figref{fig:prop}, we show the stellar wind mass loss rates. The overall trend is a rapid decrease of the mass loss rate with mass, reaching below $\dot{M}_{\text{wind}} = 10^{-9}$\Msunyr when $M_{\text{init}} < 6$\Msun (corresponding to stripped stars of masses below 2\Msun). This is the regime where we adopt the \cite{2016A&A...593A.101K} rates. 

For the higher mass stars we find an bifurcation in the wind mass loss rates' behavior. Our lowest metallicity models have relatively low temperatures and hydrogen fractions larger than $X_{\rm H,s} = 0.4$. With our current implementation this leads us to the usage of the prescriptions by \cite{2001A&A...369..574V}, originally intended for normal main sequence stars.  
Instead the more metal-rich stripped stars have surface mass fractions below $X_{\rm H,s} = 0.4$ and we use \cite{2000A&A...360..227N} instead. Here we still find a dependency on metallicity, with the metal-rich stars having stronger winds. 

We stress that the stellar wind mass loss rates of stripped stars are not well known at present, so for this work we rely on theoretical prescriptions and extrapolated algorithms. We expect the situation to improve as the number of detections increase, which would allow us to update this grid with improved input assumptions.  

\paragraph{Deviations from the smooth trends} ~\\

\noindent For stripped stars with mass $\gtrsim 6$\Msun (progenitor mass of $\gtrsim 13$\Msun), we find small deviations from the primarily smooth trends that are shown in \figref{fig:prop}. In this mass range, the layer exposed at the surface of stripped stars originate from layers inside the progenitor that are characterized by steep composition gradients induced by convective regions overlapping with the initially thick hydrogen burning shell. The location and extent of the convective region therefore impacts the surface properties of the stripped star and thus can cause the irregular behavior in \figref{fig:prop}. The effects are most pronounced at low metallicity, where the wind mass loss is low. This means that the surface composition still directly reflects the composition after mass transfer stopped and the composition will remain the same until the stellar death. We note that this behavior is model dependent as it depends on the presence and exact location of the convective region.  Fortunately the variations are relatively small and we deem this model grid suitable for our current purposes. Careful future studies of the effects of mesh refinement and assumptions concerning the treatment of overshooting as a step or exponential process are however warranted in this regime, see for example the work by \citet{2016ApJS..227...22F}.


\section{Spectral models}\label{sec:spec_models}


\begin{figure*}
\centering
\includegraphics[width=0.5\textwidth]{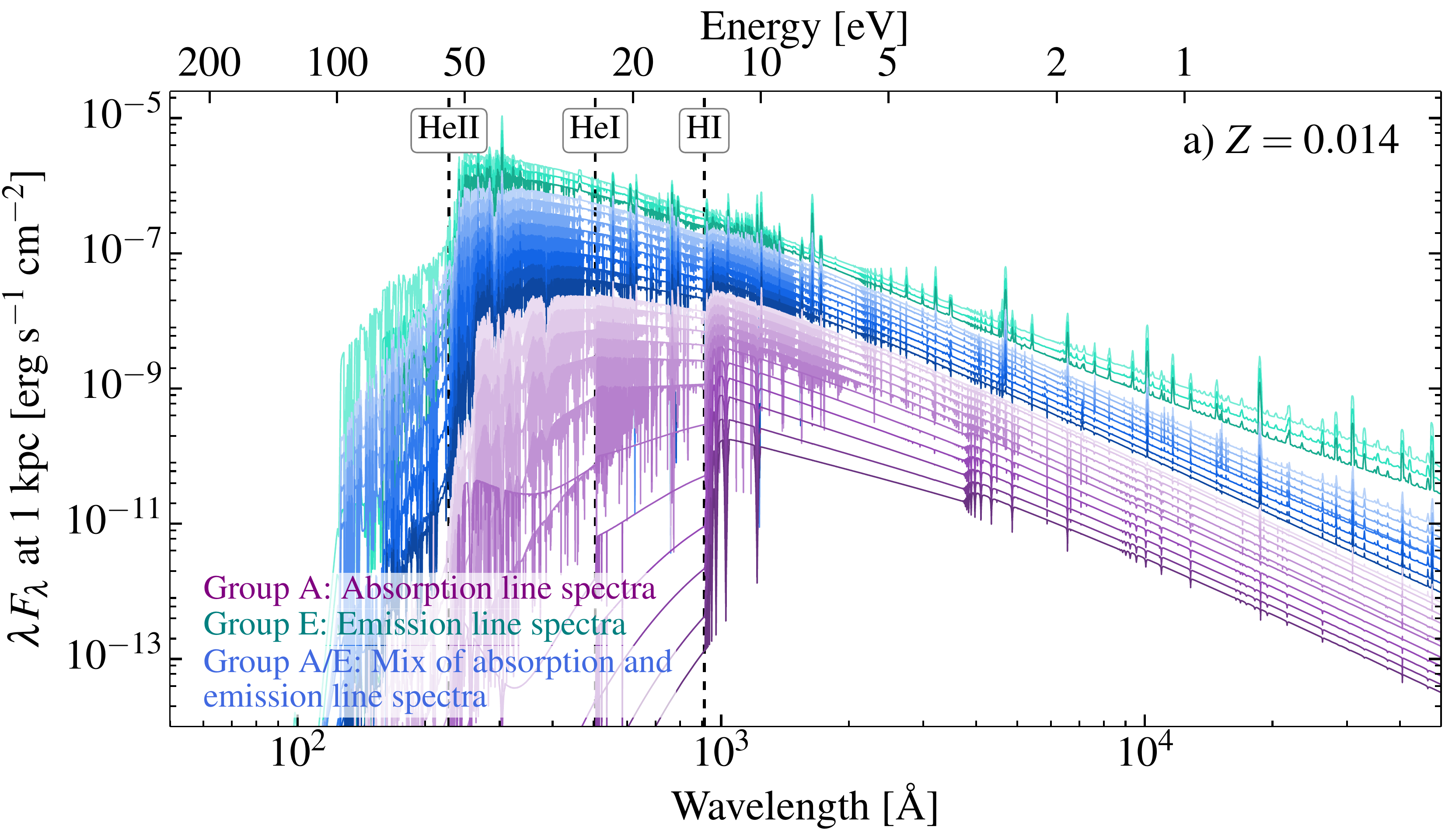}%
\includegraphics[width=0.5\textwidth]{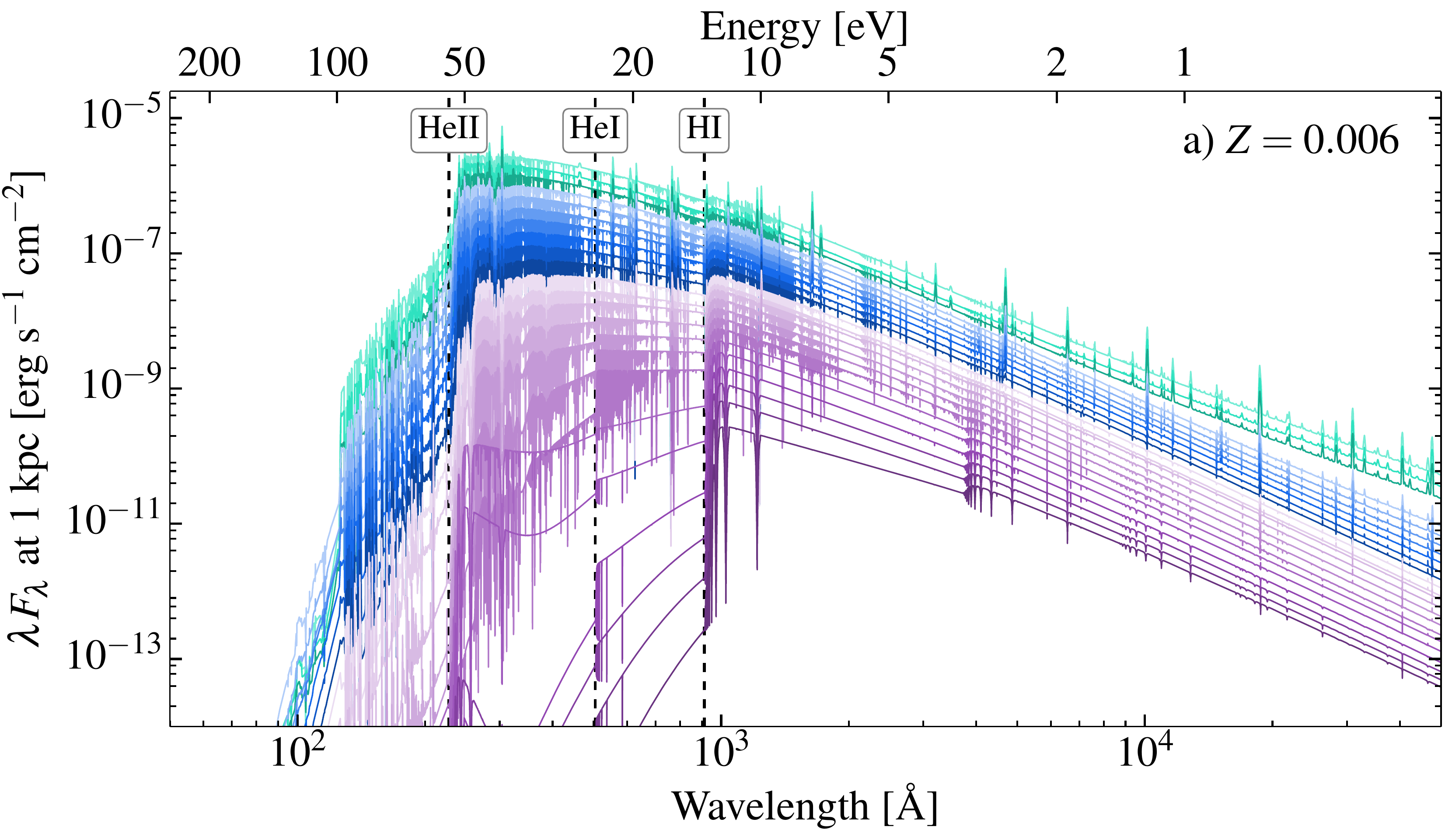}
\includegraphics[width=0.5\textwidth]{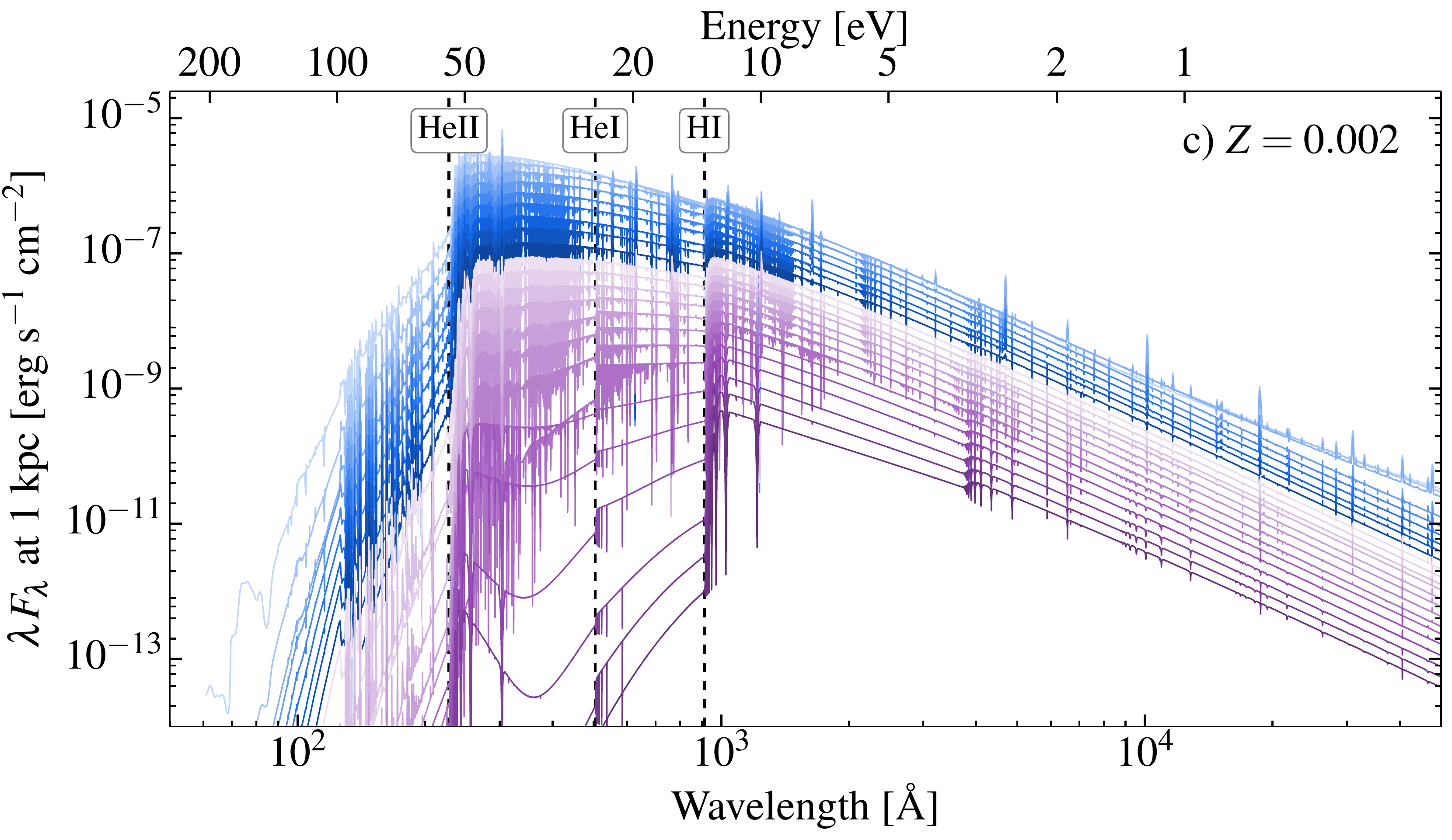}%
\includegraphics[width=0.5\textwidth]{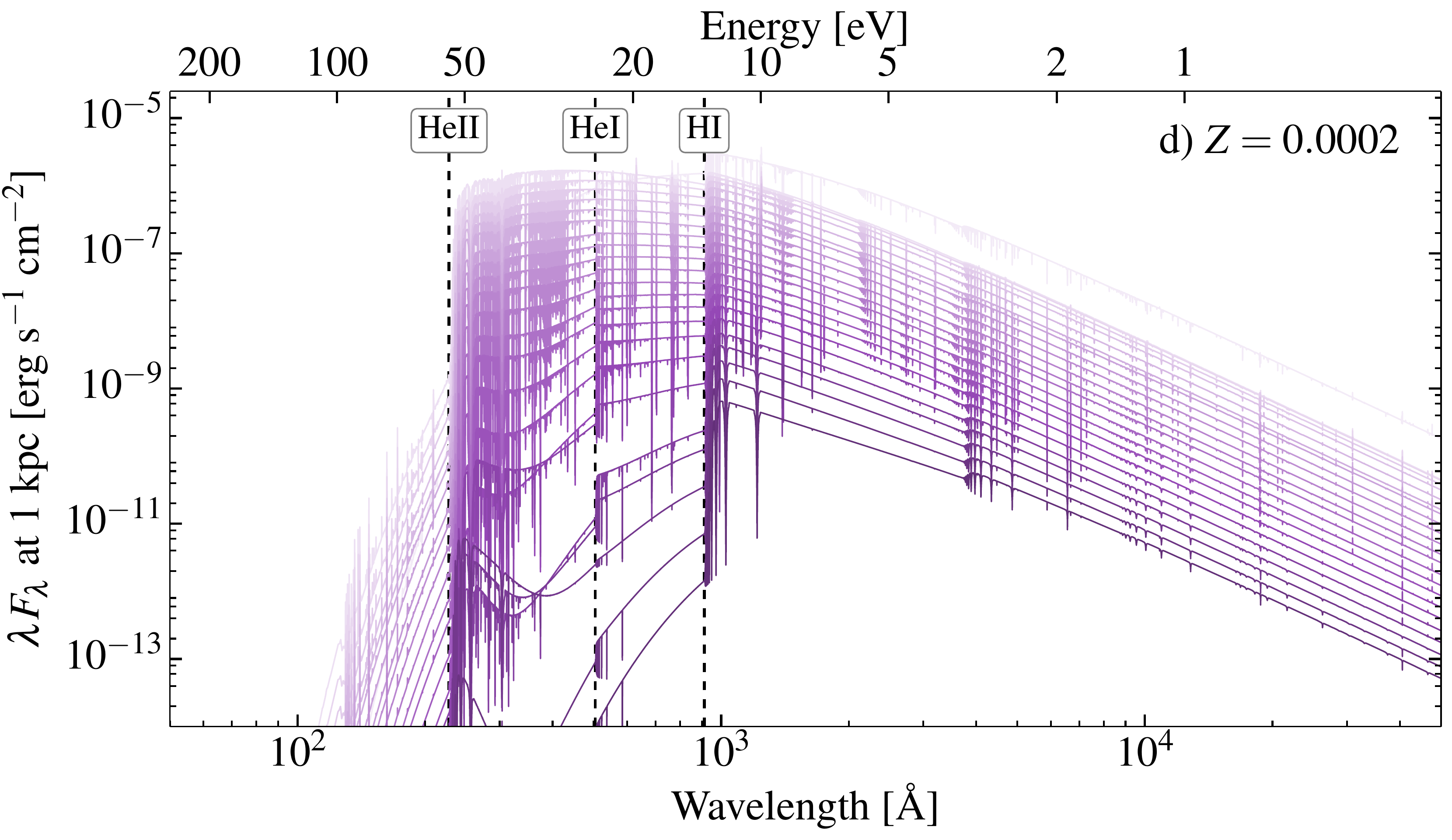}
\caption{ Spectral energy distributions of stripped stars with different metallicities.  The masses of the stripped stars range from 0.35\Msun up to 6.7\Msun. Colors indicate the different groups with morphological similarities: Group A: absorption line spectra (purple), Group E: emission line spectra (green), and Group A/E: spectra with a mix of absorption and emission lines (blue). We mark the ionization limits of HI, HeI and HeII with dashed lines.}
\label{fig:SED_all}
\end{figure*}

Here we present the full grid of spectral models, created for the stars stripped by interaction with a binary companion. All our models are publicly available for download as a service to the community.\footnote{They can be retrieved from \url{https://staff.fnwi.uva.nl/y.l.l.gotberg}. \YG{(They will be made available at a more permanent location upon publication of the paper.)}}  \figref{fig:SED_all} provides an overview of all the spectra. \rev{Panel a} shows our reference model grid, for which we assumed solar metallicity. The other panels show lower metallicities as indicated. Within each panel we show the spectra for stripped stars with masses ranging from 0.35\Msun up to 8\Msun (corresponding to progenitor stars with initial masses of 2.0 up to 18.2\Msun). 

The spectra of stripped stars are all characterized by their hard emission peaking in the far and extreme UV. The HI, HeI and HeII ionization limits are also marked for reference in \figref{fig:SED_all}. The ionizing fluxes and their implications will be discussed in a separate paper (G\"otberg et al.\ in prep.), but we already provide the emission rate of HI, HeI and HeII ionizing photons in \tabref{tab:params} (see \tabrefthree{tab:params_006}{tab:params_002}{tab:params_0002} for lower metallicity than solar). Here we focus the discussion on the morphology of the spectra. 

We find a gradual change of the overall shape spectra with mass. More massive stripped stars are brighter and hotter. This translates into a general shift of the spectra toward larger fluxes and shorter wavelengths with increasing mass. In particular, the fluxes emitted short-ward of the HI, HeI and HeII ionization limits increase with the mass of the stripped star. 

\begin{figure*}
\centering
\includegraphics[width=0.9\textwidth]{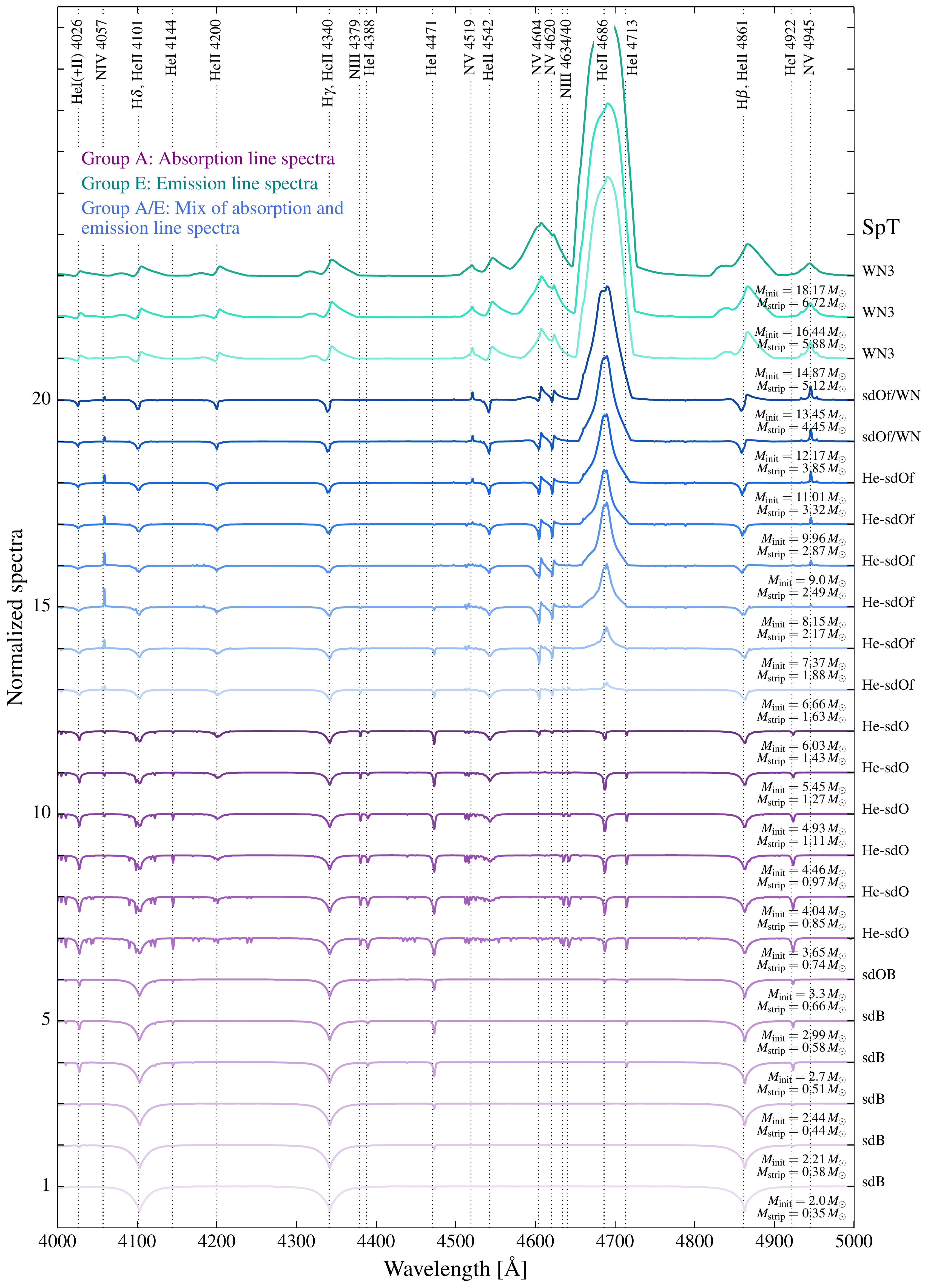}
\caption{The optical spectra of the models in the solar metallicity grid. The transition from subdwarf to O-type star to slash-star to WN star is visible. Note that as temperature increases with higher mass, lower ionization lines become weaker, while higher ionization lines become stronger. We mark the lines important for spectral classification with dotted lines and assign the spectral class to the right of the normalized spectrum.}
\label{fig:optical_spectra_014}
\end{figure*}

The spectra show a rich sequence of spectral features. The most massive stars shown in the \rev{panel a} of \figref{fig:SED_all} show emission lines. \rev{These emission lines gradually} decrease in strength for less massive stripped stars and they are absent in the spectra for our low-mass models, which instead show absorption features. The spectral features can be seen more clearly in \figref{fig:optical_spectra_014}, where we show the normalized spectra for the optical part between 4000~\AA\  and 5000~\AA\ for our solar metallicity grid. Similar diagrams for lower metallicity are provided in \appref{app:metallicity}, where we also provide the UV and IR portions of the spectra.


To organize the discussion below, we distinguish three groups that show morphological similarities. We distinguish spectra that show primarily emission features (Group~E, shown in green in \figref{fig:SED_all}), absorption  features (Group~A, shown in purple) and a transitional group that shows a mix of both (Group A/E, shown in blue). 

 
 

\rev{
\paragraph{Spectral classification scheme}~\\

\noindent The physical and wind properties of the stripped stars straddle subdwarf~OB stars and classical Wolf-Rayet stars, so in order to classify atmospheric models we follow \citet[][and references therein]{2016PASP..128h2001H} for subdwarfs, assigning sdB in instances for which HeII lines are absent, sdOB if HeII~$\lambda 4686$ is present, and sdO if other HeII lines are present. HeII~$\lambda 4686$ emission is predicted in a subset of the subdwarf~O star models, so we assign sdOf for such cases, in common with criteria used for O supergiants \citep[e.g.][]{1988A&AS...76..427M, 2002AJ....123.2754W}. We also follow the convention of using the ``He-'' prefix for models with super solar Helium abundances \citep{2007A&A...462..269S}. 

For WN stars, we adopt the ionization criteria from \citet{1996MNRAS.281..163S}, although refrain from including line strength/width information (``b'' for broad lined stars) since the full-width half maximum of HeII~$\lambda 4686$ usually indicates ``b'' whereas the equivalent width of HeII~$\lambda 5412$ does not. Formally, WN3-4(h) subclasses would result from the Pickering-Balmer decrements in the WN models. For the transition models between Of~subdwarfs and classical WN stars, we follow the approach of \citet{2011MNRAS.416.1311C} for supergiants, involving H$\beta$. We assign an Of class if H$\beta$ is observed in absorption, WN if it is purely in emission, or an intermediate Of/WN if it is observed as a P~Cygni profile. O~type luminosity classes of subdwarfs are generally adopted, owing to the high gravities of the models ($\log g_{10} > 5$). The only exception is the most massive $Z=0.0002$ model ($\log g_{10} \leq 4$) whose subclass and luminosity class follow from quantitative criteria of \citet{1988A&AS...76..427M}, although the majority of the very metal poor models possess gravities intermediate between dwarf (class V) and subdwarf (class VI).}

\subsection{Emission line spectra [Group~E]: A case for WR-like stars with atypically low masses?}

The most massive and metal-rich stripped stars in our grids have strong stellar winds and their spectra primarily show emission lines. These spectra are similar to those of observed WN type stars (nitrogen-rich Wolf-Rayet stars), which show strong lines of ionized helium and nitrogen in emission 
\rev{\citep[][]{1995A&A...302..457C, 2006A&A...457.1015H}}.
\rev{According to the classification of \citet{1996MNRAS.281..163S}, the specific spectral type is WN3.}

In our solar metallicity grid we find this behavior for stripped stars with masses larger than 5\Msun (progenitor masses above 14\Msun\rev{, see the green spectra in \figref{fig:optical_spectra_014}}). The boundary moves up in mass as metallicity decreases. In our $Z=0.006$ grid \rev{(see \figref{fig:SED_all}b and \figref{fig:optical_spectra_006})} we find spectra dominated by emission lines for stripped stars with masses larger than 6.3\Msun (progenitor masses above 16\Msun). Our mass thresholds are substantially lower than the values of $10-25\Msun$ that are typically quoted for WR stars \citep[see][and references therein]{2007ARA&A..45..177C}.

The strongest spectral features for stars in this group come from ionized helium and highly ionized nitrogen. In the optical, HeII~$\lambda 4686$ is four to six times stronger than the continuum flux. The NV~$\lambda\lambda 4604/20$ doublet shows strong emission, as an effect of the high temperature and wind mass loss rate characteristic to this group. Lower ionization stages of nitrogen are not visible. Hydrogen is present in the atmosphere, but not easily distinguishable in the optical spectra as HeII lines appear at wavelengths very close to the Balmer lines. The blend of H$\alpha$ and HeII~$\lambda 6560$ is weaker than HeII~$\lambda 4686$, but still two to three times stronger than the continuum (see \figref{fig:IR_spectra_014}). In the UV, HeII~$\lambda 1640$ is strong in emission, along with the HeII series \rev{(3202, 2733, 2511 \AA\ etc.), CIV~$\lambda 1550$, and NV~$\lambda 1240$ stellar wind lines (see \figref{fig:UV_spectra_014}). The Lyman series are also predicted, though these are usually masked by strong interstellar absorption along most sight lines.} 

We note that the spectral features for stripped stars in this group are strongly dependent on the assumed wind mass loss rate, wind speed and wind clumping factor. In particular, lower values of the mass loss rate \citep[for example as recently argued for by][]{2017A&A...607L...8V} will yield weaker emission lines \citepalias[see][]{2017A&A...608A..11G}.

\begin{figure*}
\centering
\includegraphics[width=\hsize]{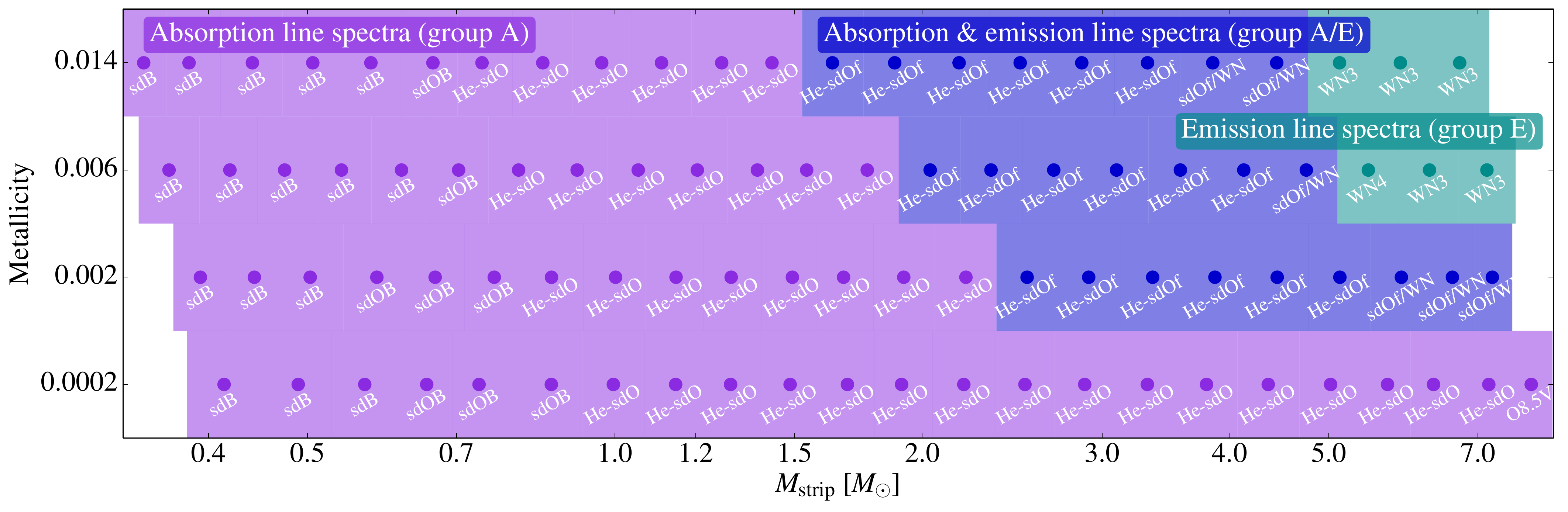}
\caption{The spectral properties of stripped stars with a range of metallicity and mass. We mark Group A: absorption line spectra in purple, Group E: emission line spectra in green, and Group A/E: a mix of absorption and emission line spectra in blue.}
\label{fig:classification_simple}
\end{figure*}

\subsection{Spectra with a mix of absorption and emission lines [Group A/E]: A new transitional spectral class?}

We identify a group of stripped stars whose spectra show a mix of absorption and emission lines. These are stars with temperatures between 40\,000 and 85\,000\,K, corresponding to masses between 1.8 and 5\Msun, in our solar metallicity grid \rev{(see the blue spectra in \figref{fig:optical_spectra_014})}. These stars have progenitor masses between 7 and 14\Msun. The mass boundaries for this transitional group shift up in mass with decreasing metallicity, as can be seen in \figref{fig:classification_simple}, where we show the transitional group in blue \rev{(see also \figref{fig:SED_all} and \appref{app:metallicity})}.

The mix of absorption and emission lines occurs because the optical continuum form in quasi-hydrostatic layers of the atmosphere together with some of the lines, which give rise to absorption features. At the same time, other lines with higher opacity form in the wind and thus create emission lines. 
For lower mass loss rates, the stellar wind outflow is transparent and emission lines cannot form. Instead, for higher mass loss rate the wind is opaque and the surface absorption features are no longer visible. 


The strongest optical spectral features in group A/E are the HeII~$\lambda 4686$ emission, the NV~$\lambda\lambda 4604/20$ seen in P~Cygni and the HeII/Balmer absorption lines. The emission features are those characteristic to hot WN type stars, while the absorption features are the same as early O-type stars. The UV shows strong emission of HeII~$\lambda 1640$, the Lyman series and several lines from NV, and CIV. The HeII series between $\sim 2000 - 3000 \AA$ is seen in absorption. The blend of HeII~$\lambda 6560$ and H$\alpha$ is in emission, but weaker compared to the group~E, with just 20-50\% above the continuum flux. 

\rev{We assign two main spectral classifications for the spectra with both emission and absorption lines: He-sdOf and sdOf/WN. Stars with lower mass loss rate than $10^{-6.6}$\Msunyr are classified as He-sdOf, while the ones with higher mass loss rate are classified as sdOf/WN. This threshold of mass loss rate corresponds to when the H$\beta$ line transitions from absorption to P~Cygni profile. We use the subdwarf~O classification since linear Stark broadening is severe in HI and HeII absorption lines, owing to their high surface gravities. This effect is moderate in lower gravity main sequence O stars. To signify emission in HeII~$\lambda 4686$ we use the notation ``f'' following e.g.\ \citet{1988A&AS...76..427M}. 
All of the stripped stars in the transitional class have super solar helium abundance and should therefore be assigned the ``He-'' prefix. However, we omit the prefix for simplicity for stars which have a WR-type classification. 
}

This characteristic mix of absorption and emission features have recently been observed in several stars during a survey for WR stars in the Large Magellanic Cloud by \citet{2014ApJ...788...83M, 2015ApJ...807...81M, 2017ApJ...837..122M}. These stars were classified as WN3/O3 type because of their morphological similarities to such a composite spectrum. The discoverers explain that a binary star combination of WN3 and O3 can not be the explanation of these stars as they are about an order of magnitude fainter than an O3-type star. Alternative evolutionary scenarios are treated in \citet{2017ApJ...841...20N} including e.g., massive O-type stars transitioning to WR type. However, this would not explain why these stars are under luminous. \citet{2018MNRAS.475..772S} show that the WN3/O3 stars are isolated from the massive O-type stars, indicating that WN3/O3 stars derive from older, lower mass stars. They discuss the possibility that the WN3/O3 stars are stripped stars with low-mass companion stars. The stripped star would in this case dominate the composite spectrum.

\subsection{Absorption line spectra [Group A]: A case for overweight subdwarf-like stars defying the canonical mass?}

The majority of spectra in our grid show primarily absorption features. For our solar metallicity grid it encompasses stripped stars with masses less than 1.7\Msun (resulting from progenitor stars with initial masses up to 7\Msun, see purple spectra in \figref{fig:optical_spectra_014}). These are all stars with very weak stellar winds. Their atmospheres are transparent and the absorption lines originate directly from the stellar surface. At lower metallicity, where the stellar winds are weaker, we find absorption line spectra up to masses of 2.3\Msun for $Z=0.006$ and $Z = 0.002$, and in our most metal poor grid we find exclusively absorption line spectra \rev{(see \figref{fig:SED_all} and \appref{app:metallicity})}. 

The spectra in this group show many similarities with those observed for hot subdwarfs \citep[e.g.][]{1990A&AS...86...53M, 2017A&A...600A..50G}. This is especially true for the lower mass models, which have the highest surface gravities ($\log_{10} (g/\text{cm\,s}^{-2}) \sim 5.5$). Their spectra show characteristic broad Balmer features due to the effects of pressure broadening (see \figref{fig:optical_spectra_014}).

Because of their high temperatures, the spectra show features of ionized elements such as HeII and NIII. The strongest optical spectral features are those of hydrogen and helium. For stripped stars with temperatures below 40\,000\,K, HeI lines dominate over HeII lines. Above this temperature the opposite occurs as an effect of the ionization state in the atmosphere. The strongest helium lines are HeI~$\lambda 4471$ and HeII~$\lambda 4686$, while for hydrogen the Balmer lines are the strongest. Stripped stars with effective temperature above 40\,000\,K show significant blending between the HeII lines and the Balmer series. The metal lines are weak in the optical spectra, the most prominent ones come from NIII. Their strength depends on the temperature of the star.  

For the lowest mass models we can see the effects of gravitational settling, which causes heavier elements to sink and hydrogen to rise to the surface. This effect is responsible for the sharp transition in the morphology taking place near 0.7\Msun in our solar metallicity grid, see \figref{fig:optical_spectra_014}. Models below this mass primarily show hydrogen features. The transition is especially striking in the UV portion of the spectra, shown in the appendix in \figref{fig:UV_spectra_014}. The more massive stars show a rich forest of metal lines, which are completely absent in our models for lower mass. We note that the effect of settling is probably exaggerated in the models we present here \rev{(see observed abundances in subdwarfs from e.g.\ \citealt{2006A&A...452..579O}, \citealt{2013A&A...549A.110G} and \citealt{2013MNRAS.434.1920N})}. This is because we are still lacking an appropriate treatment for processes such as radiative levitation and possible turbulent mixing that can (partially) counteract the effects of settling. This should be kept in mind by anyone who wishes to use these models and compare to data. 

\rev{When the metallicity is very low ($Z = 0.0002$) the stripped stars with mass above 1\Msun show the spectral morphology of O-type stars. They are however sub-luminous compared to the main sequence type star of the same spectral type (see \tabref{tab:params_0002}) and therefore we classify them as O-type subdwarfs. Due to their super-solar helium abundance on the surface we append the prefix ``He-'' to the spectral classification (see e.g.\ \figref{fig:classification_simple}). It is remarkable that stripped stars sometimes share the spectral morphology with the massive O-type main sequence stars. This despite the large difference in luminosity, which can differ with orders of magnitudes.

The most massive stripped star with metallicity $Z = 0.0002$ cannot be classified as sub-luminous as the spectral morphology and luminosity is similar to that of a late O-type main sequence star. 
}
 
\subsubsection*{Implications for interpretation of the subdwarf population}

One of our striking findings is that we obtain spectra that are indistinguishable from subdwarfs for a range of masses. First of all, this is an indication that the stable mass transfer channel provides a viable alternative mechanism to produce hot subdwarfs, as proposed originally by \citet{1976ApJ...204..488M}. The most well-known channel produces subdwarfs after a common envelope phase leaving them in close orbits. The stable channel is expected to produce subdwarfs in wide orbits (\citealt{2002MNRAS.336..449H}, see also \citealt{2013A&A...554A..54G}). This could be the explanation for the few wide-orbit subdwarf systems that have been observed \citep[cf.][]{2017A&A...605A.109V} and for subdwarfs that so far appear to be single\rev{, but may in fact be wide-orbit binaries}. We note that \citet{2002MNRAS.336..449H} discuss the stable channel, but they only considered progenitors up to about 2\Msun. Here, we show instead that donors with initial masses up to about 7\Msun can still produce subdwarf-like stars.  

Our solar metallicity grid shows that sdB type spectra may come from stripped stars with masses in the range $0.35 - 0.75\Msun$. Additionally, we find helium-rich subdwarfs with masses in the range $0.75 - 1.63\Msun$. \citet{2017A&A...600A..50G} note that the helium-rich subclasses can often only be distinguished with a proper quantitative spectral analysis and may often be misclassified as sdO stars.   

This has interesting consequences for the mass distribution of subdwarfs. It has become custom to adopt a canonical mass of $0.47 \Msun$ for subdwarfs \citep[e.g.][]{2002MNRAS.336..449H, 2012A&A...539A..12F}. The motivation behind this assumption comes in part from theoretical simulations. These show that the relatively low-mass progenitors ($\sim 1\Msun$) that evolve through the common-envelope channel tend to produce subdwarfs with a mass distribution that peaks sharply around this canonical value. Our results suggest that there may be a population of subdwarfs with a wider range of masses, including substantially ``overweight subdwarfs'' that are up to nearly 3.5 times more massive than the canonical value. \rev{There are already several subdwarfs observed to have both higher and lower masses compared to the canonical $0.47\Msun$. An interesting system is HD~49798 which harbors a $1.5 \Msun$ subdwarf with a massive and fast-spinning white dwarf companion \citep{1978A&A....70..653K, 2009Sci...325.1222M}.}

If the population of subdwarfs with masses that deviate significantly from the canonical value is substantial, this would have several important implications. The canonical value is often assumed for subdwarfs in single-lined spectroscopic binaries with an unseen companion. Assuming a value for the mass of the subdwarf allows to place a lower limit on the mass of the companion and indirectly infer something about its nature, for example whether it is a white dwarf or potentially even an neutron star \citep[e.g.][]{2015A&A...576A..44K}.  

In addition, the assumption of a single canonical value for the masses of subdwarfs also implies that they have a canonical brightness, which can be used to make a rough inference about the distance \citep[e.g.][]{2017A&A...600A..50G}. Our 0.44\Msun stripped star has a brightness of 1.1\Lsun.  However, our most massive subdwarf-like stripped star, of 1.6 \Msun, is instead nearly 100 times brighter, see \tabref{tab:params}. This means that under estimating the mass of such an object by incorrectly assuming the canonical value, would lead to under estimating its luminosity, which in turn would lead to underestimating the distance, possibly placing it 10 times closer than it is in reality.

Whether or not a population subdwarfs with abnormal masses exists will soon be verified by the Gaia satellite. \cite{2017A&A...600A..50G} compiled a sample of the roughly 5000 subdwarfs that have been identified so far. If overweight subdwarfs exist, Gaia will tell us by showing that they are much larger distances than we would have guessed based on the canonical values.

\section{Prospects for detecting stripped stars}\label{sec:observability}

The spectra we presented so far were for stripped stars in isolation. However, the majority of stripped stars is expected to have a main sequence companion \citep[e.g.][]{2011A&A...528A.131C, de-Mink+2011}.  Stripped stars that result from stable mass transfer are likely to have relatively massive companions, which outshine the stripped star in optical wavelengths. Moreover, we expect relatively wide orbits and the radial velocity variations due to the orbital motions will be very small. This can make it difficult to to detect or recognize the presence of a stripped star. The system may be readily mistaken for a single star \citep{de-Mink+2014}. 

The sample of known stripped stars in binaries is likely biased towards those with faint companions. This may be a late type main sequence star. Also white dwarf, neutron star or black holes companion are possible although they are expected to be less common \citep[e.g.][]{Dewi+2002, 2017A&A...601A..29Z}. (The formation of single stripped stars requires enhancement of stellar wind mass loss \citep[e.g.][]{DCruz+1996,Georgy+2009} or more exotic binary scenarios such as a special types of mergers  \citep{Nomoto+1993, Hall+2016} or the disruption of a binary system in which the supernova order has been reversed \citep{Pols1994a}.) 

Stripped stars with faint companions are likely the result of unstable mass transfer involving the ejection of a common envelope. For these we expect tight orbits, which makes it possible to detect radial velocity variations.  
Indeed, most of the observationally identified subdwarfs in binary systems are accompanied by white dwarfs or late type companion stars \citep[][]{2001MNRAS.326.1391M, 2011MNRAS.415.1381C, 2015A&A...576A..44K, 2017A&A...605A.109V}.

Most striking is the scarcity of known stripped stars with OB type companions. Radial velocity campaigns of young star clusters and associations suggest that the majority of massive stars have a nearby companion. This implies that about a third of all massive stars get stripped by stable mass transfer before completing helium burning  \citep{2012Sci...337..444S}. We can use this to make a very rough estimate for how common stripped stars are. Stars then spend about a tenth of their total lifetime in their central helium burning phase. This means that we expect roughly $1/3 \times 1/10 \sim 3 \%$ of all massive stars to be stripped. This assumes a constant star formation rate. If most of these stripped stars still have an early-type companion, as expected from stable mass transfer, this means that a few percent of all early B and O type stars are hiding a stripped companion. For comparison, about a thousand massive O-stars are known \citep[e.g.][]{2011ApJS..193...24S, 2014ApJS..211...10S, 2016ApJS..224....4M}. For the B-type stars, the number is about an order of magnitude higher \citep[e.g.,][]{2003AJ....125.2531R}. We therefore expect several hundred stripped stars to \rev{be} hiding in the existing surveys. Note that these surveys only cover a fraction of the massive stars in our galaxy. Ten thousand O-stars are expected for the Milky Way alone \citep{2015MNRAS.447.2322R}.

The sample of observationally confirmed binary stripped stars with B-type main sequence companions is very sparse. It includes a handful of subdwarfs ($\varphi$~Per, 59~Cyg, 60~Cyg, and FY~CMa, all with rapidly rotating Be-type companions, see \citealt{1998ApJ...493..440G, 2008ApJ...686.1280P, 2013ApJ...765....2P, 2015A&A...577A..51M, 2017ApJ...843...60W}, and \cite{abels_paper}). Only one more massive stripped star  is known \citep[the quasi-WR in HD~45166,][]{2005A&A...444..895S, 2008A&A...485..245G}. No O-type star has a confirmed binary stripped companion in the intermediate mass range, although several Wolf-Rayet plus O-type star systems are known \citep[e.g.][]{van-der-Hucht2001,2007ARA&A..45..177C, 2017MNRAS.464.2066S}. 

Detecting stripped stars in systems with bright companions should be possible with various  established observational methods as we discussed in  \citetalias{2017A&A...608A..11G}. Techniques include searches for radial velocity variations, eclipses, ultraviolet excess, searches for emission lines typical to stripped stars that pierce through the continuum spectrum of the companion. Furthermore, stripped stars may be \rev{indirectly inferred by the presence of high ionisation nebular lines (e.g.\ nebular HeII~$\lambda 4686$) in nebulae associated with normal OB stars \citep[e.g.][]{1991PASP..103..850G}.}

In this section we use our new grid of theoretical spectra to investigate two of these techniques and assess how suitable they are for revealing the presence of stripped stars. We first study the composite spectra of stripped stars and their companion for different configurations in \secref{sec:different_binaries}. We then examine the feasibility of detecting stripped stars by searching for their UV excess in \secref{sec:UVexcess}, followed by a discussion of searches for emission lines belonging to the stripped star in \secref{sec:emexcess}.

\subsection{Composite spectra of a stripped star and a main sequence companion}\label{sec:different_binaries}


\begin{figure*}
\centering
\includegraphics[width=.9\hsize]{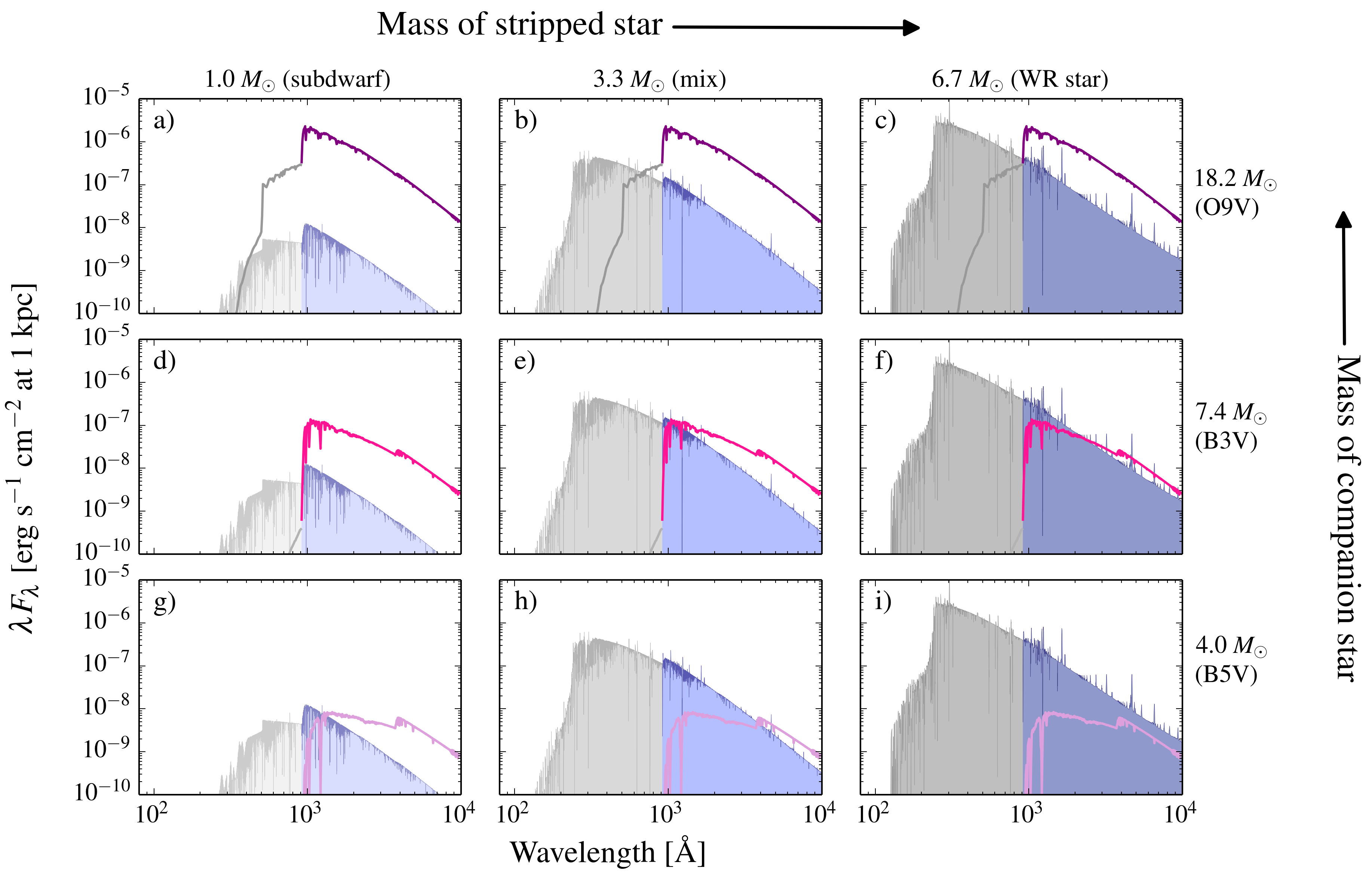}
\includegraphics[width=.9\hsize]{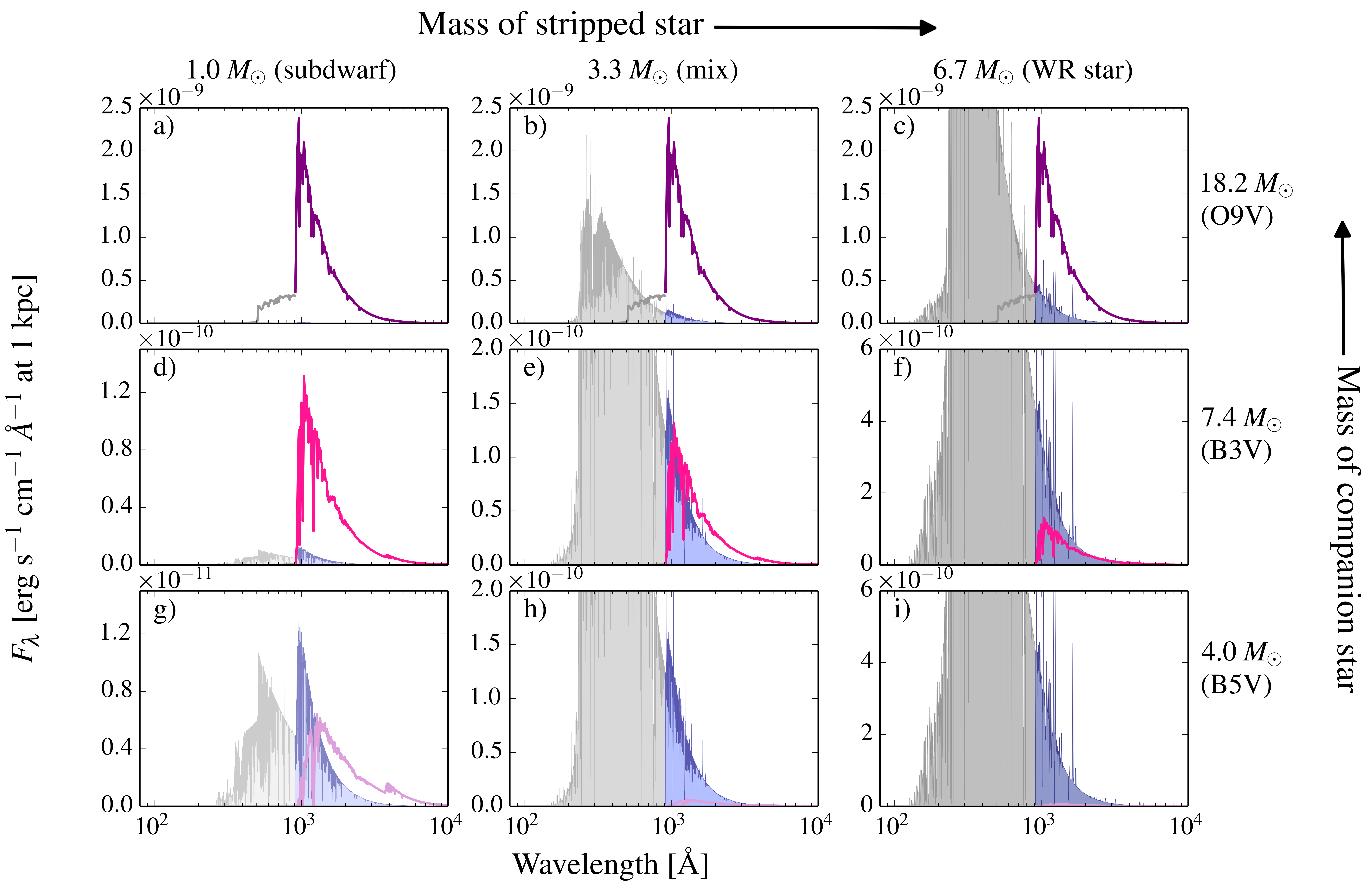}
\caption{Examples of binary systems containing a stripped star and a main sequence companion. We pick three typical stripped stars: a subdwarf (left column), a stripped star with both absorption and emission lines (middle column), and a WR star (right column). We pair each of these with three possible companions: a 4 (B5V, bottom row), 7.4 (B3V, middle row), and 18.2\Msun\ (O9V, top row) main sequence stars. The plots show the spectral energy distributions of the stripped star (blue shaded area) and its companion (solid line) in each combination. In the upper panel we use log-scale, while the bottom panel shows the same binary systems, but has linear scale. We have shaded the ionizing part of the spectra in gray shading as this part is difficult to observe due to the neutral hydrogen in the solar neighborhood. There are also no available instruments that are capable to observe in the ionizing wavelengths.}
\label{fig:grid_composites}
\end{figure*}

\begin{figure}
\centering
\includegraphics[width=\hsize]{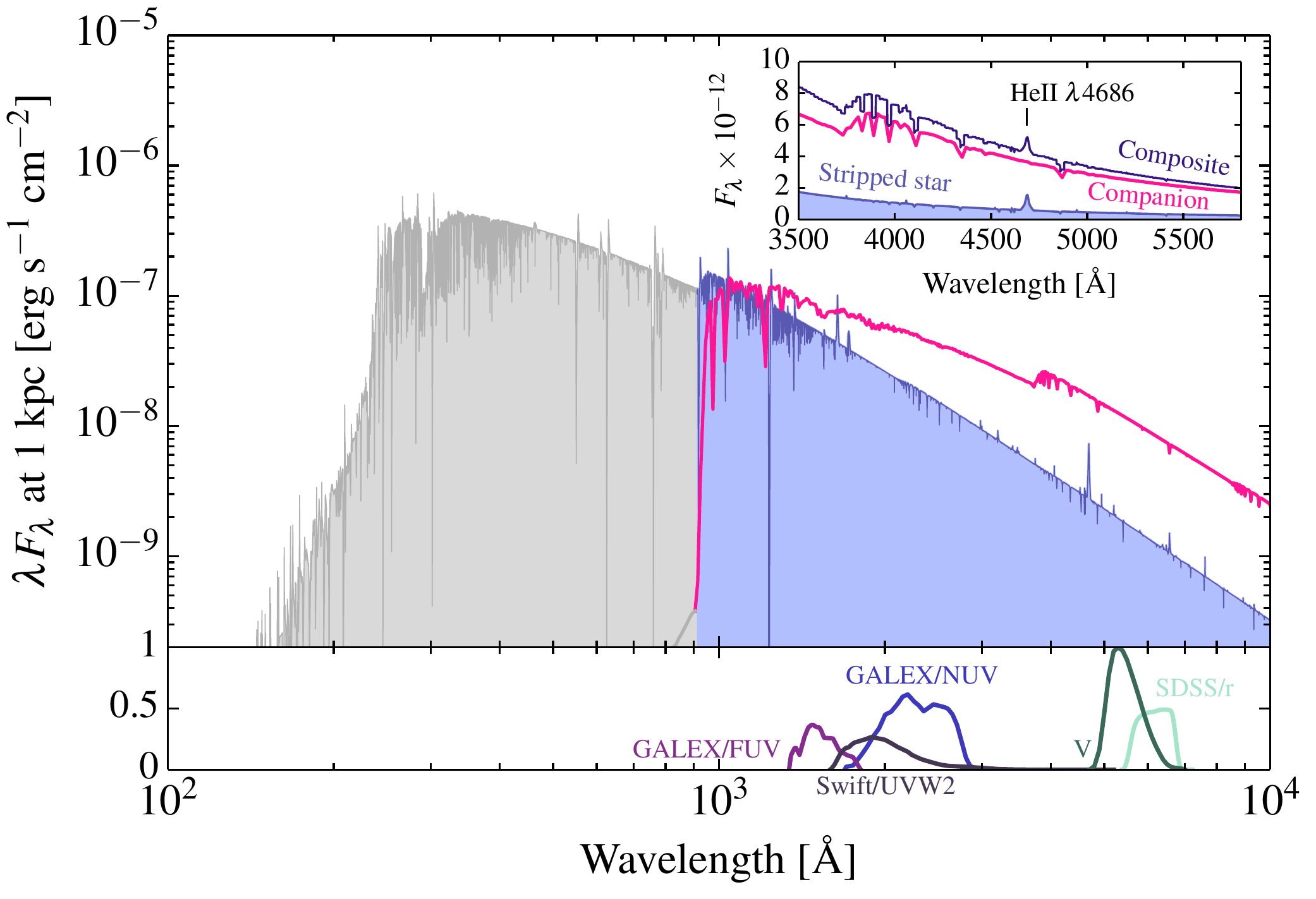}
\caption{\rev{Spectral energy distribution of a 3.3\Msun stripped star and a 7.4\Msun companion star (analog to panel e of \figref{fig:grid_composites}). We show underneath the plot several photometrical filters in the UV and optical from \textit{GALEX}, \textit{Swift} and \textit{SDSS}. These are discussed in \secref{sec:UVexcess} for detecting stripped stars via UV excess. We show a zoom-in of the optical spectrum where the composite spectrum is shown in purple. The HeII~$\lambda 4686$ emission line from the stripped star is visible in the composite spectrum. This line may be used for detecting stripped stars as discussed in \secref{sec:emexcess}.}}
\label{fig:zoom_filters}
\end{figure}

In a realistic binary system, we expect the stripped star to dominate the extreme ultra-violet part of the spectrum. The optical part of the spectrum may either (1) be dominated by the main sequence companion, or (2) the stripped star and the companion star have a comparable brightness in the optical, or (3) the stripped star dominates, depending on the configuration of the system. 
To illustrate these cases we pair three representative examples of stripped stars with three examples of possible companions, creating a total of nine examples that are shown in \figref{fig:grid_composites}. In the top nine panels we show the spectra on a standard logarithmic scale, which allows more clearly to see the spectra of both components.  In the bottom nine panels show the same systems using a linear scale, to highlight the relative contribution more clearly. 
 
The stripped star spectra shown here are taken from our solar metallicity grid. They are for a typical subdwarf (group A, shown in the left column), an intermediate mass  stripped star with both absorption and emission lines (group A/E, shown in the central column), and a Wolf-Rayet-like star (group E, shown in the right column). They are for stripped stars with  masses of 1.0, 3.3, and 6.7\Msun (corresponding to initial masses of 4.5, 11.0, and 18.2\Msun). 
We show their spectral energy distribution with a shaded background. We use gray-shading for the ionizing part of the spectrum. 
\rev{The Extreme Ultraviolet Explorer (EUVE) was sensitive to the Lyman continuum, though early-type stars were blocked by interstellar hydrogen except along rare low column density sight lines \citep[][]{1995ApJ...438..932C, 1996ApJ...460..949C}.}
The UV and optical part of the spectrum that is accessible to present-day facilities is shown with a blue-shaded background. The spectra of the companions are shown with a thick, solid pink/purple lines. 

The companions shown here are for $\sim$4.0, 7.4 and 18.2\Msun main sequence stars shown in the bottom, middle and top row respectively. To estimate their spectra we evolved single stars with the same evolutionary code and settings as described in \secref{sec:modeling} until the central hydrogen mass fraction dropped to $X_{\rm H,c}= 0.5$. This means that they are assumed to be relatively unevolved. This is expected for low-mass companions, which evolve more slowly, but also for more massive companions that may have been rejuvenated as a result of mass accretion. Based stellar properties derived from these models we assign approximate spectral types of B5V, B3V and O9V. The spectra shown here are Kurucz spectral models \citep{1992IAUS..149..225K}.

The main sequence companion star dominates the optical spectrum in most of the example systems shown in \figref{fig:grid_composites} (examples a, b, c, d, e, and g). This is the case for all combinations with an O-star. \rev{\citet{1991A&A...241..419P} found the same result when comparing spectral models of massive subdwarfs with early B-type star models.} \rev{A zoom-in of panel e, \figref{fig:grid_composites} is shown in \figref{fig:zoom_filters}.} The stripped star dominates the spectrum in only one of the example systems (example i). In this case, the stripped star is a few times brighter in the optical, even though the bolometric luminosity of the stripped star is about three orders of magnitude larger compared to that of the companion. In two of the example systems, the stripped star has similar optical brightness as the companion star (examples f and h). Then, the spectral features of both the stripped star and its companion are clearly distinguishable in the composite spectrum. 

We note that not all combinations shown here are equally likely to occur. The mass of the companion depends on the process that is responsible for stripping.  In this work we considered stripped stars resulting from Case B mass transfer, but stripped stars formed through other stripping mechanisms are expected to have similar surface properties \citep[e.g.,][]{2010ApJ...725..940Y, 2011ApJ...730...76I}, so it is worth considering the other mechanisms here and what the implications are for the mass of the companion.  

The most massive companions are expected in systems that evolve through stable conservative mass transfer, where the companion has \rev{accreted} the entire envelope of the stripped star,  $ M_{1, envelope} = M_{\rm 1, init} - M_{\rm 1, strip}$ for case~B mass transfer.  We further know that the companion should have been the initially less massive star in the system, $M_{\rm 2, init} \le M_{\rm 1, init}$.  \rev{This gives an upper limit on the mass of the companion}, $M_{2, \max} = M_{\rm 1, init}  +  M_{1, envelope}$, which is 8, 18.7, \rev{and} 30 \Msun respectively for the examples shown here.  Binary systems with stripped stars resulting from case~A mass transfer may slightly exceed this limit, since mass transfer starts before the helium core has been fully established.   

Stripped stars with low-mass companions are expected from unstable mass transfer followed by successful ejection of the common envelope. This is because (1) the companion does not significantly accrete in this scenario and (2)  unstable mass transfer preferentially occurs for systems with more extreme initial mass ratios, $M_{2, init}/M_{1, init} \lesssim q_{\rm crit}$. The threshold value a matter of debate, but it is reasonable to assume that systems with $q_{\rm crit} \lesssim 0.25$ are certainly unstable \citep[e.g.][]{2017MNRAS.471.4256V}.  Such system would produce stripped stars with main sequence companions that have a mass that is comparable or lower than the stripped star. 

Of the panels shown in \figref{fig:grid_composites} we thus expect panel e), f), g) h) and i) to be  typical cases for binary interaction with various degrees of non-conservative mass transfer. The situation in panels b) c) and d) requires rather conservative stable mass transfer.  We do not expect the situation in panel a), at least not from our current understanding of binary interaction.

\subsection{Searching for stripped stars through UV excess}\label{sec:UVexcess}

Even if a stripped star is too faint to detect \rev{at} optical wavelengths, it may introduce a detectable excess of UV radiation, compared to what is expected from the companion star alone (cf. \figref{fig:grid_composites}). 
Searches for UV excess have indeed been successful in revealing and characterizing stripped stars orbiting Be-type stars \citep[][]{1998ApJ...493..440G, 2008ApJ...686.1280P, 2013ApJ...765....2P, 2017ApJ...843...60W}. These studies used spectra taken with the International Ultraviolet Explorer (\textit{IUE}) and the Goddard High Resolution Spectrograph, (\textit{GHRS}) onboard the Hubble Space Telescope (HST). The contribution of the stripped star to the UV flux in these systems is estimated to range from a few percent up to tens of percent. 

Photometric UV surveys may be even more promising given their large sky coverage. The now decommissioned satellite \rev{Galaxy Evolution Explorer} \citep[\textit{GALEX},][]{2005ApJ...619L...1M} was designed to search for UV bright sources and all-sky surveys have been carried out \citep{2011Ap&SS.335..161B}. The survey contains deep observations using the near-ultraviolet filter (\textit{NUV}, $\lambda \sim 1800 - 2800$ \AA, up to $\sim 25$~mag). \rev{(Observations with the far-ultraviolet filter of \textit{GALEX} are also included, but these go less deep as technical issues arose during the mission.)} The resulting data is available in their open archive. Also the Swift's Ultraviolet/Optical Telescope (\textit{UVOT}, \citealt{Roming+2005}) is of interest in this context. It is still operational and also has archival data available. \rev{The field of view and sky coverage of the \textit{Swift/UVOT} is however much smaller compared to those of \textit{GALEX}.}


In the remainder of this section we use our spectral model grid to study the effectiveness of searches for UV excess to search for stripped companions using \textit{GALEX} as a case study. We note that the following results are very similar if we had adopted the corresponding filters for Swift instead.


Detecting a UV excess is possible when the distance to the star is accurately known, but this is not the case for many stars in the Galaxy. Instead, using a UV color excess, which here represents the comparison between the UV and optical flux, removes the distance dependency. Detecting a UV color excess can be done if the spectral type of the companion is known, as the UV color from the companion can be estimated. We calculate the UV color \textit{GALEX}/\textit{NUV}$-$\textit{SDSS}/\textit{r} using the filter functions with the composite spectra of each binary system and the companions alone. The $r$ filter from Sloan Digital Sky Survey (\textit{SDSS}) is a broad-band filter covering optical wavelengths ($\sim 5500 - 6800\, \AA$). \rev{We chose this filter as a demonstration, but the technique is also applicable when using other optical or infrared filters as long as the stars are detectable. The SDSS has a large database \citep[see e.g.][]{2009ApJS..182..543A} and is therefore suitable.} We then compare the color difference in magnitudes between the composite spectra and what is expected from the companion alone. In this way we can determine the UV color excess that the stripped stars introduce. \rev{Both the \textit{GALEX/NUV} and the \textit{SDSS/r} filters are shown in \figref{fig:zoom_filters} as a comparison to the binary system from panel e of \figref{fig:grid_composites}.}

\begin{figure*}
\centering
\includegraphics[width=0.7\hsize]{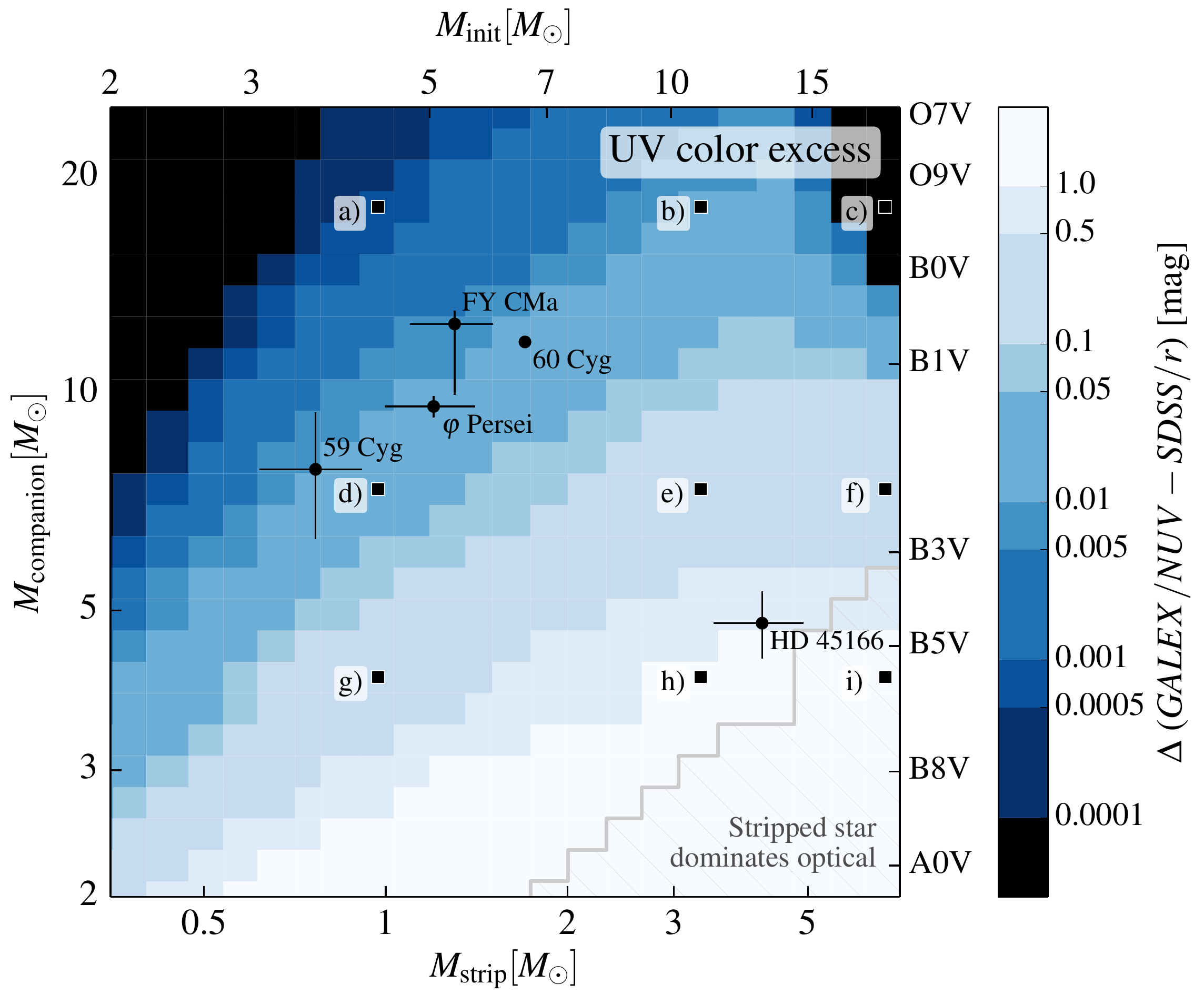}
\caption{The excess in UV color (\textit{GALEX/NUV} $-$ \textit{SDSS/r}, blue shading) of binary systems containing a stripped star compared to the UV color of the main sequence companion only. A detectable UV excess could be $\gtrsim 0.05$~mag depending on the instrument. We show binary system combinations covering wide mass ranges of stripped stars and companion stars. The stripped star dominates the optical emission in systems with low-mass companions and massive stripped stars (gray hatched region). The example systems of \figref{fig:grid_composites} are marked with black squares and labelled with corresponding panel letter. The few binaries containing B-type stars and detected stripped stars are marked and labelled in black.}
\label{fig:UVexcess}
\end{figure*}

We calculate the UV color excess for our spectral model grid of stripped stars, paired with companion stars within the mass range $2-25\Msun$. The result is shown in \figref{fig:UVexcess} using blue shading, where lighter color represent larger UV color excess. If we consider a detectable UV color excess to be above 0.05 mag, then about half of the considered binary systems have a detectable UV color excess. The example systems shown in \figref{fig:grid_composites} are marked in \figref{fig:UVexcess}, and indeed the stripped stars in the  examples a), b), c) and d) appear to have introduced a very small excess of UV radiation. The stripped stars in the examples e), f), g), h) and i) appear on the other hand detectable. The parameter space where the stripped star dominates the composite spectrum (gray hatch) is small and largely overlaps with systems that would have undergone common envelope evolution (see \secref{sec:different_binaries}). 


For reference we have over plotted the currently known observed systems with detected stripped stars in \figref{fig:UVexcess}. We expect the observations to still be be too incomplete to draw conclusions, but some interesting effects are already visible. The subdwarfs shown here all have rather massive companions. All of them would require rather conservative mass transfer to \rev{achieve} their present day mass ratios. Whether this is telling us something about the physics of the interaction of these system, or whether this is purely the result of biases in the current surveys remains to be investigated. HD~45166 is found in the opposite corner of the diagram. It is the only observed stripped star in the mass ranges considered here that shows emission lines. The location in \figref{fig:UVexcess} suggests that HD~45166 should have similar optical contribution from both stars, which indeed is the case \citep{2008A&A...485..245G}. The stripped star is clearly visible in the composite spectrum \citep{2005A&A...444..895S}. 

Most striking is the current ``zone of avoidance'' visible in \figref{fig:UVexcess}. We are currently completely lacking detections of (massive) subdwarfs with companions in the mass range 2-8 \Msun. This is remarkable, since these would be easier to detect through their UV excess. We note that  binary evolutionary models predict a particular distribution of systems in this diagram, which is sensitive to assumptions about the efficiency of mass transfer, core overshooting and the initial distributions of binary parameters. Filling this diagram with more stripped star systems (or understanding whether there are true zones of avoidance) will be of great value to test the evolutionary models.

An effect that we have not accounted here is absorption by interstellar gas and dust attenuation. UV searches for companions to massive stars within the Galactic plane is hindered by dust absorption, which is more severe at ultraviolet wavelengths than visually \citep[e.g.][]{1979MNRAS.187P..73S, 1989ApJ...345..245C}. For a standard Galactic extinction law, the ratio of far-UV to visual extinction, $A_{\text{FUV}}/A_{V} = 2.6$ so the far-UV fluxes of nearby stars with $A_{V} \sim 0.5$ ($E(B-V) \sim 0.15$) will be reduced no more than a factor of 3, but more distant stars with $A_{V} \sim 2$ ($E(B-V) \sim 0.65$) will be suppressed by a factor of 100. Consequently, UV surveys of the Galactic plane are limited to the nearest $1-2$~kpc, favoring surveys of massive stars in the low extinction Magellanic Clouds for which typically $A_V < 1$~mag. One other complication of the Galactic plane at UV wavelengths is that significant deviations from the standard extinction law are observed, affecting both the slope of the UV extinction and the strength of the 2200~\AA\ absorption feature \citep[e.g.][]{2007ApJ...663..320F}. \rev{We discuss the effects of dust attenuation on stripped star spectra and their detection methods in \appref{app:dust}.}

The four subdwarf + Be type systems have however a detected UV excess similar to the predictions from our models \citep{1998ApJ...493..440G, 2008ApJ...686.1280P, 2013ApJ...765....2P, 2017ApJ...843...60W}, which indicates that the technique may still be applicable for systems are along low density lines of sight.

\subsection{Searching for stripped stars using their emission features}\label{sec:emexcess}

\begin{figure}
\centering
\includegraphics[width=\hsize]{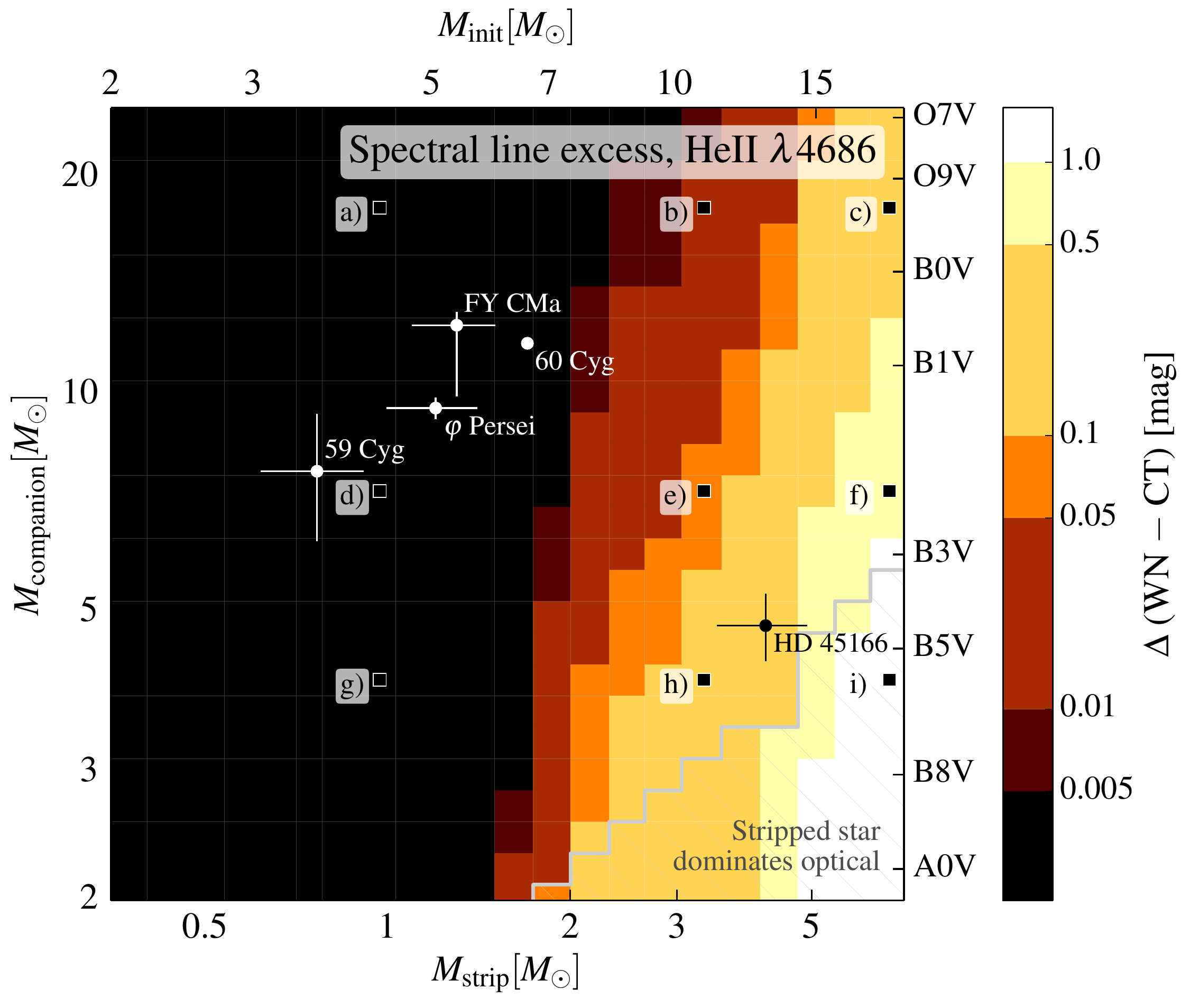}
\caption{The excess in HeII~$\lambda 4686$ emission (red shading) introduced by the stripped star compared to the expectations from the companion alone. More massive stripped stars have stronger wind mass loss rate and therefore show emission in HeII~$\lambda 4686$, which may be detectable. This figure is analog to \figref{fig:UVexcess}.}
\label{fig:HeII4686}
\end{figure}


An alternative technique to find stripped stars is to search for their emission lines, see also our discussion in \citetalias{2017A&A...608A..11G}. For certain combinations of stripped stars and companions, these emission lines may be strong enough to be detectable in the composite spectra.  The strength of the these emission lines scales directly with the (assumed) wind mass loss rates.  This technique is therefore most promising for more massive stripped stars which have stronger winds. 
This technique has already been applied to search for Wolf-Rayet type stars,  mainly using narrow-band photometry of the HeII~$\lambda 4686$ emission line \citep{1979A&A....75..120A, 1991AJ....102..716S, 1998ApJ...507..199D, 2001ApJ...550..713M, 2014ApJ...788...83M, 2015ApJ...807...81M, 2017ApJ...837..122M}. 

HeII~$\lambda 4686$ is indeed an appropriate line to search for, as it is the strongest emission feature in the optical spectrum of stripped stars that show emission lines. Also, the blend of H$\alpha$ and HeII~$\lambda 6560$ may be advantageous to use as a significant amount of archival data exists \citep[see e.g., IPHAS or VPHAS+,][respectively]{2005MNRAS.362..753D,2014MNRAS.440.2036D}. The blended line is however weaker than the HeII~$\lambda 4686$ feature. The UV-line HeII~$\lambda 1640$ is also strong and may also be advantageous in the searches for stripped stars. \rev{As a demonstration, we show a zoom of the optical range in \figref{fig:zoom_filters} of panel e, \figref{fig:grid_composites}. The HeII~$\lambda 4686$ line is clearly visible in the composite spectrum of this example system.}

Searching for this emission line can be done most efficiently with a narrow-band filters centered on the HeII~$\lambda 4686$ line. Many telescopes are equipped a filter that can be used for this. This includes the \textit{F469N} filter available for the Ultraviolet-Visible channel of the WideField Camera 3 (\textit{WFC3/UVIS}) onboard \textit{HST}. But smaller telescopes on the ground provide a more cost-effective way to conduct surveys. 

One of the most comprehensive searches for stars with excess emission $4686\, \AA$ has been conducted by \citet{2014ApJ...788...83M}, \citet{2015ApJ...807...81M} and \citet{2017ApJ...837..122M} using the 1-m Swope telescope on Las Campanas. We use their setup as a case study here. They use a narrow-band filter centered around HeII~$\lambda 4686$ (WN) and continuum filter centered at 4750 \AA\ (CT). Their filters have a 50~\AA\ bandpass (full width at half maximum, FWHM)

We use our composite models to calculate the color $WN - CT$ for each combination of stripped star with a main sequence star and also for the main sequence stars alone.  If the stripped star has strong emission of HeII~$\lambda 4686$, a color difference between the companion star alone and the binary system will be distinguishable. This color excess can be measured if the spectral type of the companion is known. If that is not the case, most companion stars will not show significant features around or in HeII~$\lambda 4686$ and therefore just detecting excess in $WN$ compared to $CT$ may suggest the presence of a stripped star. However, using simply $WN - CT$ is more approximate compared to also accounting for what is expected from the companion. 

Our results are shown in \figref{fig:HeII4686}. where we show the $WN-CT$ excess with colors, brighter representing a larger color excess.  The diagram shows that this search technique is biased towards more massive stripped stars (or stated more precisely, stripped stars that have strong enough stellar winds).  Stripped stars with pure emission line spectra (group E) should be detectable for a wide range of companion masses.  Stripped stars of the transitional class that show both absorption and emission lines (group A/E), are detectable with most B-type companions.  

The only intermediate mass stripped star known to date, HD~45166, indeed has an emission line spectrum and \figref{fig:HeII4686} predicts a detectable narrowband color excess.  
We also overplot the known subdwarf-type stripped stars.   This method is clearly not suitable to detect those since subdwarf type spectra does not show emission features. 

We note that our predictions here depend on the assumed mass loss rates, which are still uncertain  \citep[e.g.][]{2017A&A...607L...8V}. Turning this around, we believe that narrow band photometric searches combined with proper modeling of the population and the biases, may be an interesting and potentially powerful way to constrain the wind mass loss rates.


\section{Discussion of model uncertainties}\label{sec:uncertainties}

Our model predictions are subject to various uncertainties. This is largely because of the current scarcity of observed stripped stars. To obtain the results presented here, we adopt assumptions that appear reasonable at this moment, but we anticipate that our insight will improve drastically as high quality data of observed stripped stars \rev{become} available. Our calculations should be considered as a first step towards realistic model predictions. Despite the uncertainties, we believe that they are suitable enough to provide insight and to guide the design of observational searches. We briefly discuss the main uncertainties below and discuss how they may affect our findings. This should be kept in mind by anyone who wishes to use these model grids for other purposes such as direct comparisons with observations. 

\paragraph{Stellar Winds}
The treatment of stellar winds constitutes the primary uncertainty in our current work. The mass loss rates of stripped stars are not well constrained at present. The same is true for the clumping factor,  the terminal wind speed and the velocity profile that characterizes the wind outflow. The sensitivity of our findings to these assumptions is in fact something positive. It means that observed spectra of stripped stars will be extremely useful to derive empirical mass loss rates in the near future. However, for now, we rely on extrapolated recipes and theoretical estimates.  

The wind mass loss rate we predict with our current implementation agrees very well with the mass loss rate inferred for the observed  stripped star in HD~45166 \citep{2005A&A...444..895S, 2008A&A...485..245G}. This stripped star has a mass of about 4\Msun\ and is currently providing the only empirical data point in this mass regime. 
We also find a smooth transition in wind mass loss rate from the subdwarfs up to the WR stars. This gives some support that the assumptions we have adopted are not unreasonable. 

However, the mass loss rates of stripped stars are subject of debate. For example, \citet{2016ApJ...833..133T} revisited the mass loss rate of Wolf-Rayet stars and argue in favor enhanced mass loss rates, especially in the later stages. \citet{2017MNRAS.470.3970Y} argues that these new prescriptions are in better agreement with the observed dichotomy of type Ib and Ic supernova. Instead,  \citet{2017A&A...607L...8V} recently presented new theoretical models for helium stars with masses in the range 1--60\Msun that predict mass loss rates that are nearly an order of magnitude lower than what we have assumed.   

Such substantial changes in the mass loss rate (and other parameters that describe the wind)
would have important effects on the morphological characteristics of the stellar spectra. In particular, it would affect the appearance and strength of the emission lines as we showed in  Fig.~6 of \citetalias{2017A&A...608A..11G}. The substantial reduction of the wind mass loss rates as proposed by \citet{2017A&A...607L...8V} would push the transition between absorption line (group A) and emission line (group E) spectra to higher masses (and metallicities). Our main prediction of the existence of a transitional spectral class (group A/E) remains unaffected. A reduction in the wind mass loss rates only affects the the mass and metallicity at which this class occurs.  

One argument against the downward revision in wind densities proposed by \citet{2017A&A...607L...8V} is that they strongly under predict mass-loss rates of Galactic WR stars compared with empirical results. By way of example, \citet{2017A&A...607L...8V} predicts a mass-loss rate of $10^{-5.9}$~\Msunyr and high wind velocity of 3800~\kms for a 15\Msun stripped star at solar metallicity, yet \citet{2006A&A...457.1015H} derived clumping corrected mass-loss rates in the range $10^{-5.3}$ to $10^{-4.2}$~\Msunyr for hydrogen-free WN stars with $\log_{10} (L/\Lsun) \sim 5.5$, with typical wind velocities of 2000~\kms.

The impact of mass loss rate uncertainties have little effect on the overall shape of the spectral energy distribution, except for extreme cases where the outflows are optically thick and the photosphere moves outwards.  This only affects the most luminous and metal-rich models in our grid. We further note that an enhancement of the stellar wind mass loss rate can affect the emission of photons with energies in excess of 54.4 eV,  the threshold for helium ionization. This is illustrated in Fig. 4 and Fig. 11 of \citetalias{2017A&A...608A..11G}, see also \citet{1997A&A...322..598S} for a discussion.  We also explored  the impact of the assumed terminal wind velocity and volume filling factor which affects the shape and strength of the emission lines, as can be seen in  Appendix B.1 of \citetalias{2017A&A...608A..11G}. 

\rev{Because of the uncertainties in wind mass loss rate, we consider the most accurate spectral models in our grids to be those with low wind mass loss rate and thus absorption line spectra (i.e., subdwarfs). Even though many of the models with emission line features closely resemble classical WN stars and  WN3/O3 stars, we expect that they will need to be updated when observational constraints become available.}

\paragraph{Mixing and gravitational settling}

Our predictions for the surface properties of stripped stars are sensitive to the detailed assumptions about the mixing processes that occurred above the convective hydrogen burning core of the progenitor star and possible mixing in the layer above the hydrogen burning shell. These mixing processes constitute \rev{a} long-standing uncertainty in all stellar evolutionary models. In our case, they affect the mass of the resulting stripped star and the details of the chemical profile near the surface of the stripped star. 

\rev{The lowest mass stripped stars in our grid are affected by gravitational settling. This, in addition to the mixing above the convective core of the progenitor star when it still resided on the main sequence.} We account for this in our models, but we do not have a proper treatment of the effects of radiative levitation and possibly further turbulent mixing processes that can (partially) cancel the effects of gravitational settling. We performed test simulations with and without the effect of gravitational settling switched on and we find that the impact on the the stellar structure is small, but the effect on the surface abundances themselves are large. This should be kept in mind, especially when using this model grid for comparison with observations. 


\section{Summary \& Conclusions}\label{sec:summary}

In this paper we presented the first comprehensive grid of spectral models for stars stripped in binaries, expanding upon \citetalias{2017A&A...608A..11G}, which forms a series with this work. Our spectra result from radiative transfer simulations of the atmospheres, which we tailored to match our models for the stellar structure of stripped stars. To obtain the structure models we followed the evolution of progenitor stars and its interaction with a companion. We considered masses spanning 2-20\Msun for the progenitor star and metallicities ranging from solar 
down to metal-poor values that are representative for population~II stars.
We analyze the structure and spectral models to learn about the nature of stripped stars. We summarize the main findings of our paper below.  


\begin{enumerate}

\item Our stripped stars formed through stable interaction with a companion form a continuous sequence closely resembling subdwarfs at the low mass end, while those at the high mass are indistinguishable from Wolf-Rayet stars. Multiple formation channels may contribute to their formation but {\it our finding does unify subdwarfs and Wolf-Rayet stars as the possible outcome of the very same evolutionary scenario.}
\newline

\item The resulting stripped stars are characterized by very high effective temperatures ($20\,000-100\,000$~K, increasing with mass), 
high effective surface gravities ($\log_{10} g \sim 5.6-5.3$) and small radii ($0.2-1\Rsun$). Their bolometric luminosities are comparable with those of their progenitors, despite having lost nearly two thirds of their mass, following a steep mass luminosity relation ($L\propto M_{\rm strip}^{3.3}$). 
\newline

\item We identify a hybrid spectral class simultaneously showing absorption lines originating from the stellar surface as well as WR-like emission lines resulting from a semi-transparent stellar wind outflow. We find these for stars with relatively weak stellar winds, $10^{-8}-10^{-6} \Msunyr$, corresponding to stripped stars with masses ranging $2-5\Msun$. These boundaries shift up for lower metallicities and are sensitive to the assumed mass loss rates. We argue that observationally identifying such stars will be very helpful to get empirical constraints on the mass loss rates for stripped stars. 

These spectra \rev{closely resemble} 
the recently discovered class of WN3/O3 stars, at least one of which is found in an eclipsing binary \citep[][and references therein]{2017ApJ...841...20N}. 
This has raised the hypothesis that WN3/O3 are the long-sought products of envelope stripping in the intermediate mass regime, coinciding with the mass range of the progenitors of Ib/c supernova \citep{2018MNRAS.475..772S}. This hypothesis can be investigated with a sensitive radial velocity campaign, since we expect the companion responsible for stripping to still be around, although it may have low mass and reside in a relatively wide orbit.
\newline


\item We show that stable mass transfer can lead to the formation of subdwarfs with a wide range of masses ($0.35-1.7\Msun$) and luminosities ($10^{0.6}-10^{3.2}\Lsun$). This contrasts sharply with narrow mass distribution expected from formation through common envelope ejection, which peaks sharply at the canonical value 0.47\Msun. Our findings thus question the validity of adopting the canonical value, for example when making inferences about the distances or for the masses of their unseen companions.  

If indeed a substantial population of overweight subdwarfs exist, we predict that Gaia should identify these as overluminous objects after measuring the distances to the more than 5000 subdwarfs candidates  that have been identified \citep{2017A&A...600A..50G}. 
\newline

\item Mass transfer in binaries create WR stars with masses that are substantially lower than usually considered (down to $5 \Msun$ in our solar metallicity grid). This prediction is sensitive to our assumptions for the stellar wind mass loss rates, which are still poorly constrained at present.  
\newline

\item At low metallicity ($Z \leq 0.0002$), we find that the Roche-lobe stripping process is inefficient. The stripped stars can retain up to $\sim0.5\Msun$ of pure hydrogen, which is sufficient to support hydrogen burning in a shell around the core that provides up to 30\% of the total luminosity. These stars are substantially larger (up to $\sim 8\Rsun$) and cooler (down to $\sim 36\,kK$). With surface mass fractions of 0.6 in hydrogen and their weaker mass loss rates, their spectra are expected to appear as lower luminosity counterparts to bright early O-type stars.
\newline

\item Various biases make the detection of stripped stars a non-trivial endeavor. One of them is the high likelihood of the companion star to still be present and outshine the stripped star, at least in optical bands.  We explore the biases by pairing our stripped stars with possible companions and investigate the feasibility of detecting stripped stars through two  (a) their UV excess and (b) searcher for their emission features. We show that the two techniques are complementary and probe different regimes of the parameter space. The first allows to detect stripped companions around B-type stars and later. The second technique appears promising for the detection of stripped stars that have wind mass loss rates larger than about $10^{-7} \Msunyr$, which also allows for detecting stripped stars with O-type companions.  
\newline
\end{enumerate}

\noindent The models we have presented are still subject to uncertainties. For higher mass stripped stars, the main uncertainty is the mass loss rate, which has a large impact on the appearance of emission lines. For lower mass stripped stars, we consider the process of gravitational setting as the main uncertainty, which directly affects the surface abundances. Several programs to search for and characterize stripped stars are currently underway. We also expect Gaia to play a major role. 

Our models can be used for a variety of other applications. In a forthcoming separate paper we will discuss the contribution of stripped stars to the the budget of ionizing photons emitted by stellar populations. We further anticipate direct comparison with observations and possible inclusion in spectral synthesis codes. We therefore make our full grids of stellar and spectral models available as service to the community.


\begin{acknowledgements}
The authors acknowledge various people for helpful and inspiring discussion at various stages during the preparation of this manuscript, including Evan Bauer, Jared Brooks, Maria Drout, JJ Eldridge, Chris Evans, Rob Farmer, Miriam Garcia, Stephan Geier, Zhanwen Han, Edward van den Heuvel, Stephen Justam, Lex Kaper, Alex de Koter, S{\o}ren Larsen, Danny Lennon, Pablo Marchant, Colin Norman, Philipp Podsiadlowski, Onno Pols, Hugues Sana, Tomer Shenar, Nathan Smith, Elizabeth Stanway, Silvia Toonen, Jorick Vink, and the VFTS collaboration.
YG, SdM, and MR acknowledge acknowledge hospitality of the Kavli Institute for Theoretical physics, Santa Barbara, CA. Their stay was supported by the National Science Foundation under Grant No.\ NSF PHY11-25915.
This work was carried out on the Dutch national e-infrastructure with the support of SURF Cooperative. The authors acknowledges John Hillier for making his code, CMFGEN, publicly available. YG thank Martin Heemskerk for providing computing expertise and support throughout the project and Alessandro Patruno for allowing us to use the Taurus computer.
SdM has received funding  under the European Union's Horizon 2020 research and innovation programme from the European under the Marie Sklodowska-Curie (Grant Agreement No. 661502) and the European Research Council (ERC) (Grant agreement No. 715063).

\end{acknowledgements}


\bibliographystyle{aa-package/bibtex/aa.bst}
\bibliography{references_bin.bib,my_bib.bib}



\appendix

\rev{\section{Interstellar dust extinction}\label{app:dust}


\begin{figure*}
\centering
\includegraphics[width=0.8\textwidth]{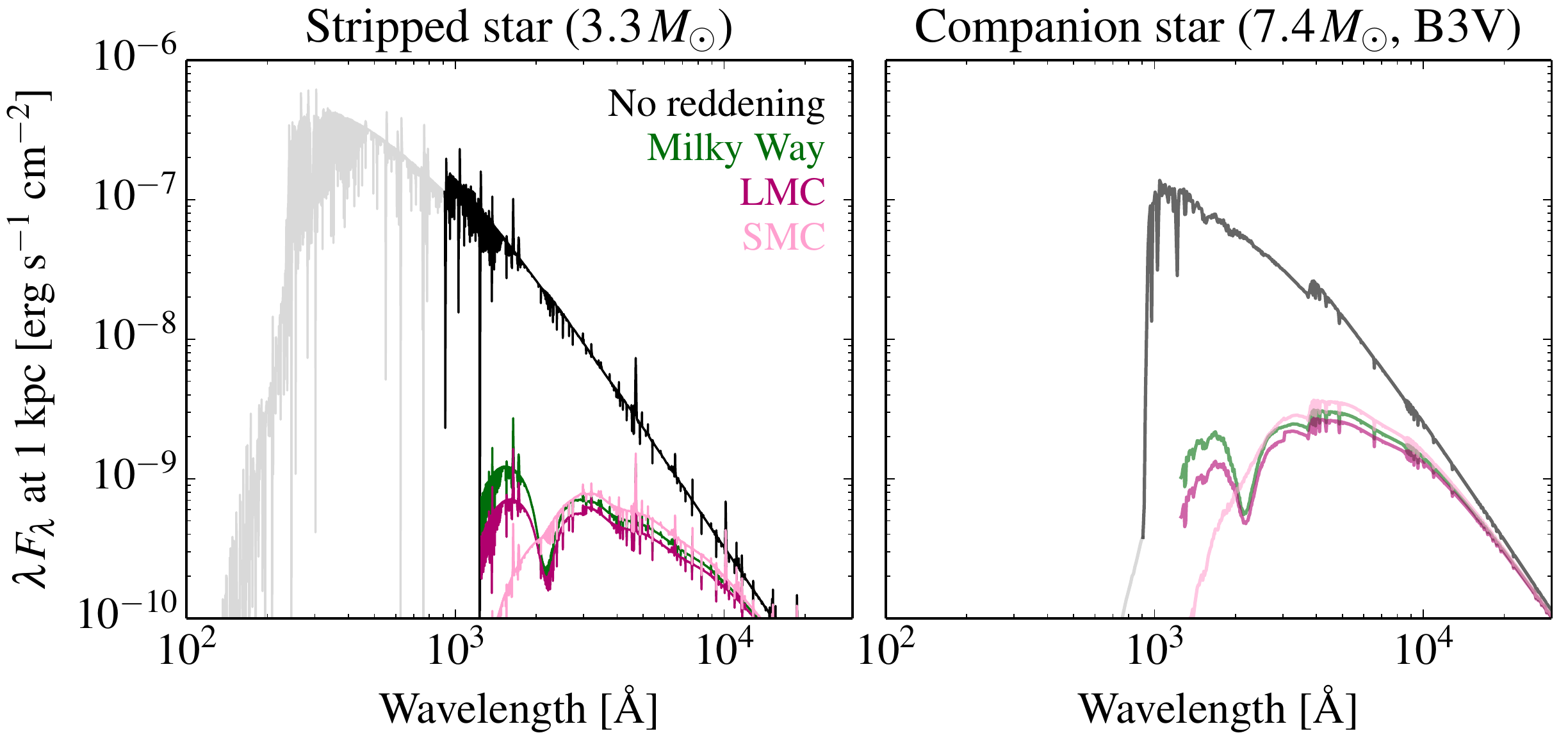}
\caption{The effect of interstellar dust reddening shown in color for a stripped star (left) and a possible companion star (right). Green, purple and pink shows the spectra reddened using extinction laws for the Milky Way, the LMC and the SMC respectively \citep[see][]{1989ApJ...345..245C, 2003ApJ...594..279G}. We have applied $E(B-V) = 0.5$~mag for all considered environments. Black shows the un-reddened spectra. We shade the ionizing part of the spectra in gray.}
\label{fig:reddened_example}
\end{figure*}

Interstellar dust attenuates the radiation of stars and shifts it towards redder wavelengths. This effect primarily makes stars fainter, especially in the ultraviolet. In this appendix, we estimate the effect of dust attenuation on the spectra of stripped stars.

We use the dust attenuation laws of \citet{1989ApJ...345..245C} and \citet{2003ApJ...594..279G} to estimate the effects of dust on the spectra of stripped stars and their companions. We account for extinction of stripped star binaries in the Milky Way, in the LMC and in the SMC. When calculating the extinction, we use for V-band extinction $A_{V} = R_{V} \times E(B-V)$ and assume a selective extinction of $R_{V} = 2.74$ (SMC), 3.41 (LMC), and 3.1~mag (Milky Way) following the observations of \citet{1989ApJ...345..245C} and \citet{2003ApJ...594..279G}. For reddening we use $E(B-V) = 0.5$~mag for all environments, which in the Milky Way is typical for a distance of $1-2$~kpc within the Galactic disk. \figref{fig:reddened_example} shows as an example how dust affects the spectra of a stripped star (3.3~\Msun) and a potential early B-type companion (7.4\Msun) in the three different environments (this system corresponds to panel e of \figref{fig:grid_composites}).  The shape of the reddened spectra varies between environment owing to differences in grain size and composition. We note that the amount of reddening, $E(B-V)$, probably is lower in the SMC and may reach much higher values in the Milky Way, compared to what we assume.

The ultraviolet flux is one to two orders of magnitudes lower when accounting for dust compared to when dust is not accounted for. This difference corresponds to a change of about four magnitudes in the \textit{GALEX} filters, and therefore puts constraints on which stripped stars will be detectable in the \textit{GALEX} data. Including dust attenuation and considering a magnitude limit for \textit{GALEX/NUV} of both 22 and 20~mag \citep{2014AdSpR..53..939S}, stripped stars with masses down to about 1.1 and 2.8~\Msun respectively are detectable in the Magellanic Clouds. For the Milky Way, single stripped stars should be detectable out to a distance of at least 2 kpc considering the crude extinction we assume. The four sdO + Be binaries \citep[$\phi$~Persei, FY~CMa, 59~Cyg, and 60~Cyg, see][respectively]{1998ApJ...493..440G, 2008ApJ...686.1280P, 2013ApJ...765....2P, 2017ApJ...843...60W} are located on distances between 200 and 500~pc and show indeed detected UV excess.

Despite strong dust attenuation, the UV color excess method described in \secref{sec:UVexcess} is not affected. The magnitude limit changes, but the introduced color difference remains the same as both the stripped star and its companion are attenuated with the same factor at each wavelength. The UV color is affected as the spectra are reddened, but the UV color excess remains the same. 
}

\section{Metallicity grids}\label{app:metallicity}

In this appendix we provide diagrams and tables for the full set of metallicities that we have considered. These include our reference grid at solar metallicity, $Z = 0.014$, as well as lower values, $Z = 0.006, 0.002$ and $0.0002$.  
 
An overview of the parameters of the grids are given in \tabrefthree{tab:params_006}{tab:params_002}{tab:params_0002}. These tables are similar to \tabref{tab:params} in the main text to which we refer for a description.  

\begin{figure*}
\centering
\includegraphics[width=0.9\textwidth]{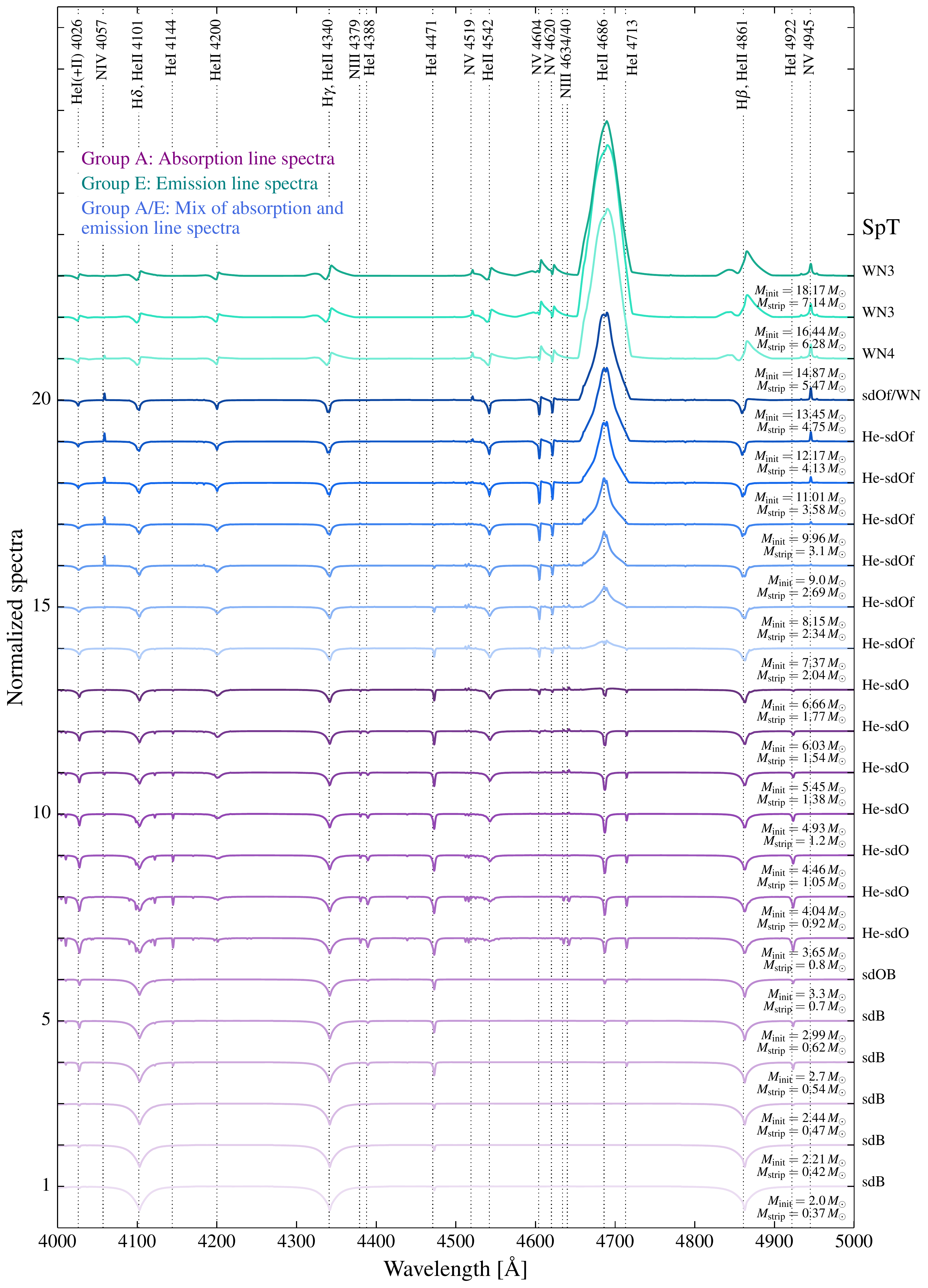}
\caption{The normalized spectra of the $Z = 0.006$}
\label{fig:optical_spectra_006}
\end{figure*}

\begin{figure*}
\centering
\includegraphics[width=0.9\textwidth]{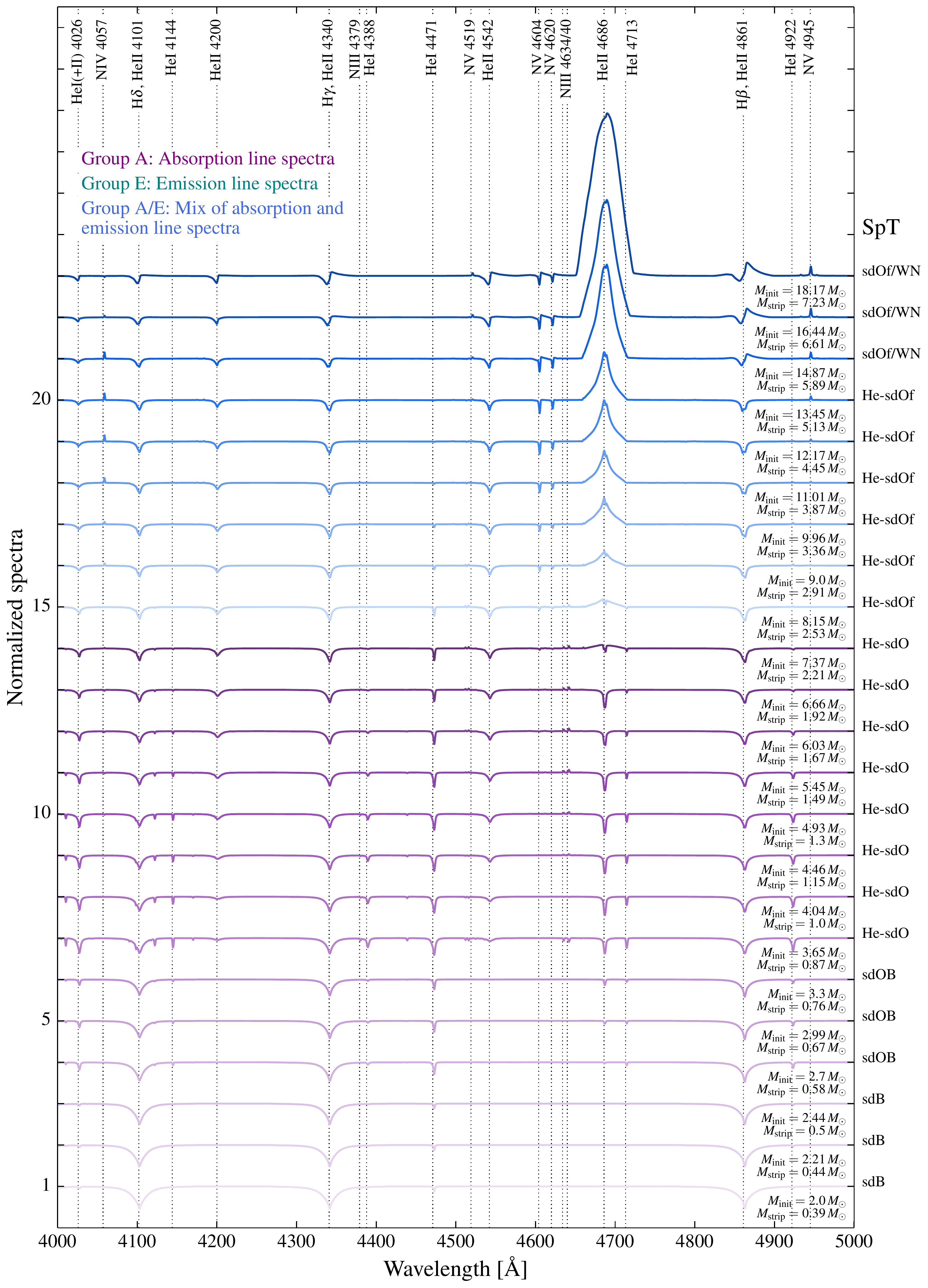}
\caption{The normalized spectra of the $Z = 0.002$}
\label{fig:optical_spectra_002}
\end{figure*}

\begin{figure*}
\centering
\includegraphics[width=0.9\textwidth]{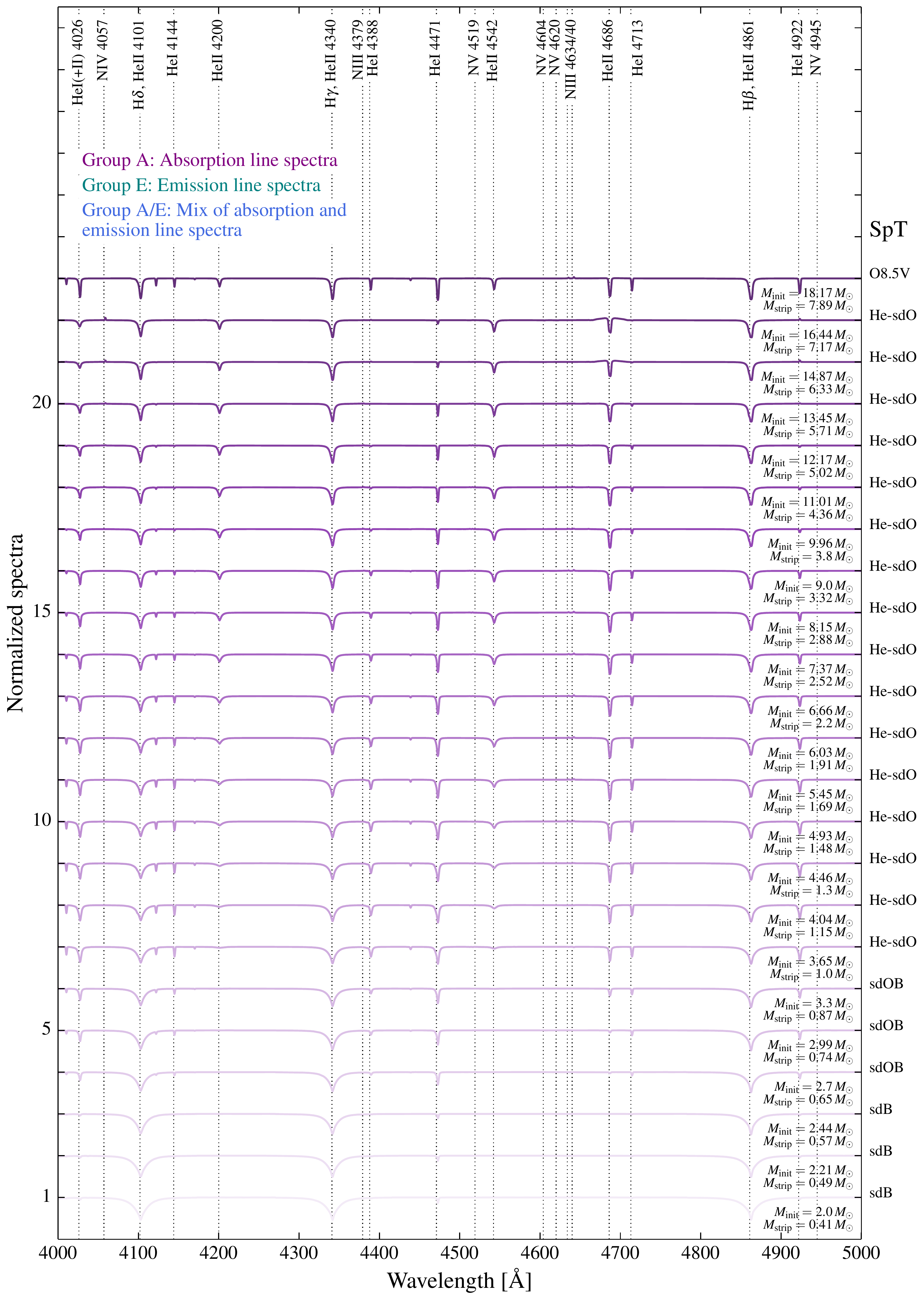}
\caption{The normalized spectra of the $Z = 0.0002$}
\label{fig:optical_spectra_0002}
\end{figure*}

\begin{figure*}
\centering
\includegraphics[width=0.9\textwidth]{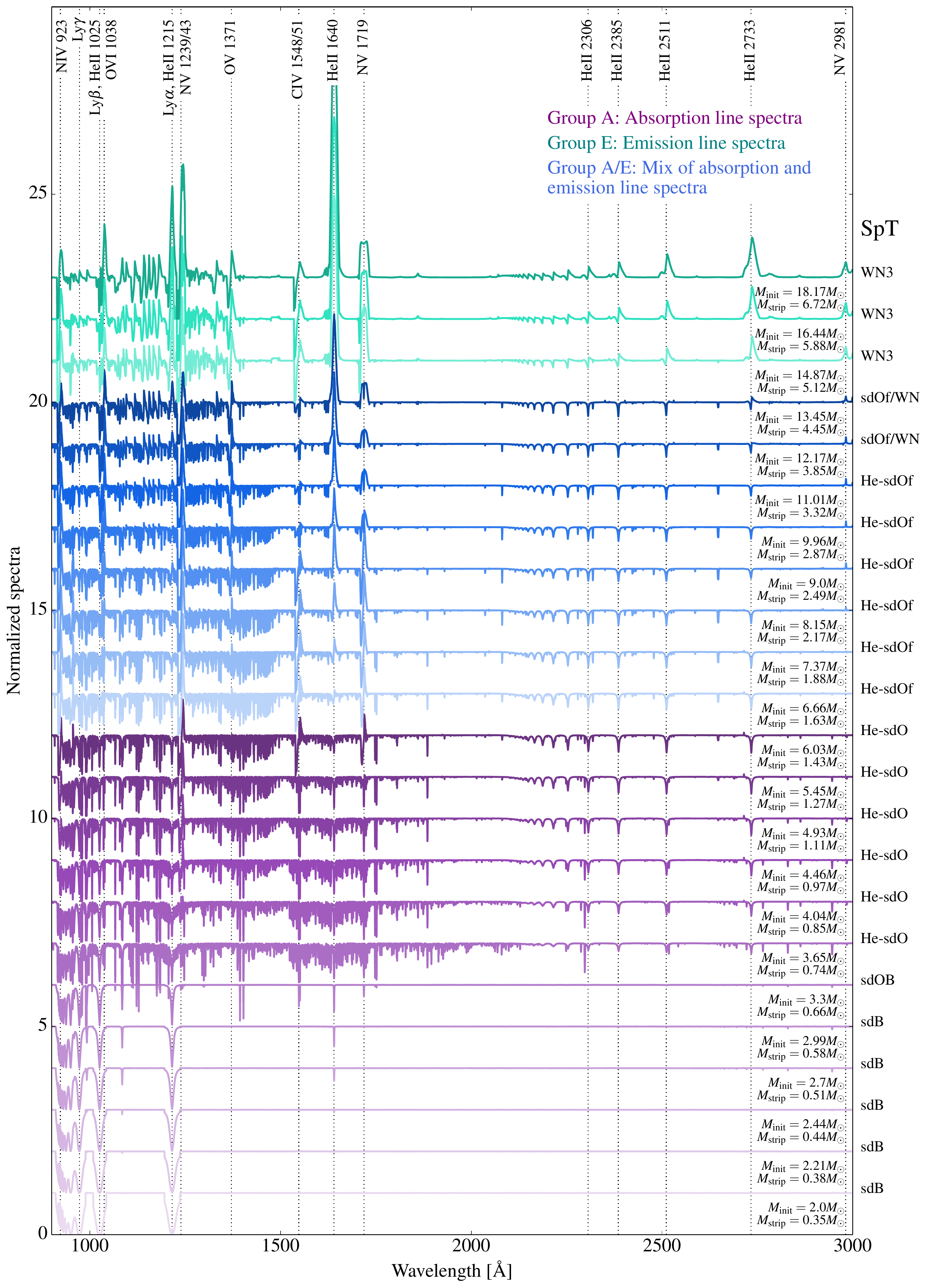}
\caption{The UV normalized spectra of the $Z = 0.014$ grid.}
\label{fig:UV_spectra_014}
\end{figure*}

\begin{figure*}
\centering
\includegraphics[width=0.9\textwidth]{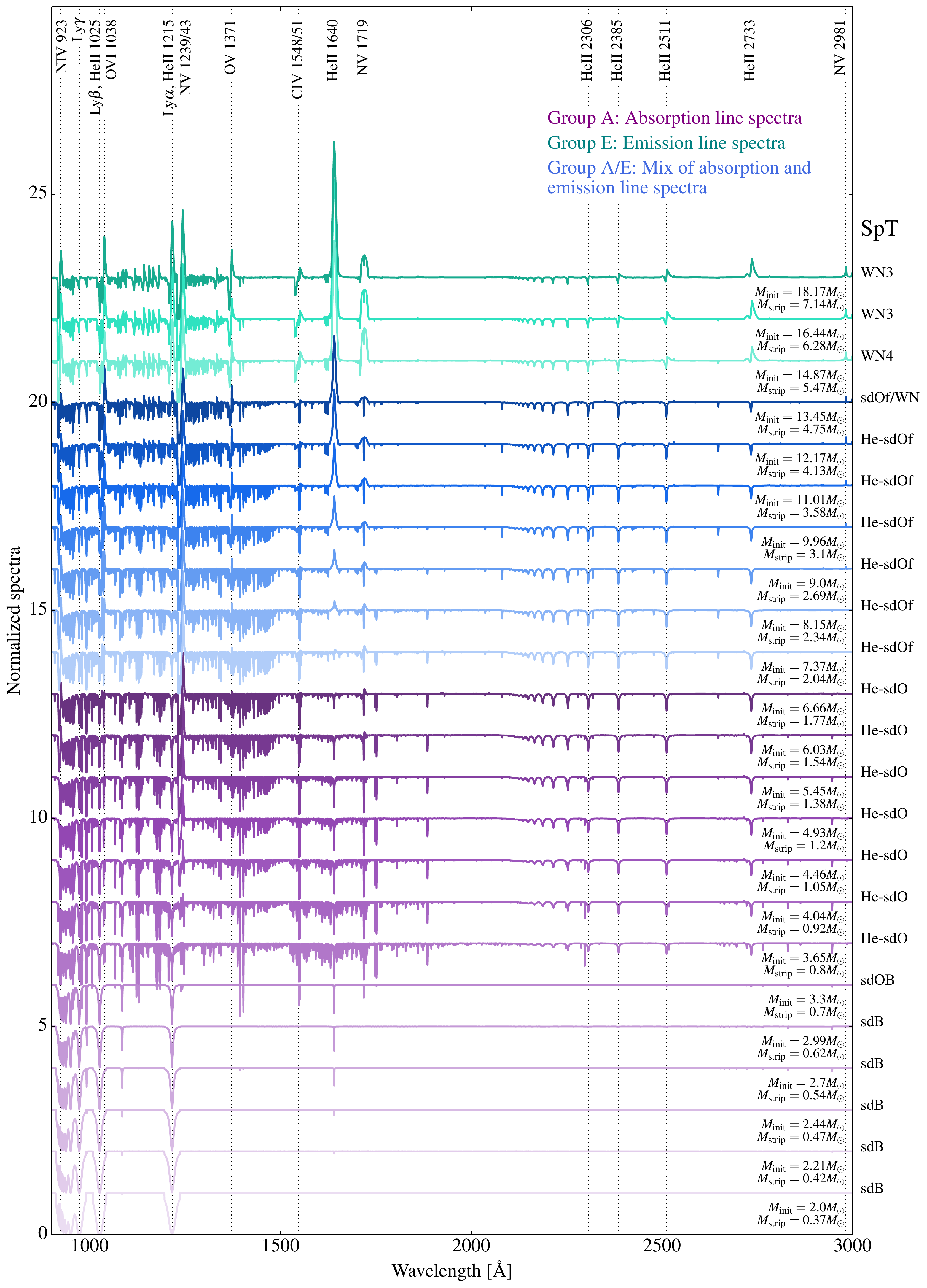}
\caption{The UV normalized spectra of the $Z = 0.006$ grid.}
\label{fig:UV_spectra_006}
\end{figure*}

\begin{figure*}
\centering
\includegraphics[width=0.9\textwidth]{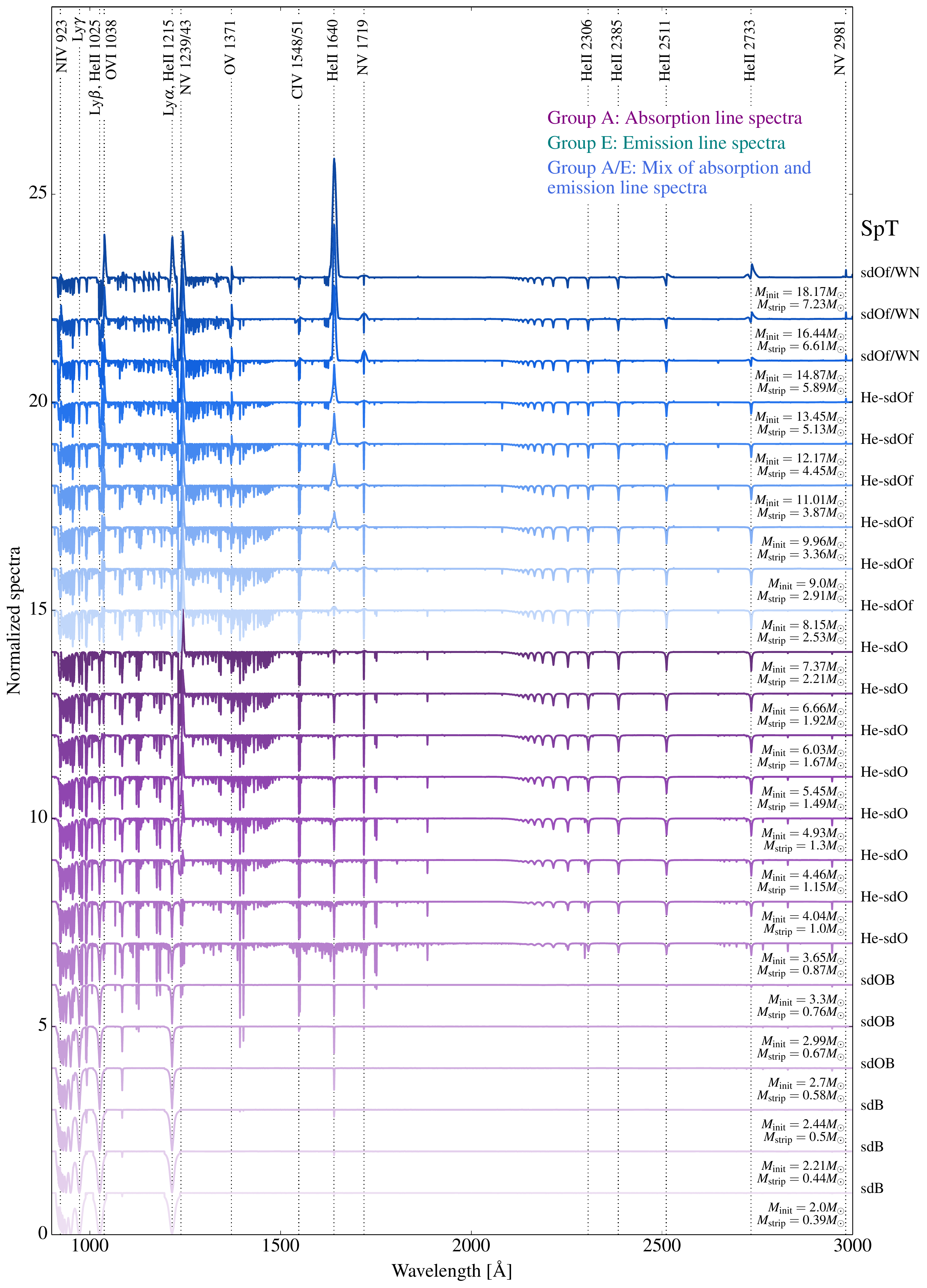}
\caption{The UV normalized spectra of the $Z = 0.002$ grid.}
\label{fig:UV_spectra_002}
\end{figure*}

\begin{figure*}
\centering
\includegraphics[width=0.9\textwidth]{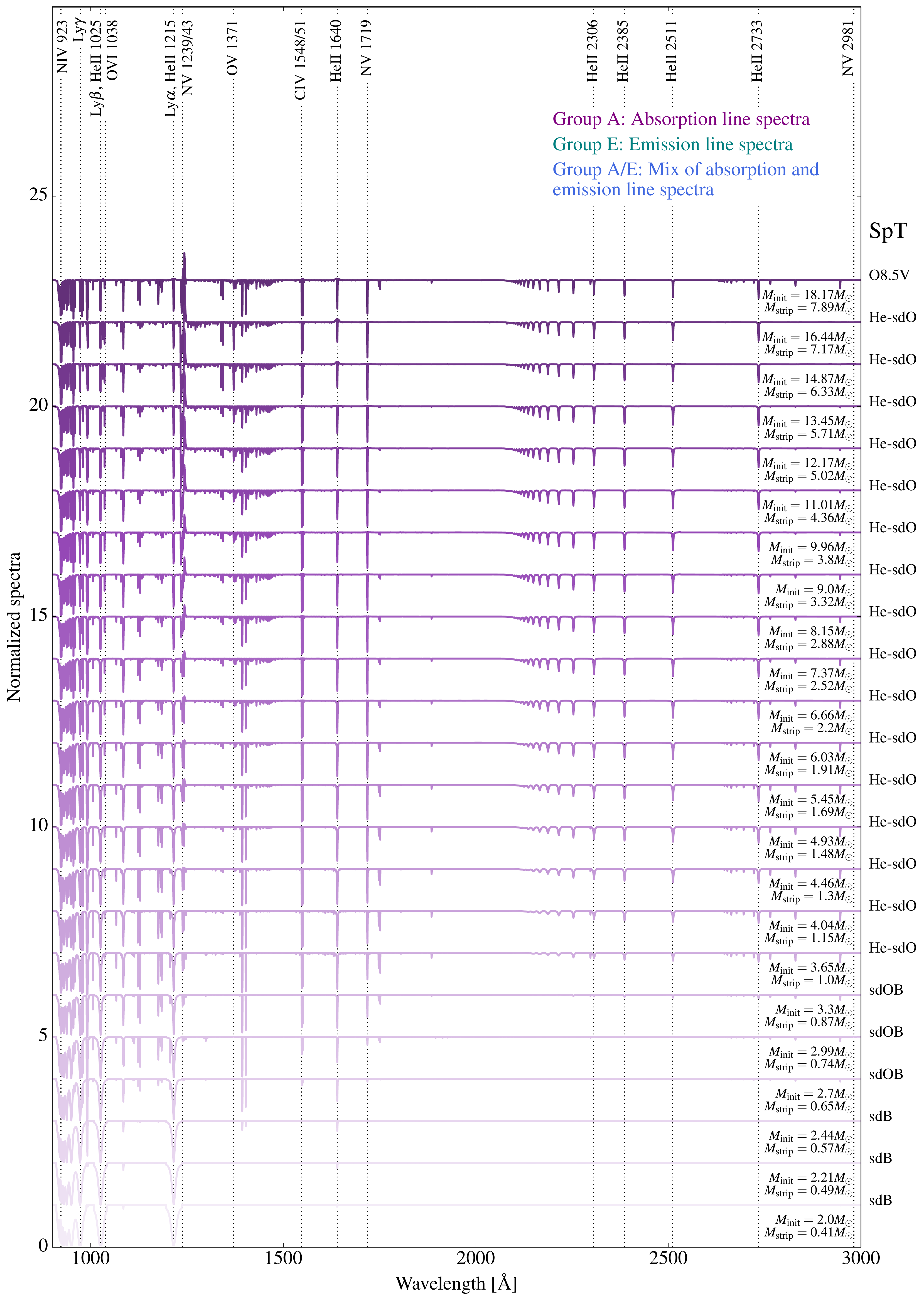}
\caption{The UV normalized spectra of the $Z = 0.0002$ grid.}
\label{fig:UV_spectra_0002}
\end{figure*}

\begin{figure*}
\centering
\includegraphics[width=0.9\textwidth]{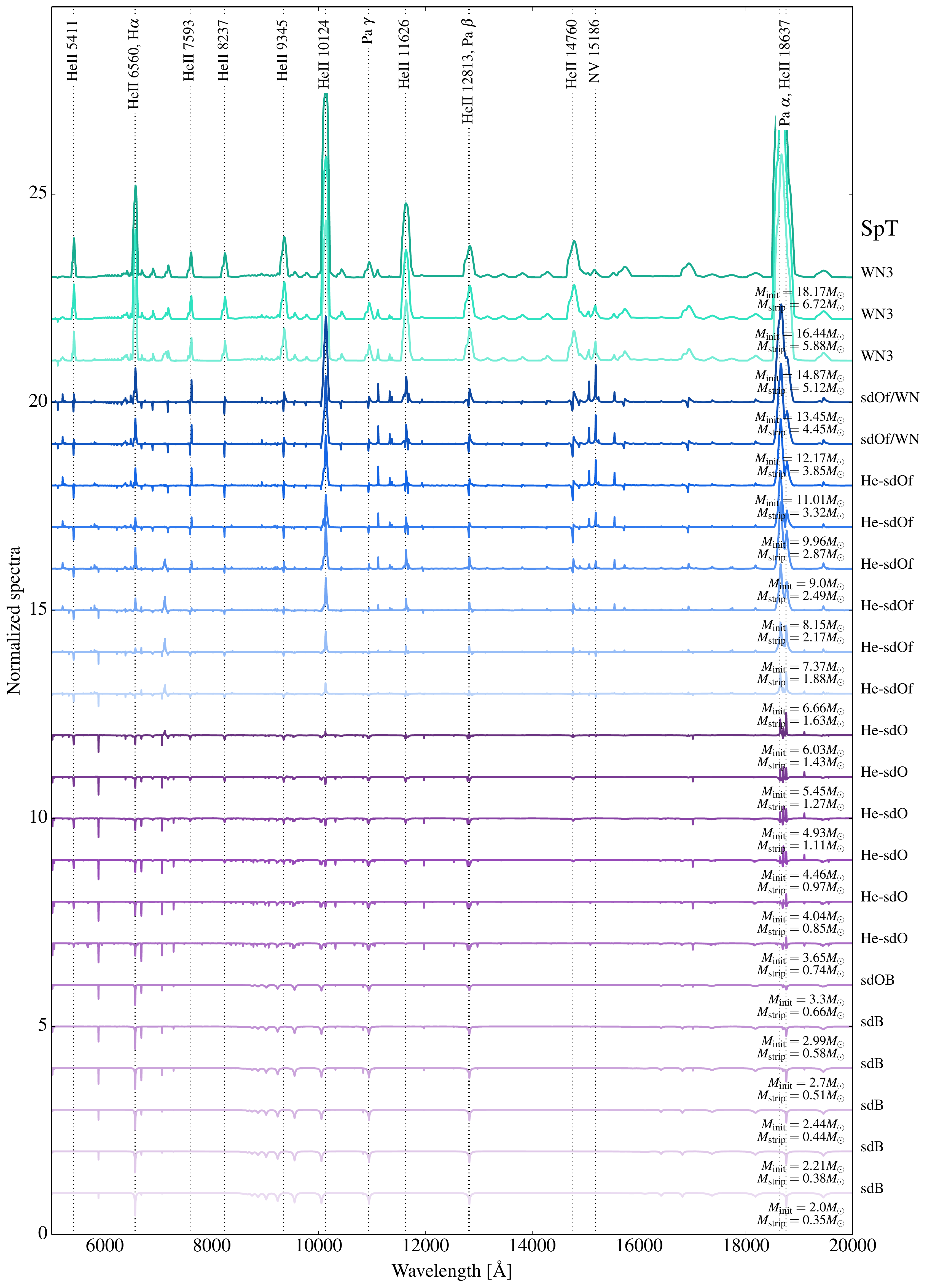}
\caption{The IR normalized spectra of the $Z = 0.014$ grid.}
\label{fig:IR_spectra_014}
\end{figure*}

\begin{figure*}
\centering
\includegraphics[width=0.9\textwidth]{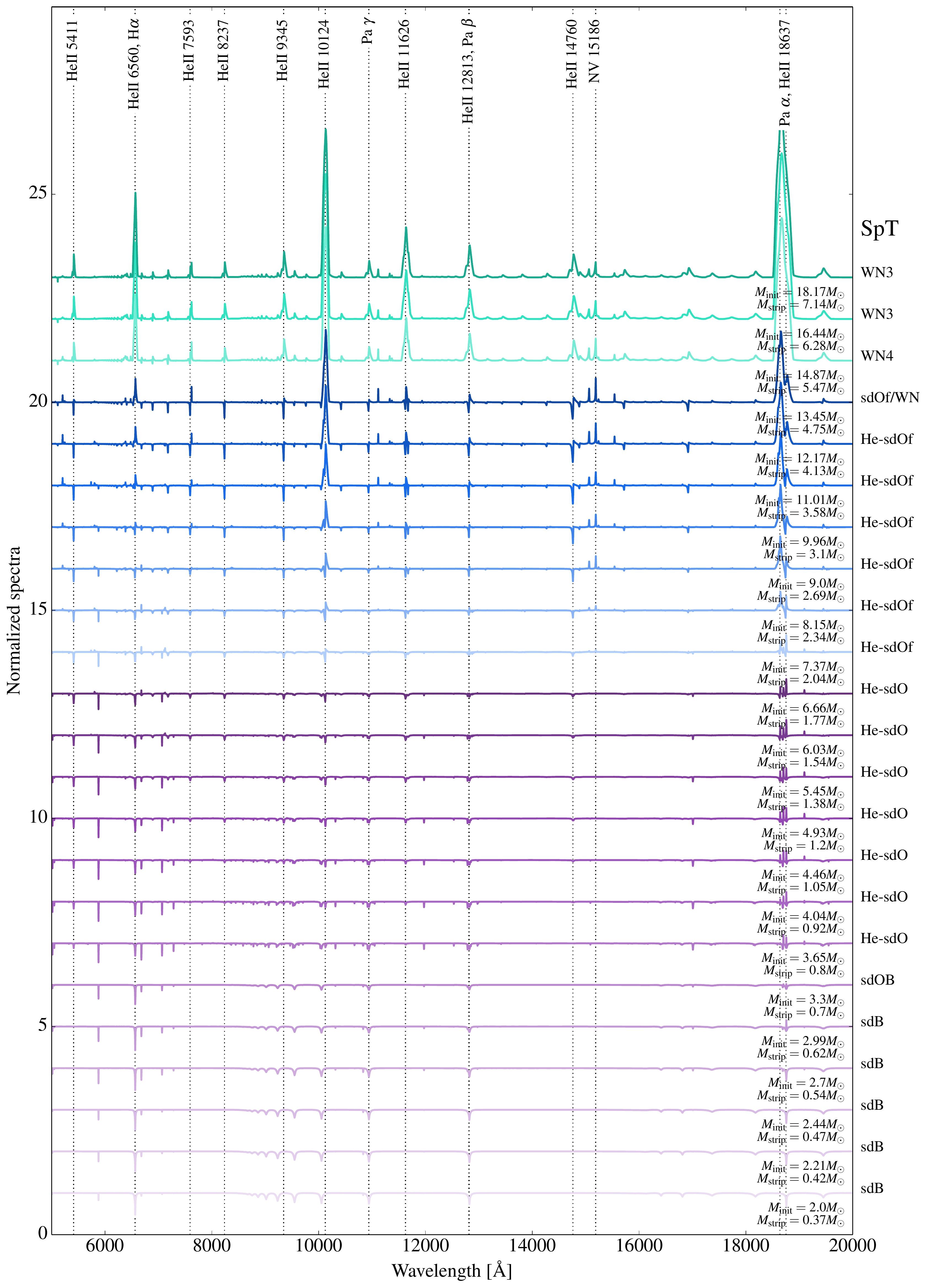}
\caption{The IR normalized spectra of the $Z = 0.006$ grid.}
\label{fig:IR_spectra_006}
\end{figure*}

\begin{figure*}
\centering
\includegraphics[width=0.9\textwidth]{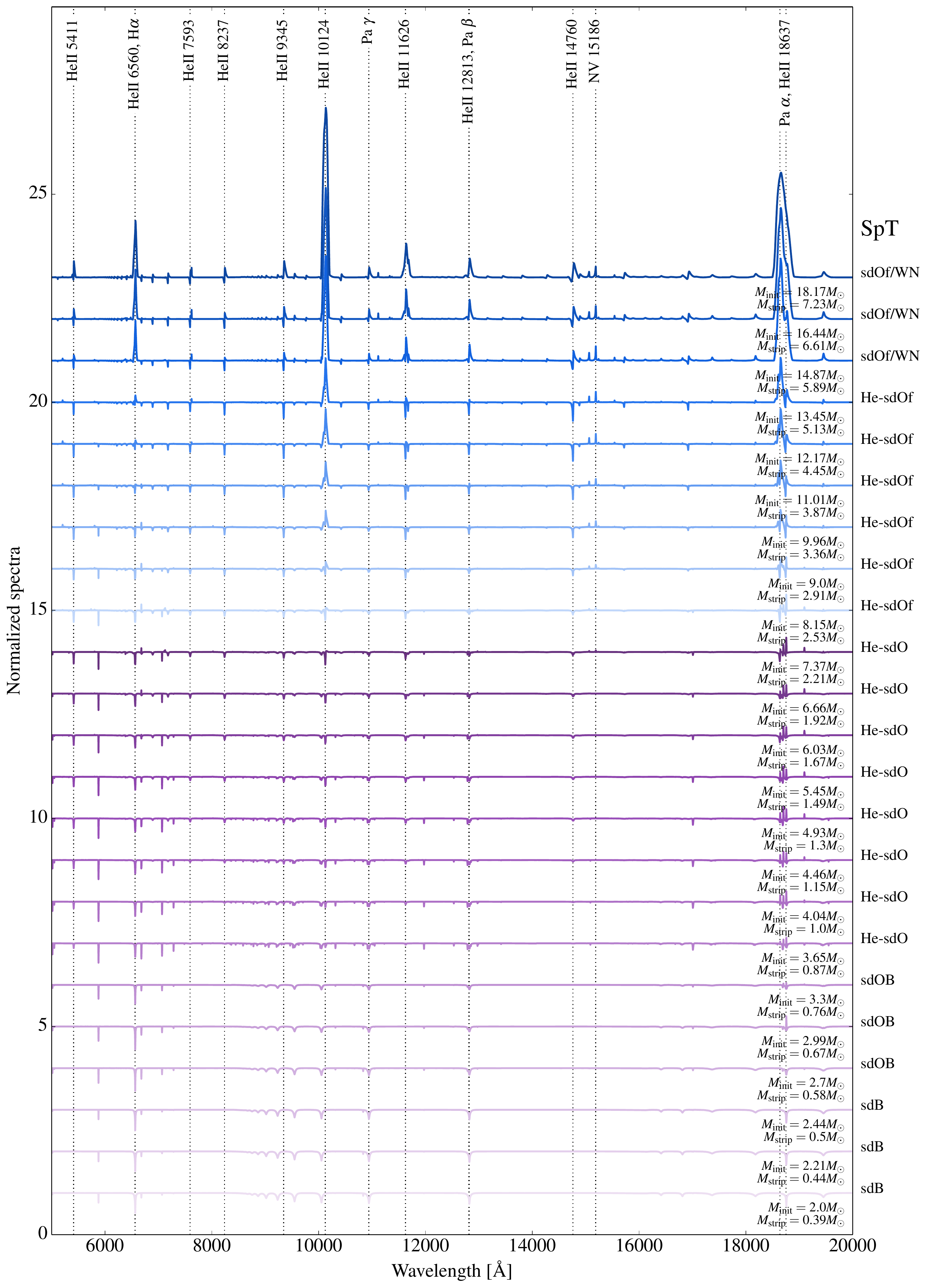}
\caption{The IR normalized spectra of the $Z = 0.002$ grid.}
\label{fig:IR_spectra_002}
\end{figure*}

\begin{figure*}
\centering
\includegraphics[width=0.9\textwidth]{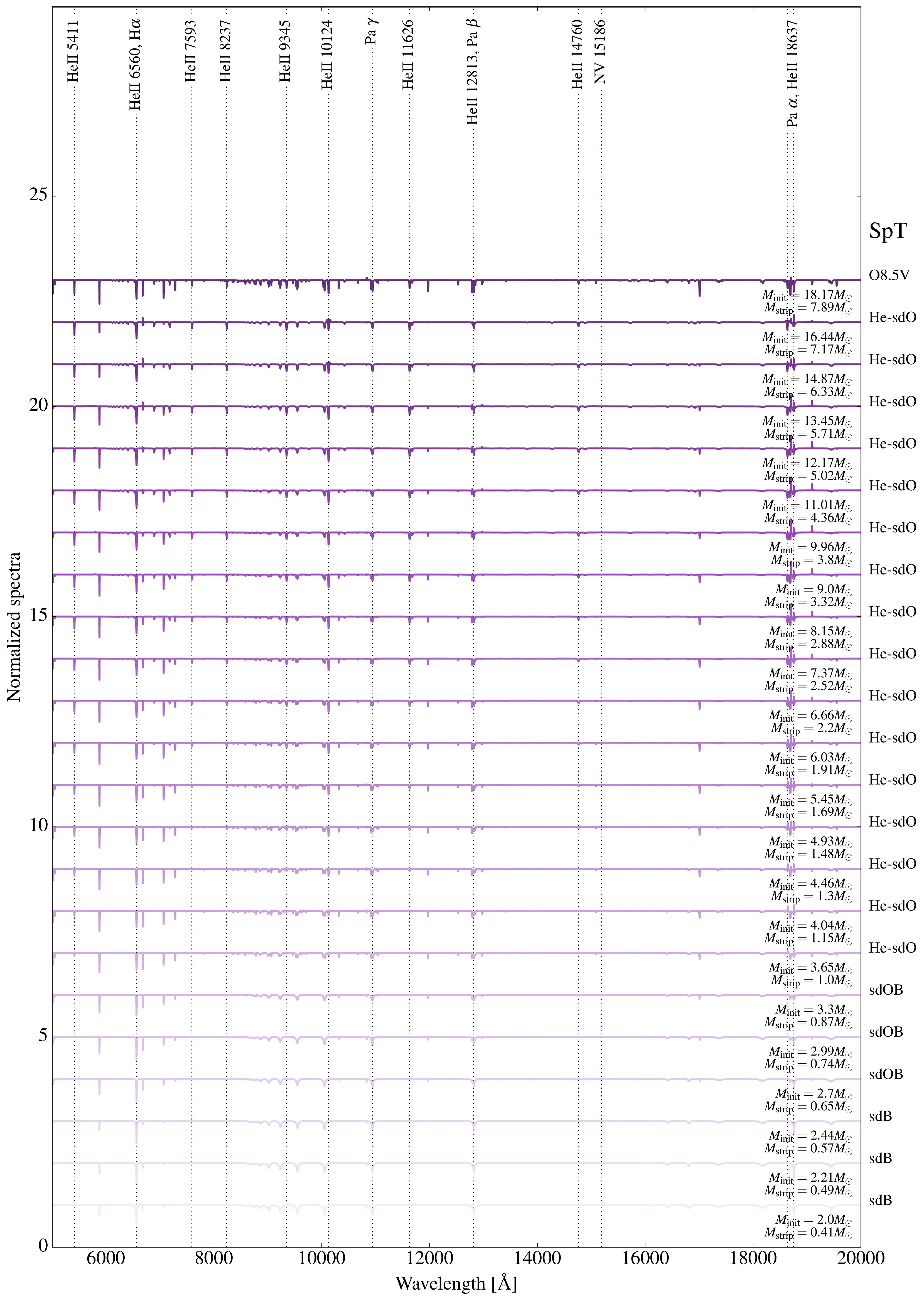}
\caption{The IR normalized spectra of the $Z = 0.0002$ grid.}
\label{fig:IR_spectra_0002}
\end{figure*}

In Figs.~\ref{fig:optical_spectra_006}--\ref{fig:IR_spectra_0002} we provide the normalized UV, optical and IR spectra. 
In Tables~\ref{tab:params_006}--\ref{tab:mag_0002} we provide estimates for absolute magnitudes of stripped stars in $U$, $B$, $V$, and the \textit{GALEX} ($NUV$ and $FUV$) and Swift ($UVW1$, $UVW2$, $UVM2$) UV filters.


\begin{sidewaystable*}
\centering
\caption{Properties of stripped stars for $Z = 0.006$.}
\label{tab:params_006}
{\small
\input{table_param_006.tex}
}
\tablefoot{See \tabref{tab:params} for a description of the parameters.}
\end{sidewaystable*}

\begin{sidewaystable*}
\centering
\caption{Properties of stripped stars for $Z = 0.002$.}
\label{tab:params_002}
{\small
\input{table_param_002.tex}
}
\tablefoot{See \tabref{tab:params} for a description of the parameters.}
\end{sidewaystable*}

\begin{sidewaystable*}
\centering
\caption{Properties of stripped stars for $Z = 0.0002$.}
\label{tab:params_0002}
{\small
\input{table_param_0002.tex}
}
\tablefoot{See \tabref{tab:params} for a description of the parameters.}
\end{sidewaystable*}


\begin{table*}
\centering
\caption{Absolute magnitudes of stripped stars in $U$, $B$, $V$, and the \textit{GALEX} (\textit{NUV} and \textit{FUV}) and Swift (UVW1, UVW2, UVM2) UV filters, $Z = 0.014$.}
\label{tab:mag_014}
{\small
\input{table_mag_014.tex}
}
\end{table*}

\begin{table*}
\centering
\caption{Absolute magnitudes of stripped stars in $U$, $B$, $V$, and the \textit{GALEX} (\textit{NUV} and \textit{FUV}) and Swift (UVW1, UVW2, UVM2) UV filters, $Z = 0.006$.}
\label{tab:mag_006}
{\small
\input{table_mag_006.tex}
}
\end{table*}

\begin{table*}
\centering
\caption{Absolute magnitudes of stripped stars in $U$, $B$, $V$, and the \textit{GALEX} (\textit{NUV} and \textit{FUV}) and Swift (UVW1, UVW2, UVM2) UV filters,  $Z = 0.002$.}
\label{tab:mag_002}
{\small
\input{table_mag_002.tex}
}
\end{table*}

\begin{table*}
\centering
\caption{Absolute magnitudes of stripped stars in $U$, $B$, $V$, and the \textit{GALEX} (\textit{NUV} and \textit{FUV}) and Swift (UVW1, UVW2, UVM2) UV filters, $Z = 0.0002$.}
\label{tab:mag_0002}
{\small
\input{table_mag_0002.tex}
}
\end{table*}

\end{document}

%% file: table_param.tex
\begin{tabular}{lcccccccccccccccccc}
\toprule\midrule 
 Group & $M_{\text{init}}$ & $P_{\mathrm{init}}$ & $M_{\text{strip}}$ & $M_{\mathrm{H, tot}}$ & $\log_{10} L$ & $\log _{10} L_H$ & $T_{\star}$ & $T_{\text{eff}}$ & $\log _{10} g _{\text{eff}}$ & $R_{\text{eff}}$ & $X_{\text{H,s}}$ & $X_{\text{He,s}}$ & $\log_{10} \dot{M}_{\text{wind}}$ & $\Gamma _e$ & $v_{\infty}$ & $\log_{10} Q_0$ & $\log_{10} Q_1$ & $\log_{10} Q_2$\\ 
 
  & [$M_{\odot}$] & [days] & [$M_{\odot}$] & [$M_{\odot}$] & [$L_{\odot}$] & [$L_{\odot}$] & [kK] & [kK] & [cm s$^{-2}$] & [$R_{\odot}$] &  &  & [$M_{\odot}$ yr$^{-1}$] &  & [km s$^{-1}$] & [s$^{-1}$] & [s$^{-1}$] & [s$^{-1}$]\\ 
 
\midrule 
A & 2.0$^{x}$$^{\dagger}$ & 3.0 & 0.35 & 0.002 & 0.6 & -1.2 & 20.5 & 20.4 & 5.58 & 0.16 & 1.0 & 0.0 & -12.0 & 0.0 & 1370 & 40.8 & 36.9 & 30.8\\ 
  & 2.21$^{\dagger}$ & 6.0 & 0.38 & 0.003 & 0.8 & -0.9 & 22.6 & 22.5 & 5.54 & 0.17 & 1.0 & 0.0 & -12.0 & 0.001 & 1380 & 41.4 & 37.8 & 30.9\\ 
  & 2.44$^{\ddagger}$ & 6.4 & 0.44 & 0.003 & 1.1 & -0.6 & 25.8 & 25.6 & 5.54 & 0.19 & 1.0 & 0.0 & -12.0 & 0.001 & 1420 & 42.2 & 39.2 & 32.0\\ 
  & 2.7$^{x}$$^{\dagger}$ & 6.8 & 0.51 & 0.004 & 1.4 & -0.3 & 28.3 & 28.2 & 5.49 & 0.21 & 0.96 & 0.04 & -12.0 & 0.002 & 1440 & 42.9 & 39.7 & 32.5\\ 
  & 2.99$^{x}$$^{\ddagger}$ & 7.3 & 0.58 & 0.004 & 1.6 & -0.0 & 31.1 & 30.9 & 5.48 & 0.23 & 0.96 & 0.04 & -12.0 & 0.002 & 1480 & 43.8 & 41.7 & 36.6\\ 
  & 3.3$^{x}$$^{\dagger}$ & 10.0 & 0.66 & 0.005 & 1.9 & 0.3 & 33.8 & 33.6 & 5.45 & 0.25 & 0.96 & 0.04 & -12.0 & 0.003 & 1510 & 44.7 & 43.8 & 38.6\\ 
  & 3.65$^{\dagger}$ & 8.2 & 0.74 & 0.006 & 2.0 & 0.5 & 36.8 & 36.7 & 5.48 & 0.26 & 0.49 & 0.5 & -12.0 & 0.003 & 1570 & 45.3 & 43.5 & 37.0\\ 
  & 4.04$^{\dagger}$ & 8.8 & 0.85 & 0.008 & 2.3 & 0.8 & 39.6 & 39.5 & 5.45 & 0.29 & 0.46 & 0.54 & -11.2 & 0.005 & 1590 & 45.8 & 44.5 & 37.7\\ 
  & 4.46$^{\dagger}$ & 9.4 & 0.97 & 0.01 & 2.5 & 1.0 & 42.1 & 41.9 & 5.4 & 0.32 & 0.46 & 0.53 & -10.6 & 0.007 & 1610 & 46.1 & 45.2 & 39.1\\ 
  & 4.93$^{\dagger}$ & 10.0 & 1.11 & 0.012 & 2.7 & 1.3 & 44.7 & 44.6 & 5.37 & 0.36 & 0.42 & 0.57 & -10.0 & 0.009 & 1630 & 46.3 & 45.6 & 40.1\\ 
  & 5.45$^{\dagger}$ & 10.7 & 1.27 & 0.014 & 2.9 & 1.4 & 47.7 & 47.5 & 5.34 & 0.4 & 0.36 & 0.63 & -9.5 & 0.012 & 1660 & 46.6 & 46.0 & 41.1\\ 
  & 6.03$^{\dagger}$ & 11.2 & 1.43 & 0.013 & 3.0 & 1.3 & 51.0 & 50.8 & 5.35 & 0.42 & 0.29 & 0.7 & -8.4 & 0.015 & 1720 & 46.8 & 46.2 & 41.4\\ \midrule 

 A/E & 6.66$^{\dagger}$ & 12.0 & 1.63 & 0.015 & 3.2 & 1.5 & 54.3 & 54.1 & 5.33 & 0.46 & 0.28 & 0.71 & -8.1 & 0.02 & 1750 & 47.0 & 46.5 & 42.6\\ 
  & 7.37$^{\dagger}$ & 13.0 & 1.88 & 0.017 & 3.4 & 1.6 & 57.8 & 57.5 & 5.31 & 0.5 & 0.25 & 0.73 & -7.8 & 0.026 & 1800 & 47.2 & 46.7 & 42.8\\ 
  & 8.15$^{\dagger}$ & 14.0 & 2.17 & 0.019 & 3.6 & 1.7 & 61.8 & 61.6 & 5.3 & 0.55 & 0.23 & 0.75 & -7.5 & 0.034 & 1850 & 47.4 & 47.0 & 43.5\\ 
  & 9.0$^{\dagger}$ & 15.2 & 2.49 & 0.02 & 3.8 & 1.8 & 65.7 & 65.5 & 5.29 & 0.59 & 0.23 & 0.76 & -7.3 & 0.044 & 1900 & 47.6 & 47.3 & 44.2\\ 
  & 9.96 & 16.6 & 2.87 & 0.023 & 3.9 & 1.8 & 69.4 & 69.1 & 5.27 & 0.65 & 0.23 & 0.76 & -7.1 & 0.057 & 1950 & 47.8 & 47.4 & 44.7\\ 
  & 11.01 & 18.3 & 3.32 & 0.023 & 4.1 & 1.8 & 74.0 & 73.7 & 5.27 & 0.7 & 0.22 & 0.77 & -6.8 & 0.074 & 2020 & 48.0 & 47.7 & 45.2\\ 
  & 12.17 & 20.2 & 3.85 & 0.023 & 4.3 & 1.8 & 78.7 & 78.3 & 5.27 & 0.76 & 0.21 & 0.78 & -6.6 & 0.094 & 2100 & 48.1 & 47.9 & 45.5\\ 
  & 13.45 & 22.4 & 4.45 & 0.021 & 4.4 & 1.6 & 83.8 & 83.4 & 5.28 & 0.8 & 0.2 & 0.79 & -6.4 & 0.116 & 2190 & 48.3 & 48.0 & 45.8\\ \midrule 

 E & 14.87 & 25.0 & 5.12 & 0.018 & 4.6 & 1.3 & 89.3 & 88.8 & 5.3 & 0.84 & 0.18 & 0.81 & -6.2 & 0.141 & 2300 & 48.4 & 48.2 & 45.1\\ 
  & 16.44 & 28.0 & 5.88 & 0.012 & 4.7 & 0.7 & 95.1 & 94.6 & 5.33 & 0.87 & 0.16 & 0.83 & -6.0 & 0.166 & 2420 & 48.6 & 48.4 & 45.7\\ 
  & 18.17 & 31.7 & 6.72 & 0.006 & 4.9 & 0.0 & 101.8 & 101.3 & 5.37 & 0.88 & 0.12 & 0.87 & -5.8 & 0.189 & 2570 & 48.7 & 48.5 & 46.3\\ 
 \bottomrule 
\end{tabular}

%% file: table_param_006.tex
\begin{tabular}{lcccccccccccccccccc}
\toprule\midrule 
 Group & $M_{\text{init}}$ & $P_{\mathrm{init}}$ & $M_{\text{strip}}$ & $M_{\mathrm{H, tot}}$ & $\log_{10} L$ & $\log _{10} L_H$ & $T_{\star}$ & $T_{\text{eff}}$ & $\log _{10} g _{\text{eff}}$ & $R_{\text{eff}}$ & $X_{\text{H,s}}$ & $X_{\text{He,s}}$ & $\log_{10} \dot{M}_{\text{wind}}$ & $\Gamma _e$ & $v_{\infty}$ & $\log_{10} Q_0$ & $\log_{10} Q_1$ & $\log_{10} Q_2$\\ 
 
  & [$M_{\odot}$] & [days] & [$M_{\odot}$] & [$M_{\odot}$] & [$L_{\odot}$] & [$L_{\odot}$] & [kK] & [kK] & [cm s$^{-2}$] & [$R_{\odot}$] &  &  & [$M_{\odot}$ yr$^{-1}$] &  & [km s$^{-1}$] & [s$^{-1}$] & [s$^{-1}$] & [s$^{-1}$]\\ 
 
\midrule 
A & 2.0$^{x}$$^{\ddagger}$ & 4.6 & 0.37 & 0.003 & 0.7 & -0.9 & 21.7 & 21.6 & 5.55 & 0.17 & 1.0 & 0.0 & -12.0 & 0.0 & 1370 & 41.2 & 37.4 & 30.8\\ 
  & 2.21$^{x}$$^{\ddagger}$ & 4.9 & 0.42 & 0.004 & 1.1 & -0.5 & 25.3 & 25.2 & 5.56 & 0.18 & 1.0 & 0.0 & -12.0 & 0.001 & 1430 & 42.1 & 38.9 & 31.4\\ 
  & 2.44$^{x}$$^{\ddagger}$ & 5.3 & 0.47 & 0.005 & 1.3 & -0.3 & 27.8 & 27.7 & 5.54 & 0.19 & 1.0 & 0.0 & -12.0 & 0.001 & 1450 & 42.7 & 40.4 & 35.6\\ 
  & 2.7$^{x}$$^{\ddagger}$ & 5.6 & 0.54 & 0.006 & 1.5 & -0.1 & 30.4 & 30.2 & 5.51 & 0.21 & 0.96 & 0.04 & -12.0 & 0.002 & 1480 & 43.5 & 41.0 & 35.2\\ 
  & 2.99$^{x}$$^{\dagger}$ & 6.0 & 0.62 & 0.007 & 1.8 & 0.2 & 32.9 & 32.8 & 5.47 & 0.24 & 0.96 & 0.04 & -12.0 & 0.003 & 1500 & 44.4 & 43.3 & 38.4\\ 
  & 3.3$^{x}$$^{\dagger}$ & 6.4 & 0.7 & 0.008 & 2.0 & 0.5 & 35.4 & 35.2 & 5.44 & 0.26 & 0.96 & 0.04 & -12.0 & 0.004 & 1520 & 45.1 & 44.3 & 39.2\\ 
  & 3.65$^{x}$$^{\dagger}$ & 6.8 & 0.8 & 0.01 & 2.2 & 0.8 & 38.6 & 38.4 & 5.44 & 0.28 & 0.52 & 0.48 & -12.0 & 0.005 & 1570 & 45.6 & 44.3 & 38.6\\ 
  & 4.04$^{x}$$^{\dagger}$ & 7.3 & 0.92 & 0.012 & 2.4 & 1.0 & 40.7 & 40.5 & 5.39 & 0.32 & 0.51 & 0.49 & -11.2 & 0.006 & 1570 & 45.9 & 45.0 & 38.2\\ 
  & 4.46$^{x}$$^{\dagger}$ & 7.8 & 1.05 & 0.015 & 2.6 & 1.3 & 43.3 & 43.1 & 5.35 & 0.36 & 0.58 & 0.42 & -10.5 & 0.009 & 1600 & 46.2 & 45.6 & 39.6\\ 
  & 4.93$^{x}$$^{\dagger}$ & 8.3 & 1.2 & 0.019 & 2.8 & 1.6 & 45.6 & 45.4 & 5.3 & 0.4 & 0.5 & 0.5 & -10.0 & 0.012 & 1600 & 46.5 & 45.9 & 40.2\\ 
  & 5.45$^{x}$$^{\dagger}$ & 8.9 & 1.38 & 0.022 & 3.0 & 1.8 & 48.2 & 48.0 & 5.27 & 0.45 & 0.39 & 0.6 & -9.5 & 0.015 & 1630 & 46.7 & 46.2 & 41.2\\ 
  & 6.03$^{x}$$^{\dagger}$ & 9.3 & 1.54 & 0.023 & 3.1 & 1.8 & 51.4 & 51.2 & 5.28 & 0.47 & 0.35 & 0.65 & -8.4 & 0.019 & 1680 & 46.9 & 46.4 & 42.4\\ 
  & 6.66$^{x}$$^{\dagger}$ & 10.0 & 1.77 & 0.027 & 3.3 & 2.0 & 54.2 & 54.0 & 5.24 & 0.53 & 0.33 & 0.66 & -8.1 & 0.025 & 1700 & 47.1 & 46.6 & 42.9\\ \midrule 

 A/E & 7.37$^{x}$$^{\dagger}$ & 10.8 & 2.04 & 0.031 & 3.5 & 2.1 & 57.2 & 57.0 & 5.21 & 0.58 & 0.29 & 0.71 & -7.8 & 0.032 & 1730 & 47.3 & 46.9 & 43.4\\ 
  & 8.15$^{x}$ & 11.7 & 2.34 & 0.035 & 3.7 & 2.3 & 60.7 & 60.4 & 5.2 & 0.64 & 0.28 & 0.71 & -7.6 & 0.041 & 1780 & 47.5 & 47.1 & 43.8\\ 
  & 9.0$^{x}$ & 12.7 & 2.69 & 0.04 & 3.9 & 2.5 & 64.0 & 63.7 & 5.17 & 0.7 & 0.29 & 0.71 & -7.4 & 0.054 & 1820 & 47.7 & 47.3 & 44.3\\ 
  & 9.96 & 13.9 & 3.1 & 0.044 & 4.0 & 2.6 & 67.8 & 67.5 & 5.16 & 0.76 & 0.28 & 0.71 & -7.2 & 0.069 & 1870 & 47.8 & 47.5 & 44.7\\ 
  & 11.01 & 15.1 & 3.58 & 0.049 & 4.2 & 2.7 & 71.6 & 71.2 & 5.15 & 0.83 & 0.28 & 0.72 & -6.9 & 0.088 & 1930 & 48.0 & 47.7 & 44.9\\ 
  & 12.17 & 16.5 & 4.13 & 0.053 & 4.4 & 2.8 & 75.5 & 75.1 & 5.14 & 0.9 & 0.27 & 0.72 & -6.7 & 0.111 & 1990 & 48.2 & 47.9 & 45.3\\ 
  & 13.45 & 18.0 & 4.75 & 0.057 & 4.5 & 2.9 & 79.2 & 78.8 & 5.14 & 0.98 & 0.27 & 0.73 & -6.5 & 0.135 & 2050 & 48.3 & 48.1 & 45.6\\ \midrule 

 E & 14.87 & 19.8 & 5.47 & 0.059 & 4.7 & 2.9 & 83.2 & 82.7 & 5.14 & 1.04 & 0.26 & 0.73 & -6.3 & 0.163 & 2130 & 48.5 & 48.3 & 45.7\\ 
  & 16.44 & 21.7 & 6.28 & 0.06 & 4.8 & 2.9 & 87.2 & 86.7 & 5.14 & 1.11 & 0.25 & 0.74 & -6.1 & 0.193 & 2210 & 48.6 & 48.4 & 46.0\\ 
  & 18.17 & 23.9 & 7.14 & 0.086 & 4.9 & 3.2 & 85.4 & 84.8 & 5.04 & 1.34 & 0.29 & 0.71 & -6.0 & 0.231 & 2150 & 48.8 & 48.5 & 46.3\\ 
 \bottomrule 
\end{tabular}

%% file: table_param_002.tex
\begin{tabular}{lcccccccccccccccccc}
\toprule\midrule 
 Group & $M_{\text{init}}$ & $P_{\mathrm{init}}$ & $M_{\text{strip}}$ & $M_{\mathrm{H, tot}}$ & $\log_{10} L$ & $\log _{10} L_H$ & $T_{\star}$ & $T_{\text{eff}}$ & $\log _{10} g _{\text{eff}}$ & $R_{\text{eff}}$ & $X_{\text{H,s}}$ & $X_{\text{He,s}}$ & $\log_{10} \dot{M}_{\text{wind}}$ & $\Gamma _e$ & $v_{\infty}$ & $\log_{10} Q_0$ & $\log_{10} Q_1$ & $\log_{10} Q_2$\\ 
 
  & [$M_{\odot}$] & [days] & [$M_{\odot}$] & [$M_{\odot}$] & [$L_{\odot}$] & [$L_{\odot}$] & [kK] & [kK] & [cm s$^{-2}$] & [$R_{\odot}$] &  &  & [$M_{\odot}$ yr$^{-1}$] &  & [km s$^{-1}$] & [s$^{-1}$] & [s$^{-1}$] & [s$^{-1}$]\\ 
 
\midrule 
A & 2.0$^{x}$$^{\ddagger}$ & 3.4 & 0.39 & 0.004 & 0.9 & -0.6 & 24.8 & 24.7 & 5.63 & 0.16 & 1.0 & 0.0 & -12.0 & 0.001 & 1460 & 41.9 & 38.6 & 31.2\\ 
  & 2.21$^{x}$$^{\ddagger}$ & 3.8 & 0.44 & 0.006 & 1.2 & -0.3 & 26.8 & 26.7 & 5.57 & 0.18 & 1.0 & 0.0 & -12.0 & 0.001 & 1450 & 42.4 & 39.7 & 33.6\\ 
  & 2.44$^{x}$$^{\ddagger}$ & 4.1 & 0.5 & 0.007 & 1.4 & -0.0 & 28.8 & 28.6 & 5.51 & 0.21 & 1.0 & 0.0 & -12.0 & 0.002 & 1450 & 43.0 & 41.3 & 36.8\\ 
  & 2.7$^{x}$$^{\dagger}$ & 4.4 & 0.58 & 0.009 & 1.7 & 0.2 & 31.9 & 31.8 & 5.48 & 0.23 & 0.96 & 0.04 & -12.0 & 0.002 & 1480 & 44.1 & 42.5 & 37.7\\ 
  & 2.99$^{x}$$^{\dagger}$ & 4.7 & 0.67 & 0.011 & 1.9 & 0.5 & 34.0 & 33.8 & 5.43 & 0.26 & 0.96 & 0.04 & -12.0 & 0.004 & 1490 & 44.8 & 43.9 & 38.9\\ 
  & 3.3$^{x}$$^{\dagger}$ & 5.1 & 0.76 & 0.013 & 2.1 & 0.8 & 36.4 & 36.2 & 5.39 & 0.29 & 0.96 & 0.04 & -12.0 & 0.005 & 1500 & 45.3 & 44.6 & 39.5\\ 
  & 3.65$^{x}$$^{\dagger}$ & 5.4 & 0.87 & 0.017 & 2.3 & 1.1 & 38.6 & 38.4 & 5.36 & 0.32 & 0.53 & 0.47 & -12.0 & 0.006 & 1520 & 45.7 & 44.5 & 39.2\\ 
  & 4.04$^{x}$$^{\dagger}$ & 5.8 & 1.0 & 0.02 & 2.5 & 1.4 & 41.2 & 41.0 & 5.31 & 0.37 & 0.59 & 0.41 & -11.3 & 0.008 & 1540 & 46.1 & 45.3 & 38.8\\ 
  & 4.46$^{x}$$^{\dagger}$ & 6.3 & 1.15 & 0.025 & 2.7 & 1.6 & 43.2 & 43.0 & 5.25 & 0.42 & 0.54 & 0.46 & -10.6 & 0.011 & 1540 & 46.4 & 45.7 & 40.1\\ 
  & 4.93$^{x}$$^{\dagger}$ & 6.8 & 1.3 & 0.031 & 2.9 & 1.9 & 45.0 & 44.8 & 5.2 & 0.47 & 0.39 & 0.61 & -9.9 & 0.013 & 1540 & 46.6 & 46.0 & 40.5\\ 
  & 5.45$^{x}$$^{\dagger}$ & 7.3 & 1.49 & 0.037 & 3.1 & 2.1 & 47.3 & 47.0 & 5.15 & 0.54 & 0.43 & 0.57 & -9.6 & 0.019 & 1550 & 46.8 & 46.3 & 41.2\\ 
  & 6.03$^{x}$$^{\dagger}$ & 7.4 & 1.67 & 0.04 & 3.3 & 2.2 & 49.8 & 49.6 & 5.14 & 0.58 & 0.36 & 0.64 & -8.5 & 0.023 & 1580 & 47.0 & 46.5 & 42.4\\ 
  & 6.66$^{x}$$^{\dagger}$ & 8.0 & 1.92 & 0.048 & 3.4 & 2.5 & 52.4 & 52.1 & 5.1 & 0.65 & 0.38 & 0.62 & -9.4 & 0.031 & 1600 & 47.2 & 46.8 & 42.1\\ 
  & 7.37$^{x}$ & 8.7 & 2.21 & 0.057 & 3.6 & 2.6 & 54.7 & 54.4 & 5.05 & 0.73 & 0.35 & 0.65 & -8.0 & 0.04 & 1620 & 47.4 & 47.0 & 43.3\\ \midrule 

 A/E & 8.15$^{x}$ & 9.5 & 2.53 & 0.067 & 3.8 & 2.9 & 57.2 & 56.8 & 5.01 & 0.83 & 0.37 & 0.63 & -7.8 & 0.053 & 1630 & 47.6 & 47.2 & 43.7\\ 
  & 9.0$^{x}$ & 10.3 & 2.91 & 0.077 & 4.0 & 3.1 & 59.7 & 59.4 & 4.97 & 0.92 & 0.37 & 0.63 & -7.6 & 0.069 & 1650 & 47.8 & 47.4 & 44.0\\ 
  & 9.96 & 11.2 & 3.36 & 0.086 & 4.1 & 3.2 & 63.0 & 62.6 & 4.96 & 1.01 & 0.35 & 0.64 & -7.3 & 0.087 & 1700 & 47.9 & 47.6 & 44.3\\ 
  & 11.01 & 12.2 & 3.87 & 0.098 & 4.3 & 3.3 & 65.5 & 65.0 & 4.92 & 1.12 & 0.36 & 0.64 & -7.1 & 0.11 & 1730 & 48.1 & 47.8 & 44.7\\ 
  & 12.17 & 13.2 & 4.45 & 0.113 & 4.5 & 3.5 & 67.9 & 67.4 & 4.89 & 1.25 & 0.36 & 0.64 & -6.9 & 0.136 & 1760 & 48.3 & 48.0 & 45.0\\ 
  & 13.45 & 14.4 & 5.13 & 0.128 & 4.6 & 3.7 & 70.2 & 69.7 & 4.86 & 1.39 & 0.36 & 0.64 & -6.7 & 0.167 & 1790 & 48.5 & 48.2 & 45.3\\ 
  & 14.87 & 15.7 & 5.89 & 0.141 & 4.8 & 3.8 & 72.8 & 72.2 & 4.84 & 1.52 & 0.35 & 0.65 & -6.5 & 0.199 & 1830 & 48.6 & 48.3 & 45.6\\ 
  & 16.44 & 17.1 & 6.61 & 0.125 & 4.9 & 3.7 & 79.9 & 79.3 & 4.96 & 1.42 & 0.32 & 0.68 & -6.4 & 0.219 & 2010 & 48.7 & 48.5 & 46.0\\ 
  & 18.17 & 18.6 & 7.23 & 0.081 & 4.9 & 3.3 & 91.0 & 90.4 & 5.14 & 1.19 & 0.24 & 0.76 & -6.2 & 0.225 & 2290 & 48.8 & 48.6 & 46.4\\ 
 \bottomrule 
\end{tabular}

%% file: table_param_0002.tex
\begin{tabular}{lcccccccccccccccccc}
\toprule\midrule 
 Group & $M_{\text{init}}$ & $P_{\mathrm{init}}$ & $M_{\text{strip}}$ & $M_{\mathrm{H, tot}}$ & $\log_{10} L$ & $\log _{10} L_H$ & $T_{\star}$ & $T_{\text{eff}}$ & $\log _{10} g _{\text{eff}}$ & $R_{\text{eff}}$ & $X_{\text{H,s}}$ & $X_{\text{He,s}}$ & $\log_{10} \dot{M}_{\text{wind}}$ & $\Gamma _e$ & $v_{\infty}$ & $\log_{10} Q_0$ & $\log_{10} Q_1$ & $\log_{10} Q_2$\\ 
 
  & [$M_{\odot}$] & [days] & [$M_{\odot}$] & [$M_{\odot}$] & [$L_{\odot}$] & [$L_{\odot}$] & [kK] & [kK] & [cm s$^{-2}$] & [$R_{\odot}$] &  &  & [$M_{\odot}$ yr$^{-1}$] &  & [km s$^{-1}$] & [s$^{-1}$] & [s$^{-1}$] & [s$^{-1}$]\\ 
 
\midrule 
A & 2.0$^{x}$$^{\ddagger}$ & 1.9 & 0.41 & 0.009 & 1.1 & -0.1 & 25.0 & 24.9 & 5.53 & 0.18 & 1.0 & 0.0 & -12.0 & 0.001 & 1400 & 42.0 & 38.8 & 31.3\\ 
  & 2.21$^{x}$$^{\ddagger}$ & 2.2 & 0.49 & 0.012 & 1.4 & 0.2 & 27.8 & 27.7 & 5.48 & 0.21 & 1.0 & 0.0 & -12.0 & 0.001 & 1420 & 42.8 & 40.5 & 35.6\\ 
  & 2.44$^{x}$$^{\dagger}$ & 2.5 & 0.57 & 0.015 & 1.6 & 0.5 & 30.1 & 30.0 & 5.42 & 0.24 & 1.0 & 0.0 & -12.0 & 0.002 & 1420 & 43.6 & 42.5 & 37.9\\ 
  & 2.7$^{x}$$^{\dagger}$ & 2.8 & 0.65 & 0.02 & 1.9 & 0.8 & 31.7 & 31.5 & 5.34 & 0.29 & 0.96 & 0.04 & -12.0 & 0.003 & 1400 & 44.2 & 42.6 & 37.6\\ 
  & 2.99$^{x}$$^{\dagger}$ & 3.1 & 0.74 & 0.026 & 2.1 & 1.0 & 32.1 & 31.9 & 5.22 & 0.35 & 0.93 & 0.07 & -12.0 & 0.005 & 1350 & 44.5 & 42.6 & 37.4\\ 
  & 3.3$^{x}$$^{\ddagger}$ & 3.4 & 0.87 & 0.032 & 2.3 & 1.4 & 35.0 & 34.8 & 5.19 & 0.39 & 0.88 & 0.12 & -12.0 & 0.007 & 1380 & 45.4 & 44.2 & 39.1\\ 
  & 3.65$^{x}$$^{\dagger}$ & 3.7 & 1.0 & 0.041 & 2.5 & 1.7 & 36.4 & 36.2 & 5.11 & 0.46 & 0.61 & 0.39 & -12.0 & 0.008 & 1370 & 45.8 & 44.3 & 38.5\\ 
  & 4.04$^{x}$$^{\dagger}$ & 4.0 & 1.15 & 0.051 & 2.7 & 2.0 & 37.8 & 37.6 & 5.02 & 0.55 & 0.54 & 0.46 & -11.3 & 0.011 & 1350 & 46.1 & 44.9 & 39.4\\ 
  & 4.46$^{x}$$^{\dagger}$ & 4.3 & 1.3 & 0.061 & 2.9 & 2.2 & 38.8 & 38.6 & 4.95 & 0.63 & 0.5 & 0.5 & -10.7 & 0.014 & 1330 & 46.4 & 45.4 & 39.9\\ 
  & 4.93$^{x}$$^{\dagger}$ & 4.7 & 1.48 & 0.074 & 3.1 & 2.4 & 39.9 & 39.6 & 4.86 & 0.75 & 0.48 & 0.52 & -10.2 & 0.019 & 1310 & 46.6 & 45.8 & 40.5\\ 
  & 5.45$^{x}$$^{\dagger}$ & 5.1 & 1.69 & 0.092 & 3.3 & 2.7 & 40.2 & 39.9 & 4.75 & 0.91 & 0.5 & 0.5 & -9.9 & 0.026 & 1270 & 46.8 & 46.1 & 40.8\\ 
  & 6.03$^{x}$$^{\dagger}$ & 5.1 & 1.91 & 0.109 & 3.4 & 2.8 & 41.5 & 41.2 & 4.69 & 1.03 & 0.48 & 0.52 & -10.0 & 0.033 & 1270 & 47.1 & 46.4 & 41.2\\ 
  & 6.66$^{x}$ & 5.6 & 2.2 & 0.13 & 3.6 & 3.1 & 42.6 & 42.3 & 4.61 & 1.21 & 0.48 & 0.52 & -9.7 & 0.044 & 1250 & 47.3 & 46.7 & 41.6\\ 
  & 7.37$^{x}$ & 6.0 & 2.52 & 0.156 & 3.8 & 3.2 & 43.2 & 42.9 & 4.52 & 1.44 & 0.49 & 0.51 & -9.4 & 0.057 & 1230 & 47.5 & 46.9 & 41.9\\ 
  & 8.15$^{x}$ & 6.6 & 2.88 & 0.183 & 4.0 & 3.4 & 44.0 & 43.6 & 4.45 & 1.68 & 0.49 & 0.51 & -9.1 & 0.073 & 1220 & 47.6 & 47.1 & 42.2\\ 
  & 9.0$^{x}$ & 7.2 & 3.32 & 0.218 & 4.1 & 3.6 & 44.6 & 44.1 & 4.36 & 1.98 & 0.5 & 0.5 & -8.8 & 0.093 & 1210 & 47.8 & 47.3 & 42.5\\ 
  & 9.96 & 7.8 & 3.8 & 0.238 & 4.3 & 3.7 & 47.4 & 46.9 & 4.38 & 2.09 & 0.48 & 0.52 & -8.6 & 0.114 & 1260 & 48.0 & 47.5 & 42.8\\ 
  & 11.01 & 8.5 & 4.36 & 0.279 & 4.4 & 3.8 & 47.6 & 47.0 & 4.29 & 2.47 & 0.49 & 0.51 & -8.3 & 0.141 & 1240 & 48.2 & 47.7 & 43.0\\ 
  & 12.17 & 9.3 & 5.02 & 0.327 & 4.6 & 4.0 & 47.7 & 47.1 & 4.21 & 2.91 & 0.49 & 0.51 & -8.1 & 0.171 & 1230 & 48.3 & 47.9 & 43.2\\ 
  & 13.45 & 10.1 & 5.71 & 0.351 & 4.7 & 4.1 & 49.4 & 48.7 & 4.2 & 3.15 & 0.5 & 0.5 & -7.9 & 0.204 & 1260 & 48.5 & 48.0 & 43.6\\ 
  & 14.87 & 10.9 & 6.33 & 0.313 & 4.8 & 4.1 & 51.3 & 50.4 & 4.2 & 3.31 & 0.69 & 0.31 & -7.7 & 0.262 & 1290 & 48.6 & 48.2 & 43.8\\ 
  & 16.44 & 8.4 & 7.17 & 0.321 & 4.9 & 4.1 & 55.1 & 54.3 & 4.27 & 3.26 & 0.55 & 0.45 & -7.6 & 0.277 & 1390 & 48.7 & 48.3 & 44.1\\ 
  & 18.17 & 6.7 & 7.89 & 0.593 & 5.0 & 4.2 & 37.2 & 36.2 & 3.56 & 7.76 & 0.59 & 0.41 & -7.5 & 0.29 & 950 & 48.6 & 47.8 & 43.0\\ 
 \bottomrule 
\end{tabular}

%% file: table_mag_014.tex
\begin{tabular}{lccccccccc}
\toprule\midrule 
 Group & $M_{\text{init}}$ & U & B & V & NUV & FUV & UVW1 & UVW2 & UVM2\\ 
 
\midrule 
A & 2.0 & 5.2 & 5.1 & 5.4 & 4.9 & 4.6 & 4.9 & 4.8 & 4.9\\ 
  & 2.21 & 4.7 & 4.7 & 5.0 & 4.4 & 4.0 & 4.4 & 4.2 & 4.4\\ 
  & 2.44 & 4.2 & 4.3 & 4.6 & 3.7 & 3.3 & 3.7 & 3.5 & 3.7\\ 
  & 2.7 & 3.7 & 3.8 & 4.2 & 3.1 & 2.6 & 3.2 & 2.9 & 3.1\\ 
  & 2.99 & 3.2 & 3.4 & 3.8 & 2.6 & 2.0 & 2.6 & 2.4 & 2.6\\ 
  & 3.3 & 2.7 & 3.0 & 3.3 & 2.0 & 1.5 & 2.1 & 1.8 & 2.0\\ 
  & 3.65 & 2.3 & 2.7 & 3.1 & 1.6 & 1.2 & 1.7 & 1.5 & 1.6\\ 
  & 4.04 & 1.9 & 2.3 & 2.7 & 1.2 & 0.8 & 1.3 & 1.1 & 1.2\\ 
  & 4.46 & 1.6 & 2.0 & 2.4 & 0.9 & 0.5 & 1.0 & 0.7 & 0.9\\ 
  & 4.93 & 1.2 & 1.6 & 2.1 & 0.6 & 0.1 & 0.6 & 0.4 & 0.6\\ 
  & 5.45 & 0.9 & 1.4 & 1.8 & 0.3 & -0.2 & 0.3 & 0.1 & 0.2\\ 
  & 6.03 & 0.8 & 1.2 & 1.6 & 0.1 & -0.4 & 0.2 & -0.1 & 0.1\\ \midrule 

 A/E & 6.66 & 0.5 & 0.9 & 1.3 & -0.2 & -0.7 & -0.2 & -0.4 & -0.2\\ 
  & 7.37 & 0.2 & 0.7 & 1.1 & -0.5 & -1.0 & -0.4 & -0.7 & -0.5\\ 
  & 8.15 & -0.0 & 0.4 & 0.8 & -0.7 & -1.3 & -0.7 & -1.0 & -0.8\\ 
  & 9.0 & -0.3 & 0.1 & 0.6 & -1.0 & -1.6 & -1.0 & -1.3 & -1.1\\ 
  & 9.96 & -0.6 & -0.1 & 0.3 & -1.3 & -1.9 & -1.2 & -1.5 & -1.3\\ 
  & 11.01 & -0.8 & -0.4 & 0.1 & -1.5 & -2.2 & -1.5 & -1.8 & -1.6\\ 
  & 12.17 & -1.1 & -0.6 & -0.1 & -1.8 & -2.5 & -1.7 & -2.1 & -1.8\\ 
  & 13.45 & -1.2 & -0.8 & -0.3 & -1.9 & -2.6 & -1.9 & -2.2 & -2.0\\ \midrule 

 E & 14.87 & -1.6 & -1.2 & -0.7 & -2.2 & -2.9 & -2.2 & -2.5 & -2.2\\ 
  & 16.44 & -1.8 & -1.6 & -1.0 & -2.4 & -3.2 & -2.4 & -2.7 & -2.4\\ 
  & 18.17 & -2.1 & -1.9 & -1.3 & -2.6 & -3.4 & -2.6 & -3.0 & -2.6\\ 
 \bottomrule 
\end{tabular}

%% file: table_mag_006.tex
\begin{tabular}{lccccccccc}
\toprule\midrule 
 Group & $M_{\text{init}}$ & U & B & V & NUV & FUV & UVW1 & UVW2 & UVM2\\ 
 
\midrule 
A & 2.0 & 4.9 & 4.9 & 5.2 & 4.6 & 4.2 & 4.6 & 4.4 & 4.6\\ 
  & 2.21 & 4.4 & 4.4 & 4.8 & 3.9 & 3.5 & 3.9 & 3.7 & 3.9\\ 
  & 2.44 & 3.9 & 4.0 & 4.4 & 3.4 & 2.9 & 3.4 & 3.2 & 3.3\\ 
  & 2.7 & 3.4 & 3.6 & 4.0 & 2.8 & 2.3 & 2.9 & 2.6 & 2.8\\ 
  & 2.99 & 2.9 & 3.1 & 3.5 & 2.2 & 1.7 & 2.3 & 2.0 & 2.2\\ 
  & 3.3 & 2.5 & 2.8 & 3.2 & 1.8 & 1.3 & 1.9 & 1.6 & 1.8\\ 
  & 3.65 & 2.0 & 2.4 & 2.8 & 1.4 & 0.9 & 1.4 & 1.2 & 1.4\\ 
  & 4.04 & 1.7 & 2.0 & 2.5 & 1.0 & 0.5 & 1.1 & 0.8 & 1.0\\ 
  & 4.46 & 1.3 & 1.7 & 2.1 & 0.7 & 0.2 & 0.7 & 0.5 & 0.7\\ 
  & 4.93 & 1.0 & 1.4 & 1.8 & 0.3 & -0.2 & 0.4 & 0.1 & 0.3\\ 
  & 5.45 & 0.7 & 1.1 & 1.5 & 0.0 & -0.5 & 0.1 & -0.2 & 0.0\\ 
  & 6.03 & 0.5 & 1.0 & 1.4 & -0.1 & -0.6 & -0.1 & -0.3 & -0.1\\ 
  & 6.66 & 0.2 & 0.7 & 1.1 & -0.4 & -1.0 & -0.4 & -0.7 & -0.5\\ \midrule 

 A/E & 7.37 & -0.0 & 0.4 & 0.8 & -0.7 & -1.3 & -0.7 & -0.9 & -0.8\\ 
  & 8.15 & -0.3 & 0.1 & 0.6 & -1.0 & -1.6 & -1.0 & -1.2 & -1.0\\ 
  & 9.0 & -0.6 & -0.2 & 0.3 & -1.3 & -1.9 & -1.2 & -1.5 & -1.3\\ 
  & 9.96 & -0.8 & -0.4 & 0.1 & -1.6 & -2.2 & -1.5 & -1.8 & -1.6\\ 
  & 11.01 & -1.1 & -0.6 & -0.1 & -1.8 & -2.4 & -1.7 & -2.0 & -1.8\\ 
  & 12.17 & -1.3 & -0.9 & -0.4 & -2.0 & -2.7 & -2.0 & -2.3 & -2.1\\ 
  & 13.45 & -1.5 & -1.1 & -0.6 & -2.3 & -2.9 & -2.2 & -2.5 & -2.3\\ \midrule 

 E & 14.87 & -1.8 & -1.4 & -0.9 & -2.5 & -3.2 & -2.5 & -2.8 & -2.5\\ 
  & 16.44 & -2.0 & -1.7 & -1.2 & -2.7 & -3.5 & -2.7 & -3.0 & -2.8\\ 
  & 18.17 & -2.5 & -2.1 & -1.6 & -3.2 & -3.9 & -3.1 & -3.4 & -3.2\\ 
 \bottomrule 
\end{tabular}

%% file: table_mag_002.tex
\begin{tabular}{lccccccccc}
\toprule\midrule 
 Group & $M_{\text{init}}$ & U & B & V & NUV & FUV & UVW1 & UVW2 & UVM2\\ 
 
\midrule 
A & 2.0 & 4.7 & 4.7 & 5.1 & 4.2 & 3.8 & 4.3 & 4.1 & 4.2\\ 
  & 2.21 & 4.2 & 4.3 & 4.6 & 3.6 & 3.2 & 3.7 & 3.5 & 3.6\\ 
  & 2.44 & 3.6 & 3.8 & 4.2 & 3.1 & 2.6 & 3.1 & 2.9 & 3.1\\ 
  & 2.7 & 3.1 & 3.3 & 3.7 & 2.4 & 1.9 & 2.5 & 2.2 & 2.4\\ 
  & 2.99 & 2.6 & 2.9 & 3.3 & 1.9 & 1.4 & 2.0 & 1.7 & 1.9\\ 
  & 3.3 & 2.2 & 2.5 & 2.9 & 1.5 & 1.0 & 1.6 & 1.3 & 1.5\\ 
  & 3.65 & 1.8 & 2.1 & 2.5 & 1.1 & 0.6 & 1.2 & 0.9 & 1.1\\ 
  & 4.04 & 1.4 & 1.8 & 2.2 & 0.7 & 0.2 & 0.8 & 0.5 & 0.7\\ 
  & 4.46 & 1.0 & 1.4 & 1.8 & 0.3 & -0.2 & 0.4 & 0.2 & 0.3\\ 
  & 4.93 & 0.7 & 1.1 & 1.5 & 0.0 & -0.5 & 0.1 & -0.1 & 0.0\\ 
  & 5.45 & 0.3 & 0.8 & 1.2 & -0.3 & -0.8 & -0.3 & -0.5 & -0.3\\ 
  & 6.03 & 0.1 & 0.6 & 1.0 & -0.5 & -1.0 & -0.5 & -0.7 & -0.5\\ 
  & 6.66 & -0.2 & 0.3 & 0.7 & -0.8 & -1.4 & -0.8 & -1.0 & -0.9\\ 
  & 7.37 & -0.4 & -0.0 & 0.4 & -1.1 & -1.7 & -1.1 & -1.3 & -1.1\\ \midrule
 
 A/E & 8.15 & -0.8 & -0.3 & 0.1 & -1.5 & -2.0 & -1.4 & -1.7 & -1.5\\ 
  & 9.0 & -1.1 & -0.7 & -0.2 & -1.8 & -2.4 & -1.7 & -2.0 & -1.8\\ 
  & 9.96 & -1.3 & -0.9 & -0.4 & -2.0 & -2.6 & -2.0 & -2.2 & -2.0\\ 
  & 11.01 & -1.6 & -1.2 & -0.8 & -2.4 & -3.0 & -2.3 & -2.6 & -2.4\\ 
  & 12.17 & -1.9 & -1.5 & -1.0 & -2.6 & -3.2 & -2.6 & -2.9 & -2.6\\ 
  & 13.45 & -2.2 & -1.8 & -1.3 & -2.9 & -3.5 & -2.8 & -3.1 & -2.9\\ 
  & 14.87 & -2.4 & -2.0 & -1.5 & -3.2 & -3.8 & -3.1 & -3.4 & -3.2\\ 
  & 16.44 & -2.4 & -2.0 & -1.5 & -3.1 & -3.8 & -3.1 & -3.4 & -3.2\\ 
  & 18.17 & -2.2 & -1.8 & -1.3 & -2.9 & -3.6 & -2.9 & -3.2 & -2.9\\ 
 \bottomrule 
\end{tabular}

%% file: table_mag_0002.tex
\begin{tabular}{lccccccccc}
\toprule\midrule 
 Group & $M_{\text{init}}$ & U & B & V & NUV & FUV & UVW1 & UVW2 & UVM2\\ 
 
\midrule 
A & 2.0 & 4.3 & 4.4 & 4.7 & 3.9 & 3.4 & 3.9 & 3.7 & 3.9\\ 
  & 2.21 & 3.7 & 3.9 & 4.2 & 3.2 & 2.7 & 3.2 & 3.0 & 3.2\\ 
  & 2.44 & 3.1 & 3.3 & 3.7 & 2.5 & 2.0 & 2.6 & 2.3 & 2.5\\ 
  & 2.7 & 2.6 & 2.8 & 3.2 & 2.0 & 1.4 & 2.0 & 1.8 & 2.0\\ 
  & 2.99 & 2.1 & 2.4 & 2.8 & 1.5 & 1.0 & 1.6 & 1.3 & 1.5\\ 
  & 3.3 & 1.6 & 1.9 & 2.3 & 1.0 & 0.4 & 1.0 & 0.8 & 0.9\\ 
  & 3.65 & 1.1 & 1.5 & 1.9 & 0.5 & -0.0 & 0.5 & 0.3 & 0.5\\ 
  & 4.04 & 0.7 & 1.0 & 1.4 & 0.0 & -0.5 & 0.1 & -0.2 & 0.0\\ 
  & 4.46 & 0.3 & 0.7 & 1.1 & -0.4 & -0.9 & -0.3 & -0.5 & -0.4\\ 
  & 4.93 & -0.1 & 0.3 & 0.7 & -0.8 & -1.2 & -0.7 & -0.9 & -0.8\\ 
  & 5.45 & -0.5 & -0.2 & 0.2 & -1.2 & -1.7 & -1.1 & -1.4 & -1.2\\ 
  & 6.03 & -0.9 & -0.5 & -0.1 & -1.5 & -2.0 & -1.4 & -1.7 & -1.5\\ 
  & 6.66 & -1.2 & -0.9 & -0.5 & -1.9 & -2.4 & -1.8 & -2.1 & -1.9\\  
  & 7.37 & -1.6 & -1.2 & -0.8 & -2.3 & -2.8 & -2.2 & -2.4 & -2.3\\ 
  & 8.15 & -2.0 & -1.6 & -1.2 & -2.6 & -3.1 & -2.6 & -2.8 & -2.6\\ 
  & 9.0 & -2.3 & -2.0 & -1.6 & -3.0 & -3.5 & -2.9 & -3.2 & -3.0\\ 
  & 9.96 & -2.5 & -2.1 & -1.7 & -3.2 & -3.7 & -3.1 & -3.4 & -3.2\\ 
  & 11.01 & -2.9 & -2.5 & -2.1 & -3.5 & -4.0 & -3.5 & -3.7 & -3.6\\ 
  & 12.17 & -3.3 & -2.9 & -2.5 & -3.9 & -4.4 & -3.8 & -4.1 & -3.9\\ 
  & 13.45 & -3.5 & -3.1 & -2.7 & -4.1 & -4.6 & -4.1 & -4.3 & -4.1\\ 
  & 14.87 & -3.6 & -3.2 & -2.8 & -4.3 & -4.8 & -4.2 & -4.5 & -4.3\\ 
  & 16.44 & -3.7 & -3.3 & -2.9 & -4.4 & -4.9 & -4.3 & -4.6 & -4.4\\ 
  & 18.17 & -4.9 & -4.6 & -4.3 & -5.5 & -5.9 & -5.5 & -5.7 & -5.5\\ 
 \bottomrule 
\end{tabular}

%% file: Gotbergetal18.bbl
\begin{thebibliography}{181}
\expandafter\ifx\csname natexlab\endcsname\relax\def\natexlab#1{#1}\fi

\bibitem[{{Abazajian} {et~al.}(2009){Abazajian}, {Adelman-McCarthy},
  {Ag{\"u}eros}, {Allam}, {Allende Prieto}, {An}, {Anderson}, {Anderson},
  {Annis}, {Bahcall}, \& et~al.}]{2009ApJS..182..543A}
{Abazajian}, K.~N., {Adelman-McCarthy}, J.~K., {Ag{\"u}eros}, M.~A., {et~al.}
  2009, \apjs, 182, 543

\bibitem[{{Almeida} {et~al.}(2017){Almeida}, {Sana}, {Taylor}, {Barb{\'a}},
  {Bonanos}, {Crowther}, {Damineli}, {de Koter}, {de Mink}, {Evans}, {Gieles},
  {Grin}, {H{\'e}nault-Brunet}, {Langer}, {Lennon}, {Lockwood}, {Ma{\'{\i}}z
  Apell{\'a}niz}, {Moffat}, {Neijssel}, {Norman}, {Ram{\'{\i}}rez-Agudelo},
  {Richardson}, {Schootemeijer}, {Shenar}, {Soszy{\'n}ski}, {Tramper}, \&
  {Vink}}]{2017A&A...598A..84A}
{Almeida}, L.~A., {Sana}, H., {Taylor}, W., {et~al.} 2017, \aap, 598, A84

\bibitem[{{Asplund} {et~al.}(2009){Asplund}, {Grevesse}, {Sauval}, \&
  {Scott}}]{2009ARA&A..47..481A}
{Asplund}, M., {Grevesse}, N., {Sauval}, A.~J., \& {Scott}, P. 2009, \araa, 47,
  481

\bibitem[{{Azzopardi} \& {Breysacher}(1979)}]{1979A&A....75..120A}
{Azzopardi}, M. \& {Breysacher}, J. 1979, \aap, 75, 120

\bibitem[{{Barkana} \& {Loeb}(2001)}]{2001PhR...349..125B}
{Barkana}, R. \& {Loeb}, A. 2001, \physrep, 349, 125

\bibitem[{{Belkus} {et~al.}(2003){Belkus}, {Van Bever}, {Vanbeveren}, \& {van
  Rensbergen}}]{2003A&A...400..429B}
{Belkus}, H., {Van Bever}, J., {Vanbeveren}, D., \& {van Rensbergen}, W. 2003,
  \aap, 400, 429

\bibitem[{{Bianchi} {et~al.}(2011){Bianchi}, {Herald}, {Efremova}, {Girardi},
  {Zabot}, {Marigo}, {Conti}, \& {Shiao}}]{2011Ap&SS.335..161B}
{Bianchi}, L., {Herald}, J., {Efremova}, B., {et~al.} 2011, \apss, 335, 161

\bibitem[{{B{\"o}hm-Vitense}(1958)}]{1958ZA.....46..108B}
{B{\"o}hm-Vitense}, E. 1958, \zap, 46, 108

\bibitem[{{Bromm} \& {Yoshida}(2011)}]{2011ARA&A..49..373B}
{Bromm}, V. \& {Yoshida}, N. 2011, \araa, 49, 373

\bibitem[{{Brott} {et~al.}(2011){Brott}, {de Mink}, {Cantiello}, {Langer}, {de
  Koter}, {Evans}, {Hunter}, {Trundle}, \& {Vink}}]{2011A&A...530A.115B}
{Brott}, I., {de Mink}, S.~E., {Cantiello}, M., {et~al.} 2011, \aap, 530, A115

\bibitem[{{Bruzual} \& {Charlot}(2003)}]{2003MNRAS.344.1000B}
{Bruzual}, G. \& {Charlot}, S. 2003, \mnras, 344, 1000

\bibitem[{{Cardelli} {et~al.}(1989){Cardelli}, {Clayton}, \&
  {Mathis}}]{1989ApJ...345..245C}
{Cardelli}, J.~A., {Clayton}, G.~C., \& {Mathis}, J.~S. 1989, \apj, 345, 245

\bibitem[{{Cassinelli} {et~al.}(1995){Cassinelli}, {Cohen}, {Macfarlane},
  {Drew}, {Lynas-Gray}, {Hoare}, {Vallerga}, {Welsh}, {Vedder}, {Hubeny}, \&
  {Lanz}}]{1995ApJ...438..932C}
{Cassinelli}, J.~P., {Cohen}, D.~H., {Macfarlane}, J.~J., {et~al.} 1995, \apj,
  438, 932

\bibitem[{{Cassinelli} {et~al.}(1996){Cassinelli}, {Cohen}, {Macfarlane},
  {Drew}, {Lynas-Gray}, {Hubeny}, {Vallerga}, {Welsh}, \&
  {Hoare}}]{1996ApJ...460..949C}
{Cassinelli}, J.~P., {Cohen}, D.~H., {Macfarlane}, J.~J., {et~al.} 1996, \apj,
  460, 949

\bibitem[{{Cassinelli} {et~al.}(2001){Cassinelli}, {Miller}, {Waldron},
  {MacFarlane}, \& {Cohen}}]{Cassinelli+2001}
{Cassinelli}, J.~P., {Miller}, N.~A., {Waldron}, W.~L., {MacFarlane}, J.~J., \&
  {Cohen}, D.~H. 2001, \apjl, 554, L55

\bibitem[{{Chen} \& {Han}(2010)}]{2010Ap&SS.329..277C}
{Chen}, X. \& {Han}, Z. 2010, \apss, 329, 277

\bibitem[{{Chini} {et~al.}(2012){Chini}, {Hoffmeister}, {Nasseri}, {Stahl}, \&
  {Zinnecker}}]{2012MNRAS.424.1925C}
{Chini}, R., {Hoffmeister}, V.~H., {Nasseri}, A., {Stahl}, O., \& {Zinnecker},
  H. 2012, \mnras, 424, 1925

\bibitem[{{Claeys} {et~al.}(2011){Claeys}, {de Mink}, {Pols}, {Eldridge}, \&
  {Baes}}]{2011A&A...528A.131C}
{Claeys}, J.~S.~W., {de Mink}, S.~E., {Pols}, O.~R., {Eldridge}, J.~J., \&
  {Baes}, M. 2011, \aap, 528, A131

\bibitem[{{Claret}(2007)}]{2007A&A...475.1019C}
{Claret}, A. 2007, \aap, 475, 1019

\bibitem[{{Cohen} {et~al.}(2014){Cohen}, {Li}, {Gayley}, {Owocki}, {Sundqvist},
  {Petit}, \& {Leutenegger}}]{Cohen+2014}
{Cohen}, D.~H., {Li}, Z., {Gayley}, K.~G., {et~al.} 2014, \mnras, 444, 3729

\bibitem[{{Copperwheat} {et~al.}(2011){Copperwheat}, {Morales-Rueda}, {Marsh},
  {Maxted}, \& {Heber}}]{2011MNRAS.415.1381C}
{Copperwheat}, C.~M., {Morales-Rueda}, L., {Marsh}, T.~R., {Maxted}, P.~F.~L.,
  \& {Heber}, U. 2011, \mnras, 415, 1381

\bibitem[{{Crowther}(2007)}]{2007ARA&A..45..177C}
{Crowther}, P.~A. 2007, \araa, 45, 177

\bibitem[{{Crowther} {et~al.}(1995){Crowther}, {Smith}, \&
  {Hillier}}]{1995A&A...302..457C}
{Crowther}, P.~A., {Smith}, L.~J., \& {Hillier}, D.~J. 1995, \aap, 302, 457

\bibitem[{{Crowther} \& {Walborn}(2011)}]{2011MNRAS.416.1311C}
{Crowther}, P.~A. \& {Walborn}, N.~R. 2011, \mnras, 416, 1311

\bibitem[{{D'Cruz} {et~al.}(1996){D'Cruz}, {Dorman}, {Rood}, \&
  {O'Connell}}]{DCruz+1996}
{D'Cruz}, N.~L., {Dorman}, B., {Rood}, R.~T., \& {O'Connell}, R.~W. 1996, \apj,
  466, 359

\bibitem[{{de Jager} {et~al.}(1988){de Jager}, {Nieuwenhuijzen}, \& {van der
  Hucht}}]{1988A&AS...72..259D}
{de Jager}, C., {Nieuwenhuijzen}, H., \& {van der Hucht}, K.~A. 1988, \aaps,
  72, 259

\bibitem[{{de Mello} {et~al.}(1998){de Mello}, {Schaerer}, {Heldmann}, \&
  {Leitherer}}]{1998ApJ...507..199D}
{de Mello}, D.~F., {Schaerer}, D., {Heldmann}, J., \& {Leitherer}, C. 1998,
  \apj, 507, 199

\bibitem[{{de Mink} \& {Belczynski}(2015)}]{2015ApJ...814...58D}
{de Mink}, S.~E. \& {Belczynski}, K. 2015, \apj, 814, 58

\bibitem[{{de Mink} {et~al.}(2011){de Mink}, {Langer}, \&
  {Izzard}}]{de-Mink+2011}
{de Mink}, S.~E., {Langer}, N., \& {Izzard}, R.~G. 2011, Bulletin de la Societe
  Royale des Sciences de Liege, 80, 543

\bibitem[{{de Mink} {et~al.}(2013){de Mink}, {Langer}, {Izzard}, {Sana}, \& {de
  Koter}}]{2013ApJ...764..166D}
{de Mink}, S.~E., {Langer}, N., {Izzard}, R.~G., {Sana}, H., \& {de Koter}, A.
  2013, \apj, 764, 166

\bibitem[{{de Mink} {et~al.}(2007){de Mink}, {Pols}, \&
  {Hilditch}}]{2007A&A...467.1181D}
{de Mink}, S.~E., {Pols}, O.~R., \& {Hilditch}, R.~W. 2007, \aap, 467, 1181

\bibitem[{{de Mink} {et~al.}(2014){de Mink}, {Sana}, {Langer}, {Izzard}, \&
  {Schneider}}]{de-Mink+2014}
{de Mink}, S.~E., {Sana}, H., {Langer}, N., {Izzard}, R.~G., \& {Schneider},
  F.~R.~N. 2014, \apj, 782, 7

\bibitem[{{Dessart} {et~al.}(2011){Dessart}, {Hillier}, {Livne}, {Yoon},
  {Woosley}, {Waldman}, \& {Langer}}]{2011MNRAS.414.2985D}
{Dessart}, L., {Hillier}, D.~J., {Livne}, E., {et~al.} 2011, \mnras, 414, 2985

\bibitem[{{Dewi} {et~al.}(2002){Dewi}, {Pols}, {Savonije}, \& {van den
  Heuvel}}]{Dewi+2002}
{Dewi}, J.~D.~M., {Pols}, O.~R., {Savonije}, G.~J., \& {van den Heuvel},
  E.~P.~J. 2002, \mnras, 331, 1027

\bibitem[{{Drew} {et~al.}(2014){Drew}, {Gonzalez-Solares}, {Greimel}, {Irwin},
  {K{\"u}pc{\"u} Yoldas}, {Lewis}, {Barentsen}, {Eisl{\"o}ffel}, {Farnhill},
  {Martin}, {Walsh}, {Walton}, {Mohr-Smith}, {Raddi}, {Sale}, {Wright},
  {Groot}, {Barlow}, {Corradi}, {Drake}, {Fabregat}, {Frew}, {G{\"a}nsicke},
  {Knigge}, {Mampaso}, {Morris}, {Naylor}, {Parker}, {Phillipps}, {Ruhland},
  {Steeghs}, {Unruh}, {Vink}, {Wesson}, \& {Zijlstra}}]{2014MNRAS.440.2036D}
{Drew}, J.~E., {Gonzalez-Solares}, E., {Greimel}, R., {et~al.} 2014, \mnras,
  440, 2036

\bibitem[{{Drew} {et~al.}(2005){Drew}, {Greimel}, {Irwin}, {Aungwerojwit},
  {Barlow}, {Corradi}, {Drake}, {G{\"a}nsicke}, {Groot}, {Hales}, {Hopewell},
  {Irwin}, {Knigge}, {Leisy}, {Lennon}, {Mampaso}, {Masheder}, {Matsuura},
  {Morales-Rueda}, {Morris}, {Parker}, {Phillipps}, {Rodriguez-Gil}, {Roelofs},
  {Skillen}, {Sokoloski}, {Steeghs}, {Unruh}, {Viironen}, {Vink}, {Walton},
  {Witham}, {Wright}, {Zijlstra}, \& {Zurita}}]{2005MNRAS.362..753D}
{Drew}, J.~E., {Greimel}, R., {Irwin}, M.~J., {et~al.} 2005, \mnras, 362, 753

\bibitem[{{Dunstall} {et~al.}(2015){Dunstall}, {Dufton}, {Sana}, {Evans},
  {Howarth}, {Sim{\'o}n-D{\'{\i}}az}, {de Mink}, {Langer}, {Ma{\'{\i}}z
  Apell{\'a}niz}, \& {Taylor}}]{2015A&A...580A..93D}
{Dunstall}, P.~R., {Dufton}, P.~L., {Sana}, H., {et~al.} 2015, \aap, 580, A93

\bibitem[{{Edelmann} {et~al.}(2003){Edelmann}, {Heber}, {Hagen}, {Lemke},
  {Dreizler}, {Napiwotzki}, \& {Engels}}]{Edelmann+2003}
{Edelmann}, H., {Heber}, U., {Hagen}, H.-J., {et~al.} 2003, \aap, 400, 939

\bibitem[{{Eldridge} {et~al.}(2013){Eldridge}, {Fraser}, {Smartt}, {Maund}, \&
  {Crockett}}]{2013MNRAS.436..774E}
{Eldridge}, J.~J., {Fraser}, M., {Smartt}, S.~J., {Maund}, J.~R., \&
  {Crockett}, R.~M. 2013, \mnras, 436, 774

\bibitem[{{Eldridge} \& {Stanway}(2009)}]{2009MNRAS.400.1019E}
{Eldridge}, J.~J. \& {Stanway}, E.~R. 2009, \mnras, 400, 1019

\bibitem[{{Eldridge} \& {Stanway}(2012)}]{2012MNRAS.419..479E}
{Eldridge}, J.~J. \& {Stanway}, E.~R. 2012, \mnras, 419, 479

\bibitem[{{Eldridge} {et~al.}(2017){Eldridge}, {Stanway}, {Xiao}, {McClelland},
  {Taylor}, {Ng}, {Greis}, \& {Bray}}]{2017PASA...34...58E}
{Eldridge}, J.~J., {Stanway}, E.~R., {Xiao}, L., {et~al.} 2017, \pasa, 34, e058

\bibitem[{{Farmer} {et~al.}(2016){Farmer}, {Fields}, {Petermann}, {Dessart},
  {Cantiello}, {Paxton}, \& {Timmes}}]{2016ApJS..227...22F}
{Farmer}, R., {Fields}, C.~E., {Petermann}, I., {et~al.} 2016, \apjs, 227, 22

\bibitem[{{Feldmeier} {et~al.}(1997){Feldmeier}, {Puls}, \&
  {Pauldrach}}]{Feldmeier+1997}
{Feldmeier}, A., {Puls}, J., \& {Pauldrach}, A.~W.~A. 1997, \aap, 322, 878

\bibitem[{{Filippenko}(1997)}]{1997ARA&A..35..309F}
{Filippenko}, A.~V. 1997, \araa, 35, 309

\bibitem[{{Fitzpatrick} \& {Massa}(2007)}]{2007ApJ...663..320F}
{Fitzpatrick}, E.~L. \& {Massa}, D. 2007, \apj, 663, 320

\bibitem[{{Fontaine} {et~al.}(2012){Fontaine}, {Brassard}, {Charpinet},
  {Green}, {Randall}, \& {Van Grootel}}]{2012A&A...539A..12F}
{Fontaine}, G., {Brassard}, P., {Charpinet}, S., {et~al.} 2012, \aap, 539, A12

\bibitem[{{Fontaine} {et~al.}(2008){Fontaine}, {Brassard}, {Green}, {Chayer},
  {Charpinet}, {Andersen}, \& {Portouw}}]{2008A&A...486L..39F}
{Fontaine}, G., {Brassard}, P., {Green}, E.~M., {et~al.} 2008, \aap, 486, L39

\bibitem[{{Fryer}(1999)}]{1999ApJ...522..413F}
{Fryer}, C.~L. 1999, \apj, 522, 413

\bibitem[{{Garnett} {et~al.}(1991){Garnett}, {Kennicutt}, {Chu}, \&
  {Skillman}}]{1991PASP..103..850G}
{Garnett}, D.~R., {Kennicutt}, Jr., R.~C., {Chu}, Y.-H., \& {Skillman}, E.~D.
  1991, \pasp, 103, 850

\bibitem[{{Geier}(2013)}]{2013A&A...549A.110G}
{Geier}, S. 2013, \aap, 549, A110

\bibitem[{{Geier} {et~al.}(2013){Geier}, {Marsh}, {Wang}, {Dunlap}, {Barlow},
  {Schaffenroth}, {Chen}, {Irrgang}, {Maxted}, {Ziegerer}, {Kupfer},
  {Miszalski}, {Heber}, {Han}, {Shporer}, {Telting}, {G{\"a}nsicke},
  {{\O}stensen}, {O'Toole}, \& {Napiwotzki}}]{2013A&A...554A..54G}
{Geier}, S., {Marsh}, T.~R., {Wang}, B., {et~al.} 2013, \aap, 554, A54

\bibitem[{{Geier} {et~al.}(2017){Geier}, {{\O}stensen}, {Nemeth}, {Gentile
  Fusillo}, {G{\"a}nsicke}, {Telting}, {Green}, \&
  {Schaffenroth}}]{2017A&A...600A..50G}
{Geier}, S., {{\O}stensen}, R.~H., {Nemeth}, P., {et~al.} 2017, \aap, 600, A50

\bibitem[{{Georgy} {et~al.}(2009){Georgy}, {Meynet}, {Walder}, {Folini}, \&
  {Maeder}}]{Georgy+2009}
{Georgy}, C., {Meynet}, G., {Walder}, R., {Folini}, D., \& {Maeder}, A. 2009,
  \aap, 502, 611

\bibitem[{{Gies} {et~al.}(1998){Gies}, {Bagnuolo}, {Ferrara}, {Kaye},
  {Thaller}, {Penny}, \& {Peters}}]{1998ApJ...493..440G}
{Gies}, D.~R., {Bagnuolo}, Jr., W.~G., {Ferrara}, E.~C., {et~al.} 1998, \apj,
  493, 440

\bibitem[{{Gordon} {et~al.}(2003){Gordon}, {Clayton}, {Misselt}, {Landolt}, \&
  {Wolff}}]{2003ApJ...594..279G}
{Gordon}, K.~D., {Clayton}, G.~C., {Misselt}, K.~A., {Landolt}, A.~U., \&
  {Wolff}, M.~J. 2003, \apj, 594, 279

\bibitem[{{G{\"o}tberg} {et~al.}(2017){G{\"o}tberg}, {de Mink}, \&
  {Groh}}]{2017A&A...608A..11G}
{G{\"o}tberg}, Y., {de Mink}, S.~E., \& {Groh}, J.~H. 2017, \aap, 608, A11

\bibitem[{{Gr{\"a}fener} {et~al.}(2002){Gr{\"a}fener}, {Koesterke}, \&
  {Hamann}}]{2002A&A...387..244G}
{Gr{\"a}fener}, G., {Koesterke}, L., \& {Hamann}, W.-R. 2002, \aap, 387, 244

\bibitem[{{Graur} {et~al.}(2017){Graur}, {Bianco}, {Modjaz}, {Shivvers},
  {Filippenko}, {Li}, \& {Smith}}]{2017ApJ...837..121G}
{Graur}, O., {Bianco}, F.~B., {Modjaz}, M., {et~al.} 2017, \apj, 837, 121

\bibitem[{{Grevesse} \& {Sauval}(1998)}]{1998SSRv...85..161G}
{Grevesse}, N. \& {Sauval}, A.~J. 1998, \ssr, 85, 161

\bibitem[{{Groh} {et~al.}(2014){Groh}, {Meynet}, {Ekstr{\"o}m}, \&
  {Georgy}}]{2014A&A...564A..30G}
{Groh}, J.~H., {Meynet}, G., {Ekstr{\"o}m}, S., \& {Georgy}, C. 2014, \aap,
  564, A30

\bibitem[{{Groh} {et~al.}(2013){Groh}, {Meynet}, {Georgy}, \&
  {Ekstr{\"o}m}}]{2013A&A...558A.131G}
{Groh}, J.~H., {Meynet}, G., {Georgy}, C., \& {Ekstr{\"o}m}, S. 2013, \aap,
  558, A131

\bibitem[{{Groh} {et~al.}(2008){Groh}, {Oliveira}, \&
  {Steiner}}]{2008A&A...485..245G}
{Groh}, J.~H., {Oliveira}, A.~S., \& {Steiner}, J.~E. 2008, \aap, 485, 245

\bibitem[{{Hall} \& {Jeffery}(2016)}]{Hall+2016}
{Hall}, P.~D. \& {Jeffery}, C.~S. 2016, \mnras, 463, 2756

\bibitem[{{Hamann} \& {Gr{\"a}fener}(2003)}]{Hamann+2003}
{Hamann}, W.-R. \& {Gr{\"a}fener}, G. 2003, \aap, 410, 993

\bibitem[{{Hamann} {et~al.}(2006){Hamann}, {Gr{\"a}fener}, \&
  {Liermann}}]{2006A&A...457.1015H}
{Hamann}, W.-R., {Gr{\"a}fener}, G., \& {Liermann}, A. 2006, \aap, 457, 1015

\bibitem[{{Han} {et~al.}(2010){Han}, {Podsiadlowski}, \&
  {Lynas-Gray}}]{2010Ap&SS.329...41H}
{Han}, Z., {Podsiadlowski}, P., \& {Lynas-Gray}, A. 2010, \apss, 329, 41

\bibitem[{{Han} {et~al.}(2007){Han}, {Podsiadlowski}, \&
  {Lynas-Gray}}]{2007MNRAS.380.1098H}
{Han}, Z., {Podsiadlowski}, P., \& {Lynas-Gray}, A.~E. 2007, \mnras, 380, 1098

\bibitem[{{Han} {et~al.}(2002){Han}, {Podsiadlowski}, {Maxted}, {Marsh}, \&
  {Ivanova}}]{2002MNRAS.336..449H}
{Han}, Z., {Podsiadlowski}, P., {Maxted}, P.~F.~L., {Marsh}, T.~R., \&
  {Ivanova}, N. 2002, \mnras, 336, 449

\bibitem[{{Heber}(2016)}]{2016PASP..128h2001H}
{Heber}, U. 2016, \pasp, 128, 082001

\bibitem[{{Heger} {et~al.}(2003){Heger}, {Fryer}, {Woosley}, {Langer}, \&
  {Hartmann}}]{2003ApJ...591..288H}
{Heger}, A., {Fryer}, C.~L., {Woosley}, S.~E., {Langer}, N., \& {Hartmann},
  D.~H. 2003, \apj, 591, 288

\bibitem[{{Hillier}(1990)}]{1990A&A...231..116H}
{Hillier}, D.~J. 1990, \aap, 231, 116

\bibitem[{{Hillier} \& {Miller}(1998)}]{1998ApJ...496..407H}
{Hillier}, D.~J. \& {Miller}, D.~L. 1998, \apj, 496, 407

\bibitem[{{Hillier} \& {Miller}(1999)}]{Hillier+1999}
{Hillier}, D.~J. \& {Miller}, D.~L. 1999, \apj, 519, 354

\bibitem[{{Hopkins} {et~al.}(2014){Hopkins}, {Kere{\v s}}, {O{\~n}orbe},
  {Faucher-Gigu{\`e}re}, {Quataert}, {Murray}, \&
  {Bullock}}]{2014MNRAS.445..581H}
{Hopkins}, P.~F., {Kere{\v s}}, D., {O{\~n}orbe}, J., {et~al.} 2014, \mnras,
  445, 581

\bibitem[{{Hu} {et~al.}(2011){Hu}, {Tout}, {Glebbeek}, \&
  {Dupret}}]{2011MNRAS.418..195H}
{Hu}, H., {Tout}, C.~A., {Glebbeek}, E., \& {Dupret}, M.-A. 2011, \mnras, 418,
  195

\bibitem[{{Hut}(1981)}]{1981A&A....99..126H}
{Hut}, P. 1981, \aap, 99, 126

\bibitem[{{Ivanova}(2011)}]{2011ApJ...730...76I}
{Ivanova}, N. 2011, \apj, 730, 76

\bibitem[{{Kalogera} {et~al.}(2007){Kalogera}, {Belczynski}, {Kim},
  {O'Shaughnessy}, \& {Willems}}]{2007PhR...442...75K}
{Kalogera}, V., {Belczynski}, K., {Kim}, C., {O'Shaughnessy}, R., \& {Willems},
  B. 2007, \physrep, 442, 75

\bibitem[{{Kippenhahn} {et~al.}(1980){Kippenhahn}, {Ruschenplatt}, \&
  {Thomas}}]{1980A&A....91..175K}
{Kippenhahn}, R., {Ruschenplatt}, G., \& {Thomas}, H.-C. 1980, \aap, 91, 175

\bibitem[{{Kippenhahn} \& {Weigert}(1967)}]{1967ZA.....65..251K}
{Kippenhahn}, R. \& {Weigert}, A. 1967, \zap, 65, 251

\bibitem[{{Kobulnicky} \& {Fryer}(2007)}]{2007ApJ...670..747K}
{Kobulnicky}, H.~A. \& {Fryer}, C.~L. 2007, \apj, 670, 747

\bibitem[{{Kobulnicky} {et~al.}(2014){Kobulnicky}, {Kiminki}, {Lundquist},
  {Burke}, {Chapman}, {Keller}, {Lester}, {Rolen}, {Topel}, {Bhattacharjee},
  {Smullen}, {Vargas {\'A}lvarez}, {Runnoe}, {Dale}, \&
  {Brotherton}}]{2014ApJS..213...34K}
{Kobulnicky}, H.~A., {Kiminki}, D.~C., {Lundquist}, M.~J., {et~al.} 2014,
  \apjs, 213, 34

\bibitem[{{Krti{\v c}ka} {et~al.}(2016){Krti{\v c}ka}, {Kub{\'a}t}, \& {Krti{\v
  c}kov{\'a}}}]{2016A&A...593A.101K}
{Krti{\v c}ka}, J., {Kub{\'a}t}, J., \& {Krti{\v c}kov{\'a}}, I. 2016, \aap,
  593, A101

\bibitem[{{Kudritzki} \& {Simon}(1978)}]{1978A&A....70..653K}
{Kudritzki}, R.~P. \& {Simon}, K.~P. 1978, \aap, 70, 653

\bibitem[{{Kupfer} {et~al.}(2015){Kupfer}, {Geier}, {Heber}, {{\O}stensen},
  {Barlow}, {Maxted}, {Heuser}, {Schaffenroth}, \&
  {G{\"a}nsicke}}]{2015A&A...576A..44K}
{Kupfer}, T., {Geier}, S., {Heber}, U., {et~al.} 2015, \aap, 576, A44

\bibitem[{{Kurucz}(1992)}]{1992IAUS..149..225K}
{Kurucz}, R.~L. 1992, in IAU Symposium, Vol. 149, The Stellar Populations of
  Galaxies, ed. B.~{Barbuy} \& A.~{Renzini}, 225

\bibitem[{{Lamers} \& {Cassinelli}(1999)}]{1999isw..book.....L}
{Lamers}, H.~J.~G.~L.~M. \& {Cassinelli}, J.~P. 1999, {Introduction to Stellar
  Winds}, 452

\bibitem[{{Langer}(1991)}]{1991A&A...252..669L}
{Langer}, N. 1991, \aap, 252, 669

\bibitem[{{Langer} {et~al.}(1983){Langer}, {Fricke}, \&
  {Sugimoto}}]{1983A&A...126..207L}
{Langer}, N., {Fricke}, K.~J., \& {Sugimoto}, D. 1983, \aap, 126, 207

\bibitem[{{Leitherer} {et~al.}(2014){Leitherer}, {Ekstr{\"o}m}, {Meynet},
  {Schaerer}, {Agienko}, \& {Levesque}}]{2014ApJS..212...14L}
{Leitherer}, C., {Ekstr{\"o}m}, S., {Meynet}, G., {et~al.} 2014, \apjs, 212, 14

\bibitem[{{Leitherer} {et~al.}(1999){Leitherer}, {Schaerer}, {Goldader},
  {Delgado}, {Robert}, {Kune}, {de Mello}, {Devost}, \&
  {Heckman}}]{1999ApJS..123....3L}
{Leitherer}, C., {Schaerer}, D., {Goldader}, J.~D., {et~al.} 1999, \apjs, 123,
  3

\bibitem[{{Li} {et~al.}(2012){Li}, {Zhang}, \& {Liu}}]{2012MNRAS.424..874L}
{Li}, Z., {Zhang}, L., \& {Liu}, J. 2012, \mnras, 424, 874

\bibitem[{{Ma} {et~al.}(2016){Ma}, {Hopkins}, {Kasen}, {Quataert},
  {Faucher-Gigu{\`e}re}, {Kere{\v s}}, {Murray}, \&
  {Strom}}]{2016MNRAS.459.3614M}
{Ma}, X., {Hopkins}, P.~F., {Kasen}, D., {et~al.} 2016, \mnras, 459, 3614

\bibitem[{{Maeder}(1976)}]{1976A&A....47..389M}
{Maeder}, A. 1976, \aap, 47, 389

\bibitem[{{Ma{\'{\i}}z Apell{\'a}niz} {et~al.}(2016){Ma{\'{\i}}z
  Apell{\'a}niz}, {Sota}, {Arias}, {Barb{\'a}}, {Walborn},
  {Sim{\'o}n-D{\'{\i}}az}, {Negueruela}, {Marco}, {Le{\~a}o}, {Herrero},
  {Gamen}, \& {Alfaro}}]{2016ApJS..224....4M}
{Ma{\'{\i}}z Apell{\'a}niz}, J., {Sota}, A., {Arias}, J.~I., {et~al.} 2016,
  \apjs, 224, 4

\bibitem[{{Marchant} {et~al.}(2017){Marchant}, {Langer}, {Podsiadlowski},
  {Tauris}, {de Mink}, {Mandel}, \& {Moriya}}]{2017A&A...604A..55M}
{Marchant}, P., {Langer}, N., {Podsiadlowski}, P., {et~al.} 2017, \aap, 604,
  A55

\bibitem[{{Martin} {et~al.}(2005){Martin}, {Fanson}, {Schiminovich},
  {Morrissey}, {Friedman}, {Barlow}, {Conrow}, {Grange}, {Jelinsky},
  {Milliard}, {Siegmund}, {Bianchi}, {Byun}, {Donas}, {Forster}, {Heckman},
  {Lee}, {Madore}, {Malina}, {Neff}, {Rich}, {Small}, {Surber}, {Szalay},
  {Welsh}, \& {Wyder}}]{2005ApJ...619L...1M}
{Martin}, D.~C., {Fanson}, J., {Schiminovich}, D., {et~al.} 2005, \apjl, 619,
  L1

\bibitem[{{Massey} \& {Duffy}(2001)}]{2001ApJ...550..713M}
{Massey}, P. \& {Duffy}, A.~S. 2001, \apj, 550, 713

\bibitem[{{Massey} {et~al.}(2015){Massey}, {Neugent}, \&
  {Morrell}}]{2015ApJ...807...81M}
{Massey}, P., {Neugent}, K.~F., \& {Morrell}, N. 2015, \apj, 807, 81

\bibitem[{{Massey} {et~al.}(2017){Massey}, {Neugent}, \&
  {Morrell}}]{2017ApJ...837..122M}
{Massey}, P., {Neugent}, K.~F., \& {Morrell}, N. 2017, \apj, 837, 122

\bibitem[{{Massey} {et~al.}(2014){Massey}, {Neugent}, {Morrell}, \&
  {Hillier}}]{2014ApJ...788...83M}
{Massey}, P., {Neugent}, K.~F., {Morrell}, N., \& {Hillier}, D.~J. 2014, \apj,
  788, 83

\bibitem[{{Mathys}(1988)}]{1988A&AS...76..427M}
{Mathys}, G. 1988, \aaps, 76, 427

\bibitem[{{Maxted} {et~al.}(2001){Maxted}, {Heber}, {Marsh}, \&
  {North}}]{2001MNRAS.326.1391M}
{Maxted}, P.~F.~L., {Heber}, U., {Marsh}, T.~R., \& {North}, R.~C. 2001,
  \mnras, 326, 1391

\bibitem[{{Mengel} {et~al.}(1976){Mengel}, {Norris}, \&
  {Gross}}]{1976ApJ...204..488M}
{Mengel}, J.~G., {Norris}, J., \& {Gross}, P.~G. 1976, \apj, 204, 488

\bibitem[{{Mereghetti} {et~al.}(2009){Mereghetti}, {Tiengo}, {Esposito}, {La
  Palombara}, {Israel}, \& {Stella}}]{2009Sci...325.1222M}
{Mereghetti}, S., {Tiengo}, A., {Esposito}, P., {et~al.} 2009, Science, 325,
  1222

\bibitem[{{Moe} \& {Di Stefano}(2017)}]{2017ApJS..230...15M}
{Moe}, M. \& {Di Stefano}, R. 2017, \apjs, 230, 15

\bibitem[{{Moehler} {et~al.}(1990){Moehler}, {Richtler}, {de Boer}, {Dettmar},
  \& {Heber}}]{1990A&AS...86...53M}
{Moehler}, S., {Richtler}, T., {de Boer}, K.~S., {Dettmar}, R.~J., \& {Heber},
  U. 1990, \aaps, 86, 53

\bibitem[{{Mourard} {et~al.}(2015){Mourard}, {Monnier}, {Meilland}, {Gies},
  {Millour}, {Benisty}, {Che}, {Grundstrom}, {Ligi}, {Schaefer}, {Baron},
  {Kraus}, {Zhao}, {Pedretti}, {Berio}, {Clausse}, {Nardetto}, {Perraut},
  {Spang}, {Stee}, {Tallon-Bosc}, {McAlister}, {ten Brummelaar}, {Ridgway},
  {Sturmann}, {Sturmann}, {Turner}, \& {Farrington}}]{2015A&A...577A..51M}
{Mourard}, D., {Monnier}, J.~D., {Meilland}, A., {et~al.} 2015, \aap, 577, A51

\bibitem[{{Naslim} {et~al.}(2013){Naslim}, {Jeffery}, {Hibbert}, \&
  {Behara}}]{2013MNRAS.434.1920N}
{Naslim}, N., {Jeffery}, C.~S., {Hibbert}, A., \& {Behara}, N.~T. 2013, \mnras,
  434, 1920

\bibitem[{{Neugent} {et~al.}(2017){Neugent}, {Massey}, {Hillier}, \&
  {Morrell}}]{2017ApJ...841...20N}
{Neugent}, K.~F., {Massey}, P., {Hillier}, D.~J., \& {Morrell}, N. 2017, \apj,
  841, 20

\bibitem[{{Nomoto} {et~al.}(1993){Nomoto}, {Suzuki}, {Shigeyama}, {Kumagai},
  {Yamaoka}, \& {Saio}}]{Nomoto+1993}
{Nomoto}, K., {Suzuki}, T., {Shigeyama}, T., {et~al.} 1993, \nat, 364, 507

\bibitem[{{Nugis} \& {Lamers}(2000)}]{2000A&A...360..227N}
{Nugis}, T. \& {Lamers}, H.~J.~G.~L.~M. 2000, \aap, 360, 227

\bibitem[{{O'Connor} \& {Ott}(2011)}]{2011ApJ...730...70O}
{O'Connor}, E. \& {Ott}, C.~D. 2011, \apj, 730, 70

\bibitem[{{O'Toole} \& {Heber}(2006)}]{2006A&A...452..579O}
{O'Toole}, S.~J. \& {Heber}, U. 2006, \aap, 452, 579

\bibitem[{{Owocki} {et~al.}(1988){Owocki}, {Castor}, \&
  {Rybicki}}]{1988ApJ...335..914O}
{Owocki}, S.~P., {Castor}, J.~I., \& {Rybicki}, G.~B. 1988, \apj, 335, 914

\bibitem[{{Owocki} {et~al.}(2013){Owocki}, {Sundqvist}, {Cohen}, \&
  {Gayley}}]{Owocki+2013}
{Owocki}, S.~P., {Sundqvist}, J.~O., {Cohen}, D.~H., \& {Gayley}, K.~G. 2013,
  \mnras, 429, 3379

\bibitem[{{Packet}(1981)}]{1981A&A...102...17P}
{Packet}, W. 1981, \aap, 102, 17

\bibitem[{{Paczy{\'n}ski}(1971)}]{1971ARA&A...9..183P}
{Paczy{\'n}ski}, B. 1971, \araa, 9, 183

\bibitem[{{Paquette} {et~al.}(1986){Paquette}, {Pelletier}, {Fontaine}, \&
  {Michaud}}]{1986ApJS...61..177P}
{Paquette}, C., {Pelletier}, C., {Fontaine}, G., \& {Michaud}, G. 1986, \apjs,
  61, 177

\bibitem[{{Paxton} {et~al.}(2011){Paxton}, {Bildsten}, {Dotter}, {Herwig},
  {Lesaffre}, \& {Timmes}}]{2011ApJS..192....3P}
{Paxton}, B., {Bildsten}, L., {Dotter}, A., {et~al.} 2011, \apjs, 192, 3

\bibitem[{{Paxton} {et~al.}(2013){Paxton}, {Cantiello}, {Arras}, {Bildsten},
  {Brown}, {Dotter}, {Mankovich}, {Montgomery}, {Stello}, {Timmes}, \&
  {Townsend}}]{2013ApJS..208....4P}
{Paxton}, B., {Cantiello}, M., {Arras}, P., {et~al.} 2013, \apjs, 208, 4

\bibitem[{{Paxton} {et~al.}(2015){Paxton}, {Marchant}, {Schwab}, {Bauer},
  {Bildsten}, {Cantiello}, {Dessart}, {Farmer}, {Hu}, {Langer}, {Townsend},
  {Townsley}, \& {Timmes}}]{2015ApJS..220...15P}
{Paxton}, B., {Marchant}, P., {Schwab}, J., {et~al.} 2015, \apjs, 220, 15

\bibitem[{{Paxton} {et~al.}(2017){Paxton}, {Schwab}, {Bauer}, {Bildsten},
  {Blinnikov}, {Duffell}, {Farmer}, {Goldberg}, {Marchant}, {Sorokina},
  {Thoul}, {Townsend}, \& {Timmes}}]{2017arXiv171008424P}
{Paxton}, B., {Schwab}, J., {Bauer}, E.~B., {et~al.} 2017, ArXiv e-prints
  [\eprint[arXiv]{1710.08424}]

\bibitem[{{Peters} {et~al.}(2008){Peters}, {Gies}, {Grundstrom}, \&
  {McSwain}}]{2008ApJ...686.1280P}
{Peters}, G.~J., {Gies}, D.~R., {Grundstrom}, E.~D., \& {McSwain}, M.~V. 2008,
  \apj, 686, 1280

\bibitem[{{Peters} {et~al.}(2013){Peters}, {Pewett}, {Gies}, {Touhami}, \&
  {Grundstrom}}]{2013ApJ...765....2P}
{Peters}, G.~J., {Pewett}, T.~D., {Gies}, D.~R., {Touhami}, Y.~N., \&
  {Grundstrom}, E.~D. 2013, \apj, 765, 2

\bibitem[{{Podsiadlowski} {et~al.}(1992){Podsiadlowski}, {Joss}, \&
  {Hsu}}]{1992ApJ...391..246P}
{Podsiadlowski}, P., {Joss}, P.~C., \& {Hsu}, J.~J.~L. 1992, \apj, 391, 246

\bibitem[{{Pols}(1994)}]{Pols1994a}
{Pols}, O.~R. 1994, \aap, 290, 119

\bibitem[{{Pols} {et~al.}(1991){Pols}, {Cote}, {Waters}, \&
  {Heise}}]{1991A&A...241..419P}
{Pols}, O.~R., {Cote}, J., {Waters}, L.~B.~F.~M., \& {Heise}, J. 1991, \aap,
  241, 419

\bibitem[{{Pols} {et~al.}(1998){Pols}, {Schr{\"o}der}, {Hurley}, {Tout}, \&
  {Eggleton}}]{1998MNRAS.298..525P}
{Pols}, O.~R., {Schr{\"o}der}, K.-P., {Hurley}, J.~R., {Tout}, C.~A., \&
  {Eggleton}, P.~P. 1998, \mnras, 298, 525

\bibitem[{{Puls} {et~al.}(2008){Puls}, {Vink}, \&
  {Najarro}}]{2008A&ARv..16..209P}
{Puls}, J., {Vink}, J.~S., \& {Najarro}, F. 2008, \aapr, 16, 209

\bibitem[{{Ram{\'{\i}}rez-Agudelo} {et~al.}(2015){Ram{\'{\i}}rez-Agudelo},
  {Sana}, {de Mink}, {H{\'e}nault-Brunet}, {de Koter}, {Langer}, {Tramper},
  {Gr{\"a}fener}, {Evans}, {Vink}, {Dufton}, \& {Taylor}}]{2015A&A...580A..92R}
{Ram{\'{\i}}rez-Agudelo}, O.~H., {Sana}, H., {de Mink}, S.~E., {et~al.} 2015,
  \aap, 580, A92

\bibitem[{{Reed}(2003)}]{2003AJ....125.2531R}
{Reed}, B.~C. 2003, \aj, 125, 2531

\bibitem[{{Renzo} {et~al.}(2017){Renzo}, {Ott}, {Shore}, \& {de
  Mink}}]{2017A&A...603A.118R}
{Renzo}, M., {Ott}, C.~D., {Shore}, S.~N., \& {de Mink}, S.~E. 2017, \aap, 603,
  A118

\bibitem[{{Richardson} {et~al.}(2016){Richardson}, {Allen}, {Baldwin},
  {Hewett}, {Ferland}, {Crider}, \& {Meskhidze}}]{2016MNRAS.458..988R}
{Richardson}, C.~T., {Allen}, J.~T., {Baldwin}, J.~A., {et~al.} 2016, \mnras,
  458, 988

\bibitem[{{Richer} {et~al.}(1998){Richer}, {Michaud}, {Rogers}, {Iglesias},
  {Turcotte}, \& {LeBlanc}}]{1998ApJ...492..833R}
{Richer}, J., {Michaud}, G., {Rogers}, F., {et~al.} 1998, \apj, 492, 833

\bibitem[{{Ritter}(1988)}]{1988A&A...202...93R}
{Ritter}, H. 1988, \aap, 202, 93

\bibitem[{{Roming} {et~al.}(2005){Roming}, {Kennedy}, {Mason}, {Nousek}, {Ahr},
  {Bingham}, {Broos}, {Carter}, {Hancock}, {Huckle}, {Hunsberger}, {Kawakami},
  {Killough}, {Koch}, {McLelland}, {Smith}, {Smith}, {Soto}, {Boyd},
  {Breeveld}, {Holland}, {Ivanushkina}, {Pryzby}, {Still}, \&
  {Stock}}]{Roming+2005}
{Roming}, P.~W.~A., {Kennedy}, T.~E., {Mason}, K.~O., {et~al.} 2005, \ssr, 120,
  95

\bibitem[{{Rosslowe} \& {Crowther}(2015)}]{2015MNRAS.447.2322R}
{Rosslowe}, C.~K. \& {Crowther}, P.~A. 2015, \mnras, 447, 2322

\bibitem[{{Sana} {et~al.}(2012){Sana}, {de Mink}, {de Koter}, {Langer},
  {Evans}, {Gieles}, {Gosset}, {Izzard}, {Le Bouquin}, \&
  {Schneider}}]{2012Sci...337..444S}
{Sana}, H., {de Mink}, S.~E., {de Koter}, A., {et~al.} 2012, Science, 337, 444

\bibitem[{{Sander} {et~al.}(2015){Sander}, {Shenar}, {Hainich},
  {G{\'{\i}}menez-Garc{\'{\i}}a}, {Todt}, \& {Hamann}}]{Sander+2015}
{Sander}, A., {Shenar}, T., {Hainich}, R., {et~al.} 2015, \aap, 577, A13

\bibitem[{{Schaerer} \& {de Koter}(1997)}]{1997A&A...322..598S}
{Schaerer}, D. \& {de Koter}, A. 1997, \aap, 322, 598

\bibitem[{{Schootemeijer} {et~al.}(2017, subm.){Schootemeijer}, {G{\"o}tberg},
  {de Mink}, {Gies}, \& {Zapartas}}]{abels_paper}
{Schootemeijer}, A., {G{\"o}tberg}, Y., {de Mink}, S.~E., {Gies}, D.~R., \&
  {Zapartas}, E. 2017, subm., \aap

\bibitem[{{Schroder} {et~al.}(1997){Schroder}, {Pols}, \&
  {Eggleton}}]{1997MNRAS.285..696S}
{Schroder}, K.-P., {Pols}, O.~R., \& {Eggleton}, P.~P. 1997, \mnras, 285, 696

\bibitem[{{Seaton}(1979)}]{1979MNRAS.187P..73S}
{Seaton}, M.~J. 1979, \mnras, 187, 73P

\bibitem[{{Shara} {et~al.}(2017){Shara}, {Crawford}, {Vanbeveren}, {Moffat},
  {Zurek}, \& {Crause}}]{2017MNRAS.464.2066S}
{Shara}, M.~M., {Crawford}, S.~M., {Vanbeveren}, D., {et~al.} 2017, \mnras,
  464, 2066

\bibitem[{{Shara} {et~al.}(1991){Shara}, {Smith}, {Potter}, \&
  {Moffat}}]{1991AJ....102..716S}
{Shara}, M.~M., {Smith}, L.~F., {Potter}, M., \& {Moffat}, A.~F.~J. 1991, \aj,
  102, 716

\bibitem[{{Simons} {et~al.}(2014){Simons}, {Thilker}, {Bianchi}, \&
  {Wyder}}]{2014AdSpR..53..939S}
{Simons}, R., {Thilker}, D., {Bianchi}, L., \& {Wyder}, T. 2014, Advances in
  Space Research, 53, 939

\bibitem[{{Smartt}(2009)}]{2009ARA&A..47...63S}
{Smartt}, S.~J. 2009, \araa, 47, 63

\bibitem[{{Smith} {et~al.}(1996){Smith}, {Shara}, \&
  {Moffat}}]{1996MNRAS.281..163S}
{Smith}, L.~F., {Shara}, M.~M., \& {Moffat}, A.~F.~J. 1996, \mnras, 281, 163

\bibitem[{{Smith}(2014)}]{2014ARA&A..52..487S}
{Smith}, N. 2014, \araa, 52, 487

\bibitem[{{Smith} {et~al.}(2018){Smith}, {G{\"o}tberg}, \& {de
  Mink}}]{2018MNRAS.475..772S}
{Smith}, N., {G{\"o}tberg}, Y., \& {de Mink}, S.~E. 2018, \mnras, 475, 772

\bibitem[{{Sota} {et~al.}(2014){Sota}, {Ma{\'{\i}}z Apell{\'a}niz}, {Morrell},
  {Barb{\'a}}, {Walborn}, {Gamen}, {Arias}, \& {Alfaro}}]{2014ApJS..211...10S}
{Sota}, A., {Ma{\'{\i}}z Apell{\'a}niz}, J., {Morrell}, N.~I., {et~al.} 2014,
  \apjs, 211, 10

\bibitem[{{Sota} {et~al.}(2011){Sota}, {Ma{\'{\i}}z Apell{\'a}niz}, {Walborn},
  {Alfaro}, {Barb{\'a}}, {Morrell}, {Gamen}, \& {Arias}}]{2011ApJS..193...24S}
{Sota}, A., {Ma{\'{\i}}z Apell{\'a}niz}, J., {Walborn}, N.~R., {et~al.} 2011,
  \apjs, 193, 24

\bibitem[{{Stanway} {et~al.}(2016){Stanway}, {Eldridge}, \&
  {Becker}}]{2016MNRAS.456..485S}
{Stanway}, E.~R., {Eldridge}, J.~J., \& {Becker}, G.~D. 2016, \mnras, 456, 485

\bibitem[{{Steiner} \& {Oliveira}(2005)}]{2005A&A...444..895S}
{Steiner}, J.~E. \& {Oliveira}, A.~S. 2005, \aap, 444, 895

\bibitem[{{Stevenson} {et~al.}(2017){Stevenson}, {Vigna-G{\'o}mez}, {Mandel},
  {Barrett}, {Neijssel}, {Perkins}, \& {de Mink}}]{2017NatCo...814906S}
{Stevenson}, S., {Vigna-G{\'o}mez}, A., {Mandel}, I., {et~al.} 2017, Nature
  Communications, 8, 14906

\bibitem[{{Stroeer} {et~al.}(2007){Stroeer}, {Heber}, {Lisker}, {Napiwotzki},
  {Dreizler}, {Christlieb}, \& {Reimers}}]{2007A&A...462..269S}
{Stroeer}, A., {Heber}, U., {Lisker}, T., {et~al.} 2007, \aap, 462, 269

\bibitem[{{Sukhbold} {et~al.}(2016){Sukhbold}, {Ertl}, {Woosley}, {Brown}, \&
  {Janka}}]{2016ApJ...821...38S}
{Sukhbold}, T., {Ertl}, T., {Woosley}, S.~E., {Brown}, J.~M., \& {Janka}, H.-T.
  2016, \apj, 821, 38

\bibitem[{{Tauris} {et~al.}(2017){Tauris}, {Kramer}, {Freire}, {Wex}, {Janka},
  {Langer}, {Podsiadlowski}, {Bozzo}, {Chaty}, {Kruckow}, {van den Heuvel},
  {Antoniadis}, {Breton}, \& {Champion}}]{2017ApJ...846..170T}
{Tauris}, T.~M., {Kramer}, M., {Freire}, P.~C.~C., {et~al.} 2017, \apj, 846,
  170

\bibitem[{{Tauris} \& {van den Heuvel}(2006)}]{2006csxs.book..623T}
{Tauris}, T.~M. \& {van den Heuvel}, E.~P.~J. 2006, {Formation and evolution of
  compact stellar X-ray sources}, ed. W.~H.~G. {Lewin} \& M.~{van der Klis},
  623--665

\bibitem[{{Thoul} {et~al.}(1994){Thoul}, {Bahcall}, \&
  {Loeb}}]{1994ApJ...421..828T}
{Thoul}, A.~A., {Bahcall}, J.~N., \& {Loeb}, A. 1994, \apj, 421, 828

\bibitem[{{Tramper} {et~al.}(2016){Tramper}, {Sana}, \& {de
  Koter}}]{2016ApJ...833..133T}
{Tramper}, F., {Sana}, H., \& {de Koter}, A. 2016, \apj, 833, 133

\bibitem[{{van Bever} \& {Vanbeveren}(1998)}]{van-Bever+1998}
{van Bever}, J. \& {Vanbeveren}, D. 1998, \aap, 334, 21

\bibitem[{{Van Bever} \& {Vanbeveren}(2003)}]{2003A&A...400...63V}
{Van Bever}, J. \& {Vanbeveren}, D. 2003, \aap, 400, 63

\bibitem[{{van den Heuvel} {et~al.}(2017){van den Heuvel}, {Portegies Zwart},
  \& {de Mink}}]{2017MNRAS.471.4256V}
{van den Heuvel}, E.~P.~J., {Portegies Zwart}, S.~F., \& {de Mink}, S.~E. 2017,
  \mnras, 471, 4256

\bibitem[{{van der Hucht}(2001)}]{van-der-Hucht2001}
{van der Hucht}, K.~A. 2001, \nar, 45, 135

\bibitem[{{Vanbeveren} {et~al.}(2007){Vanbeveren}, {Van Bever}, \&
  {Belkus}}]{2007ApJ...662L.107V}
{Vanbeveren}, D., {Van Bever}, J., \& {Belkus}, H. 2007, \apjl, 662, L107

\bibitem[{{Vink}(2017)}]{2017A&A...607L...8V}
{Vink}, J.~S. 2017, \aap, 607, L8

\bibitem[{{Vink} {et~al.}(2001){Vink}, {de Koter}, \&
  {Lamers}}]{2001A&A...369..574V}
{Vink}, J.~S., {de Koter}, A., \& {Lamers}, H.~J.~G.~L.~M. 2001, \aap, 369, 574

\bibitem[{{Vos} {et~al.}(2017){Vos}, {{\O}stensen}, {Vu{\v c}kovi{\'c}}, \&
  {Van Winckel}}]{2017A&A...605A.109V}
{Vos}, J., {{\O}stensen}, R.~H., {Vu{\v c}kovi{\'c}}, M., \& {Van Winckel}, H.
  2017, \aap, 605, A109

\bibitem[{{Walborn} {et~al.}(2002){Walborn}, {Howarth}, {Lennon}, {Massey},
  {Oey}, {Moffat}, {Skalkowski}, {Morrell}, {Drissen}, \&
  {Parker}}]{2002AJ....123.2754W}
{Walborn}, N.~R., {Howarth}, I.~D., {Lennon}, D.~J., {et~al.} 2002, \aj, 123,
  2754

\bibitem[{{Wang} {et~al.}(2017){Wang}, {Gies}, \&
  {Peters}}]{2017ApJ...843...60W}
{Wang}, L., {Gies}, D.~R., \& {Peters}, G.~J. 2017, \apj, 843, 60

\bibitem[{{Wang} {et~al.}(2018){Wang}, {Gies}, \&
  {Peters}}]{2018ApJ...853..156W}
{Wang}, L., {Gies}, D.~R., \& {Peters}, G.~J. 2018, \apj, 853, 156

\bibitem[{{Yoon}(2017)}]{2017MNRAS.470.3970Y}
{Yoon}, S.-C. 2017, \mnras, 470, 3970

\bibitem[{{Yoon} {et~al.}(2017){Yoon}, {Dessart}, \&
  {Clocchiatti}}]{2017ApJ...840...10Y}
{Yoon}, S.-C., {Dessart}, L., \& {Clocchiatti}, A. 2017, \apj, 840, 10

\bibitem[{{Yoon} {et~al.}(2010){Yoon}, {Woosley}, \&
  {Langer}}]{2010ApJ...725..940Y}
{Yoon}, S.-C., {Woosley}, S.~E., \& {Langer}, N. 2010, \apj, 725, 940

\bibitem[{{Zapartas} {et~al.}(2017){Zapartas}, {de Mink}, {Izzard}, {Yoon},
  {Badenes}, {G{\"o}tberg}, {de Koter}, {Neijssel}, {Renzo}, {Schootemeijer},
  \& {Shrotriya}}]{2017A&A...601A..29Z}
{Zapartas}, E., {de Mink}, S.~E., {Izzard}, R.~G., {et~al.} 2017, \aap, 601,
  A29

\bibitem[{{Zhang} {et~al.}(2004){Zhang}, {Han}, {Li}, \&
  {Hurley}}]{2004A&A...415..117Z}
{Zhang}, F., {Han}, Z., {Li}, L., \& {Hurley}, J.~R. 2004, \aap, 415, 117

\bibitem[{{Zhang} {et~al.}(2015){Zhang}, {Li}, {Cheng}, {Wang}, {Kang},
  {Zhuang}, \& {Han}}]{2015MNRAS.447L..21Z}
{Zhang}, F., {Li}, L., {Cheng}, L., {et~al.} 2015, \mnras, 447, L21

\bibitem[{{Zhang} {et~al.}(2012){Zhang}, {Li}, {Zhang}, {Kang}, \&
  {Han}}]{2012MNRAS.421..743Z}
{Zhang}, F., {Li}, L., {Zhang}, Y., {Kang}, X., \& {Han}, Z. 2012, \mnras, 421,
  743

\end{thebibliography}
